\documentclass[12pt,rotate]{article}

\usepackage{graphicx} 

\begin{document}

\title{
Meson and Baryon Spectroscopy on a Lattice.
}

\author{
Craig McNeile
}

\maketitle

\begin{abstract}
  I review the results of hadron spectroscopy calculations from
  lattice QCD for an intended audience of low energy hadronic
  physicists.  I briefly introduce the ideas of numerical lattice QCD.
  The various systematic errors, such as the lattice spacing and
  volume dependence, in lattice QCD calculations are discussed.  In
  addition to the discussion of the properties of ground state
  hadrons, I also review the small amount of work done on the
  spectroscopy of excited hadrons and the effect of electromagnetic
  fields on hadron masses. I also discuss the attempts
  to understand the physical mechanisms behind hadron mass splittings.
\end{abstract}
\section{INTRODUCTION}

QCD at low energies is hard to solve, perhaps too hard for mere
mortals to solve, even when assisted with the latest supercomputers. 
QCD is the theory that describes the interactions
of quarks and gluons. QCD has been well tested in high energy
scattering experiments where perturbation theory is valid. However,
QCD should also describe nuclear physics and the mass spectrum of
hadrons.  Hadron masses depend on the coupling ($g$) like $M \sim
e^{-1/g^2}$ hence perturbation theory can't be used to compute the
masses of hadrons such as the proton.

The only technique that offers any prospect of computing masses and
matrix elements non-perturbatively, from first principles, is lattice
QCD. In lattice QCD, QCD is transcribed to a lattice and the resulting
equations are solved numerically on a computer.  The computation of
the hadron spectrum using lattice QCD started in the early
80's~\cite{Weingarten:1982jy,Hamber:1981zn}.  The modern era in
lattice QCD calculations of the hadron spectrum started with the
results of the GF11 group~\cite{Butler:1993ki,Butler:1994em}.  The
GF11 group were the first to try to quantify the systematic errors in
taking the continuum and infinite volume limits.

The goal of a ``numerical solution'' to QCD is not
some kind of weird and misguided reductionist  quest.
Our inability to solve QCD has many profound consequences.  A major
goal of particle physics is to look for evidence for physics beyond
the standard model of particle physics. One way of doing this is to
extract the basic parameters of the standard model and look for
relations between them that suggest deeper structure. 
To test the quark sector of the standard model requires that  matrix
elements are computed from QCD~\cite{Beneke:2002ks}. 
The problem of solving QCD is symbolically summarised 
by the errors on the quark masses.
For example, the allowed range on the strange quark mass in the
particle data
table~\cite{Hagiwara:2002fs} is 80 to 155 MeV; a range of almost 100\%.
The value of top quark mass, quoted in the particle data table,  
is $174.3  \pm 5.1$ GeV
As the mass of the quark increases its relative error decreases.
The dynamics of QCD becomes simpler as the mass of the quarks gets
heavier.
Wittig has reviewed the latest results for the light quark masses
from lattice QCD~\cite{Wittig:2002ux}
Irrespective of applications of solutions to QCD to searches
for physics beyond the standard model, QCD is a fascinating theory
in its own right. QCD does allow us to test our meagre tools for
extracting non-perturbative physics from a field theory.

In this review I will focus on the results from lattice gauge theory
for the masses of the light mesons and baryons. 
I will not discuss flavour singlet mesons as these have been
reviewed by Michael~\cite{Michael:2003ai,Michael:2001rv}.
There has been much
work on the spectroscopy of hadrons that include heavy 
quarks~\cite{Davies:1997hv,Davies:2002cx,McNeile:2002uy}, however
I will not discuss this work. The treatment of heavy quarks (charm and bottom)
on the lattice has a different  set of problems and opportunities
over those for light quarks.  Although the spectroscopy 
of hadrons with heavy quarks in them can naturally be 
reviewed separately from light quark spectroscopy, the physics
of heavy hadrons does depend on the light quarks in the sea.
In particular the hyperfine splittings are known to have an important
dependence on the  sea quarks~\cite{McNeile:2002uy}.

Until recently, the computation of the light hadron spectrum used to
be just a test of the calculational tools of lattice QCD. The light
hadron spectrum was only really good for providing the quark masses
and estimates of the systematic errors.  However, the experimental
program at places such as the 
Jefferson lab~\cite{Rossi:2003np,Burkert:2002nr,Burkert:2001nv}
has asked for a new set of
quantities from lattice QCD. In particular the computation of the
spectrum of the $N^{\star}$'s is now a goal of lattice QCD
calculations.

As the aim of the review is to focus more on the results of lattice
calculations, I shall mostly treat lattice calculations as a
black box that produces physical numbers. However,
 ``errors are the kings'' of lattice QCD calculations
because the quality and usefulness of a result usually depends 
on the size of its error bar, hence I will discuss the systematic
errors in lattice calculations. 
Most of systematic errors in lattice QCD calculations can be understood 
using standard field theory techniques. 
I have also included an appendix~\ref{cmn:se:technicalDETAILS}.
on  some of the  ``technical tricks'' that
are important for lattice QCD insiders, but of limited interest
to consumers of lattice results. However, it is useful to know 
some of the jargon and issues, as they do effect the quality of 
the final results.

There are a number of text books on lattice QCD.  For example the
books by Montvay and Munster~\cite{Montvay:1994cy},
Rothe~\cite{Rothe:1997kp}, Smit~\cite{Smit:2002ug} 
and Creutz~\cite{Creutz:1984mg}
provide
important background information.  The large review articles by
Gupta~\cite{Gupta:1997nd}, Davies~\cite{Davies:2002cx} 
and Kronfeld~\cite{Kronfeld:2002pi}
also contain
pertinent information.  The annual lattice conference is a snap-shot
of what is happening in the lattice field every year.  The contents of
the proceedings of the lattice conference have been put on the hep-lat
archive for the past couple of
years~\cite{Davies:1998ur,DeGrand:1998gr,Mueller-Preussker:2002cp}.
The reviews of the baryon spectroscopy from lattice QCD 
by Bali~\cite{Bali:2003wj} 
and Edwards~\cite{Edwards:2002uz} describe a different perspective
on the field to  mine.
There used to be a plenary reviews 
specifically
on hadron spectroscopy at the
lattice
conference~\cite{Gottlieb:1997hy,Mawhinney:2000fw,Aoki:2000kp,Toussaint:2001zc,Kaneko:2001ux}.
The subject of hadron spectroscopy has now been split into a number of
smaller topics, such as quark masses.

If the reader wants to play a bit with some lattice QCD code, then the
papers by Di
Pierro~\cite{DiPierro:2000nt,DiPierro:1998kb,DiPierro:2000bd}, contain
some exercises and pointers to source code. The MILC collaboration
also make their code publicly available (try putting ``MILC
collaboration'' into a search engine).

\section{BASIC LATTICE GAUGE THEORY} \label{cmn:sec:basicLGT}

In this section, I briefly describe the main elements of numerical
lattice QCD calculations.  Quantum Chromodynamics (QCD) is the quantum
field theory that describes the interactions of elementary particles
called quarks and gluons.  The key aspect of a quantum field theory is
the creation and destruction of particles. This type of dynamics is
crucial to QCD and one of the reasons that it is a hard theory to
solve.

In principle, because we know the Lagrangian for QCD, the quantum
field theory formalism should allow us to compute any quantity.  The
best starting point for solving QCD on the computer is the path
integral formalism.  The problem of computing bound state properties
from QCD is reduced to evaluating equation~\ref{eq:reallyQCD}.
\begin{eqnarray}
\langle B \rangle & = & \frac{1}{{\cal Z}} \int dU  \int d\psi  \int d\overline{\psi}
\;
B
e^{-S_F - S_G }
\label{eq:reallyQCD} \\
{\cal Z} & =&  \int dU  \int d\psi  \int d\overline{\psi}
e^{-S_F - S_G }
\end{eqnarray}
where $S_F$ and $S_G$ are the actions for the 
fermion and gauge fields respectively.
The path integral is defined in Euclidean space for the
convergence of the measure.  

The fields in equation~\ref{eq:reallyQCD} fluctuate on
all distance scales. 
The short distance fluctuations need to be regulated. For computations
of non-perturbative quantities a lattice is introduced with
a lattice spacing that regulates short distance fluctuations.
The lattice regulator is useful both for numerical calculations,
as well for formal work~\cite{Glimm:1987ng} (theorem proving), because it provides
a specific representation of the path integral~\ref{eq:reallyQCD}.

A four dimensional grid of space-time points is introduced. A typical
size in lattice QCD calculations is $24^3 \;48$.  
The introduction of a hyper-cubic lattice
breaks Lorentz invariance, however this is restored as the continuum
limit is taken. The lattice actions do have a well defined
hyper-cubic symmetry group~\cite{Mandula:1983ut,Mandula:1984wb}.
The continuum QCD Lagrangian is transcribed
to the lattice using ``clever'' finite difference techniques.

In the standard lattice QCD formulation, 
the decision has been made to keep
gauge invariance explicit.
The quark fields are put on the sites 
of the lattice.
The gauge fields connect adjacent lattice points.
The connection between the gauge fields in lattice QCD
and the fields used in perturbative calculations is 
made via:
\begin{equation}
U_{\mu}(x) = e^{ -g i A_{\mu}(x)  } 
\label{eq:Connect}
\end{equation}
The gauge invariant objects are either products of gauge links
between quark and anti-quark fields, or products of gauge links 
that form closed paths. All gauge invariant operators in numerical
lattice QCD calculations are built out of such objects.
For example the lattice version of the 
gauge action 
is constructed from simple products of links called plaquettes.
\begin{equation}
U_P(x;\mu\nu) = U_{\mu}(x) U_{\nu}(x+\hat{\mu}) 
U_{\mu}(x+\hat{\nu})^{\dagger}
U_{\nu}(x)^{\dagger}
\end{equation}

The Wilson gauge action 
\begin{equation}
S_G = -\beta ( \sum_{p} \frac{1}{2N_c}  \mbox{Tr} (U_P + U_P^\dagger   ) -1  )
\label{eq:Wgaugeaction}
\end{equation}
is written in terms of plaquettes. $N_c$ is the number of colours.
The action in~\ref{eq:Wgaugeaction} can be 
expanded in the lattice spacing using equation~\ref{eq:Connect},
to get the continuum gauge action.
\begin{equation}
S_G = a^4 \frac{\beta}{4N_c} \sum_x  \mbox{Tr} (F_{\mu \nu} F^{\mu \nu} )
+ O(a^6)
\end{equation}
The coupling is related to $\beta$  via
\begin{equation}
\beta = \frac{2 N_c} {g^2}
\label{eq:bareCOUPLING}
\end{equation}
The coupling in equation~\ref{eq:bareCOUPLING} is known as the bare
coupling. More physical definitions of the
coupling~\cite{Lepage:1993xa} are typically used in perturbative
calculations.

The fermion action is generically written as
\begin{equation}
S_F = \overline{\psi} M \psi
\label{eq:Faction}
\end{equation}
where $M$ is called the fermion operator, a lattice approximation to
the Dirac operator.

One approximation to the fermion operator on the lattice is the Wilson
operator.  There are many new lattice fermion actions, however
the basic ideas can still be seen from the Wilson fermion operator.
\begin{equation}
S_{f}^{W}= \sum_{x}
(
\kappa \sum_{\mu }
\{
 \overline{\psi}_{x}(\gamma_{\mu }-1)U_{\mu }(x)\psi_{x+\mu }
-\overline{\psi}_{x+\mu}(\gamma_{\mu }+1) U_{\mu }^{\dagger}(x)\psi
_{x}
\}
+
\overline{\psi}_{x}\psi_{x}
)
\label{eq:wilsonFERMION}
\end{equation}
The Wilson action can be expanded in the lattice
spacing to obtain the continuum Dirac action with
lattice spacing corrections.
to the Dirac Lagrangian. 
The $\kappa$ parameter is called the hopping 
parameter. It is a simple rescaling factor 
that is related to the quark mass via
\begin{equation}
\kappa = \frac{1}{2 (4+m)}
\label{eq:kappaTOmass}
\end{equation}
at tree level.  
An expansion in $\kappa$ is not useful for light quarks,
because of problems with convergence, however
for a few specialised applications, it is convenient to expand 
in terms of $\kappa$~\cite{Henty:1992cw}.
The fermion operator in equation~\ref{eq:wilsonFERMION}
contains a term called the 
Wilson term that is required to remove fermion
doubling~\cite{Montvay:1994cy,Rothe:1997kp,Smit:2002ug}. 
The Wilson term explicitly breaks chiral symmetry so
equation~\ref{eq:kappaTOmass} gets renormalised.  Currently, there is a lot
of research effort in designing lattice QCD actions for fermions
with better theoretical properties.
I briefly describe some of these
developments in the appendix~\ref{cmn:se:technicalDETAILS}.

Most lattice QCD calculations obtain hadron masses from the time sliced
correlator $c(t)$.
\begin{equation}
c(t) = \langle \sum_{\underline{x}} e^{i p x} O(\underline{x} , t )  O(0 , 0)^\dagger
       \rangle
\label{eq:timeSLICEDCORR}
\end{equation}
where the average is defined in equation~\ref{eq:reallyQCD}.

Any gauge invariant combination of quark fields and gauge links can be
used as interpolating operators ($O(\underline{x},t)$) in 
equation~\ref{eq:timeSLICEDCORR}.
An example for an interpolating operator in equation~\ref{eq:timeSLICEDCORR}
for the $\rho$ meson would be
\begin{equation}
O(\underline{x},t)_i = \overline{\psi_1}(\underline{x},t)) 
\gamma_i 
\psi_2(\underline{x},t)) 
\label{eq:localOPER}
\end{equation}
The interpolating operator in equation~\ref{eq:localOPER} has the 
same $J^{PC}$ quantum numbers as the $\rho$.

The operator in equation~\ref{eq:localOPER} is local, as the 
quark and anti-quarks are at the same location. It has been
found to better to use operators that build in some kind
of ``wave function'' between the quark and anti-quarks.
In lattice-QCD-speak we talk about ``smearing'' the operator.
An extended operator such as 
\begin{equation}
O(\underline{x},0) = 
\sum_{\underline{r}} f(\underline{r})
\overline{\psi}_1(\underline{x},0)) 
\gamma_i 
\psi_2(\underline{x}+\underline{r},0)) 
\label{eq:NONlocalOPER}
\end{equation}
might have a better overlap to the $\rho$ meson than the 
local operator in equation~\ref{eq:NONlocalOPER}.  The $f(\underline{r})$
function is a wave-function like function.
The function $f$ is designed to give a better signal,
but the final results should be independent of the
choice of $f$.
Typical choices for $f$ might be hydrogenic wave functions.
Unfortunately equation~\ref{eq:NONlocalOPER} is not gauge 
invariant, hence vanishes by Elitzur's theorem~\cite{Elitzur:1975im}.
One way to use an operator such as equation~\ref{eq:NONlocalOPER}
is to fix the gauge. A  popular choice for spectroscopy calculations
is Coulomb's gauge.
On the lattice Coulomb's gauge is implemented by 
maximising
\begin{equation}
F = \sum_x \sum_{i=\hat{x}}^{\hat{z}} (  U_i(x) + U_i^\dagger(x))
\end{equation}
The  gauge fixing
conditions that are typically used in lattice QCD 
spectroscopy calculations
are either: Coulomb or Landau gauge.
Different gauges
are used in other types of lattice calculations.

Non-local operators can be measured in lattice QCD
calculations, 
\begin{equation}
O(\underline{x},0) = 
\sum_{\underline{r}} 
\overline{\psi}_1(\underline{x},0)) 
\gamma_i 
\psi_2(\underline{x}+\underline{r},0)) 
\label{eq:NONlocalOPERWVAVE}
\end{equation}
however these do not have any use in 
phenomenology~\cite{Velikson:1985qw,Gottlieb:1985xq,Duncan:1993eb,Hecht:1993ps,Hecht:1992uq},
because experiments are usually not sensitive to 
single $\overline{\psi}\psi$ Fock states.

There are also gauge invariant non-local operators that
are used in calculations.
\begin{equation}
O(\underline{x},t) = 
\overline{\psi}_1(\underline{x},t)
\gamma_i 
F(\underline{x}, \underline{x}+\underline{r})
\psi_2(\underline{x}+\underline{r},t)
\label{eq:FUZZEDNONlocalOPER}
\end{equation}
where 
$F(\underline{x}, \underline{x}+\underline{r})$
is a product of gauge links  between 
the quark and anti-quark.

There are many different paths between the quark and 
anti-quark.
It has been found useful to fuzz~\cite{Lacock:1995qx}.
the gauge links by adding bended paths:
\begin{equation}
U_{new} = P_{SU(3)} (c U_{old} + \sum_1^4 U_{u-bend}  )
\end{equation}
where $P_{SU(3)}$ projects onto SU(3).
There are other techniques for building up 
gauge fields between the quark and anti-quark fields,
based on 
computing a 
scalar propagator~\cite{Billoire:1985yn,Allton:1993wc}.

\begin{figure}
\def\filename{./prop.eps}
\begin{center}
\includegraphics[scale=0.6]{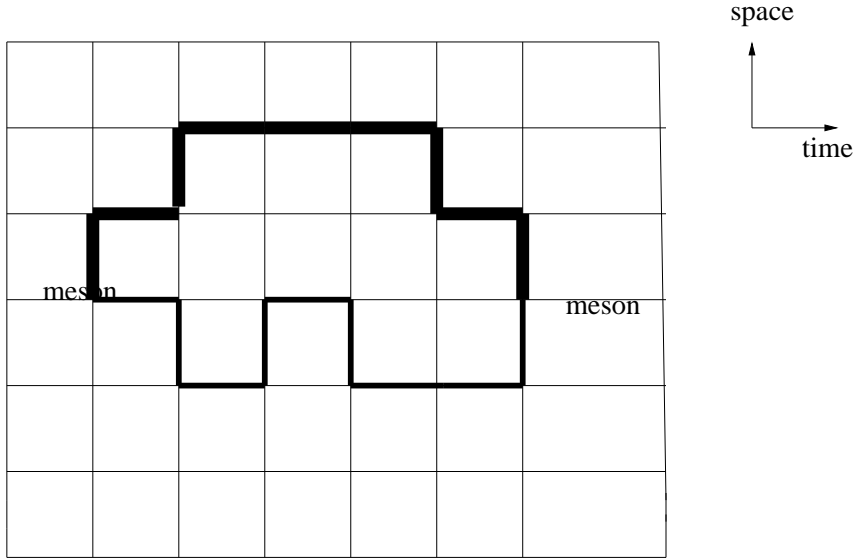}
\end{center}
   \caption{
Physical picture of a meson propagator
on the lattice. The thick lines are 
the quark propagators.
 }
\label{eq:MesonPROP}
\end{figure}

The path integral in equation~\ref{eq:reallyQCD} is evaluated using
algorithms that are  generalisations of the Monte Carlo methods
used to compute low dimensional integrals~\cite{Negele:1988vy}.
The physical picture for equation~\ref{eq:timeSLICEDCORR} is that a
hadron is created at time 0, from where it propagates to the time t,
where it is destroyed. This is shown in figure~\ref{eq:MesonPROP}.
Equation ~\ref{eq:timeSLICEDCORR} can be thought of as a meson propagator.
Duncan et al.~\cite{Duncan:2001cc}
have used the meson propagator representation to extract
couplings in a chiral Lagrangian. 

The algorithms, usually based on importance sampling, 
produce $N$ samples of the gauge fields on the lattice.
Each gauge field is a snapshot of the vacuum.
The QCD vacuum is a complicated structure. There
is a community of people who are trying to describe
the QCD vacuum in terms of objects such as a 
liquid of instantons (for example~\cite{Negele:1998ev}).
The lattice QCD community are starting to create
publicly available 
gauge 
configurations~\cite{McNeile:2000qm,Davies:2002mu}.
This is particularly important due to the high computational cost of 
unquenched calculations. An archive of gauge configurations has been
running at NERSC for many years.

The correlator $c(t)$ is a function of the 
light quark 
propagators ($M(U(i))_{q}^{-1}$) averaged
over the samples of the gauge fields.
\begin{equation}
c(t) \sim  \frac{1}{N}\sum
_{i}^{N}f(M(U(i))^{-1},M(U(i))^{-1}) 
\label{eq:algorEXPLAIN}
\end{equation}
The
quark propagator $M^{-1}$ is the inverse of the fermion operator.  
The quark propagator depends on the gauge configurations
from the gauge field dependence of the fermion action
(see~\ref{eq:wilsonFERMION} for example).

The correlator in equation~\ref{eq:algorEXPLAIN} is essentially
computed in two stages. First the samples of the gauge configurations
are generated. These are the building blocks of most
lattice QCD calculations. The correlators for specific
processes are computed by calculating the hadron correlator
for each gauge configuration and then averaging over each
gauge configuration.

To better explain the idea behind computing
quark propagators in lattice QCD calculations, it
is helpful to consider the computation of the 
propagator in perturbation theory.
The starting point of perturbative calculations 
is when the quarks do not interact with gluons. 
This corresponds to quarks moving in gauge potentials that
are gauge transforms from the unit configuration
\begin{equation}
A_\mu(x) = 0  \Rightarrow U_{\mu}(x) = 1
\label{eq:FreeField}
\end{equation}
Under these conditions
the quark propagator can be computed
analytically from the fermion operator
using Fourier transforms (see ~\cite{Rothe:1997kp} 
for derivation).
\begin{equation}
M^{unit} = \frac{1}{ \sum_{i=0}^4 \gamma_\mu  \sin(p_\mu)  + M(p)  }
\label{eq:freeProp}
\end{equation}
\begin{equation}
M(p) = m + \frac{2}{a} \sum_{\mu} sin^2(a p_\mu/1)
\end{equation}
Although in principle the propagator in equation~\ref{eq:freeProp}
could be used as a basis of a perturbative expansion of 
equation~\ref{eq:algorEXPLAIN}, the physical 
masses depend on the coupling like
$M \sim e^{-1/g^2}$ (see section~\ref{se:LatticeSpacing}), 
so hadron masses can not be computed using this approach.
The quark propagator on its own is not gauge invariant. Quark
propagators are typically built into gauge invariant hadron
operators. However, in a fixed gauge quark propagators can be 
computed and studied. This is useful for the attempts to calculate
hadron spectroscopy using Dyson-Schwinger 
equations~\cite{Maris:2003vk,Roberts:1994dr}.

In lattice QCD calculations the gauge fields have complicated
space-time dependence so the quark propagator is inverted numerically
using variants of the conjugate gradient algorithms~\cite{Frommer:1997ta}.
Weak coupling perturbation theory on the lattice is important for
determining weak matrix elements, quark masses and the strong
coupling.

The sum over $\underline{x}$ in the time slice correlator
(equation~\ref{eq:timeSLICEDCORR}) projects onto a specific momentum
at the sink. Traditionally for computational reasons, the spatial
origin had to be fixed either at a point, or with a specific wave
function distribution between the quarks.  Physically, it would be
clearly better to project out onto a specific momentum at the origin.
The number of spatial positions at the origin
is related to the cost of the
calculation. There are new lattice techniques called ``all-to-all''
that can be used to compute a quark propagator from any point to
another without spending prohibitive amounts of computer
time~\cite{deDivitiis:1996qx,Michael:1998sg,Duncan:2001ta}.

There have been some studies of the 
point to point correlator~\cite{Chu:1993cn,Hands:1995cj,DeGrand:2001tm}
\begin{equation}
\langle 
O(\underline{x} , t )  O(0 , 0)^\dagger
\rangle
\label{eq:pointTOpoint}
\end{equation}
It was pointed out by Shuryak~\cite{Shuryak:1993kg} that the correlator in
equation~\ref{eq:pointTOpoint} may be more easily compared
to experiment. Also, the correlator~\ref{eq:pointTOpoint} could
give some information on the density of states.
In practise the point to point correlators were quite noisy.

The physics from the 
time sliced correlator
is extracted using a 
fit model~\cite{Montvay:1994cy}:
\begin{equation}
c(t) = a_0 exp( -m_0 t ) + a_1 exp( -m_1 t ) + \cdots
\label{eq:fitmodel}
\end{equation}
where $m_0$ ($m_1$) is the ground (first excited) state
mass and the dots represent higher excitations.
There are simple corrections to equation~\ref{eq:fitmodel}
for the finite size of the lattice in the time direction.
In practise as recently emphasised by Lepage et
al.~\cite{Lepage:2001ym}, 
fitting masses from equation~\ref{eq:fitmodel}
is nontrivial. The fit region in time has to be 
selected as well as the number of exponentials.
The situation is roughly analogous to the choice 
of ``cuts'' in experimental particle physics.
The correlators are correlated in time, hence
a correlated $\chi^2$ should be minimised.
With limited statistics it can be hard to estimate the 
underlying covariance matrix making the $\chi^2/dof$ test
nontrivial~\cite{Michael:1994yj,Michael:1995sz}.
As ensemble sizes have increased the problems resulting
from poorly estimated covariance matrices have decreased.

It is usually better to measure a matrix of correlators
so that variational techniques can be used to extract the
masses and amplitudes~\cite{Michael:1989jr,Luscher:1990ck,Allton:1993wc}.
\begin{equation}
c(t)_{A \; B} = 
\langle \sum_{\underline{x}} e^{i p x} O(\underline{x} , t )_A  O(0 , 0)_B^{\dagger}
\rangle
\label{cmn:eq:timeSLICEDvary}
\end{equation}
The correlator in equation~\ref{cmn:eq:timeSLICEDvary} is analysed
using the fit model in equation~\ref{cmn:eq:timeSLICEDvaryanal}.
\begin{equation}
c(t)_{A \; B} = \sum_{N=1 .. N_0} X_{A \; N } e^{-m_N } X_{N \; A }
\label{cmn:eq:timeSLICEDvaryanal}
\end{equation}
The matrix $X$ is independent of time and in general
has no special structure. The masses can be extracted from
equation~\ref{cmn:eq:timeSLICEDvaryanal} either by fitting or
by reorganising the 
problem as a generalised eigenvalue problem~\cite{McNeile:2000xx}.
Obtaining the matrix structure on the right hand of
equation~\ref{cmn:eq:timeSLICEDvaryanal} is a non-trivial test 
of the multi-exponential fit.
One piece of ``folk wisdom'' is that the largest mass extracted from
the fit is some average over the truncated states, and hence is
unphysical. This may explain why the excited state masses obtained
from~\cite{Allton:1993wc} in calculation with only two basis
states were higher than any physical state

The main (minor) disadvantage of variational techniques is
that they require more computer time, because additional quark
propagators have to be computed, if the basis functions are 
``smearing functions''. If different local  interpolating
operators are used as basis states, as is done for studies 
of the Roper resonance (see section~\ref{cmn:se:baryonexcite}),
then there is no additional cost.
The amount of computer time
depends linearly on the number of basis states. Also the basis
states have to be ``significantly'' different to gain any benefit.
Apart from a few specialised applications~\cite{Draper:1994qj}
 the efficacy of a basis
state is not obvious until the calculations is done.

Although in principle excited state masses can be extracted from a
multiple exponential fit, in practise this is a numerically
non-trivial
task, because of the noise in the data from the calculation.
There is a physical argument~\cite{Lepage:1992ui,DeGrand:1992yx} 
that explains the 
signal to noise ratio. The variance of the correlator in 
equation~\ref{eq:timeSLICEDCORR} is:
\begin{equation}
\sigma_O^2 = \frac{1}{N} 
(
\langle (O(t)O(0))^2  \rangle
-c(t)^2
\end{equation}
The square of the operator will couple to two (three) pion for a 
meson (baryon). Hence the noise to signal  ratio for mesons is 
\begin{equation}
\frac{\sigma_M(t) }{c(t) }
\sim
e^{(m_M - m_\pi)t}
\end{equation}
and 
noise to signal ratio for baryons
is
\begin{equation}
\frac{\sigma_B(t) }{c(t) }
\sim
e^{(m_B - 3/2 m_\pi)t}
\end{equation}
The signal to noise ratio will get worse as the 
mass of the hadron increases.

\begin{figure}
\def\filename{meff_proton.ps}
\begin{center}
\includegraphics[scale=0.6]{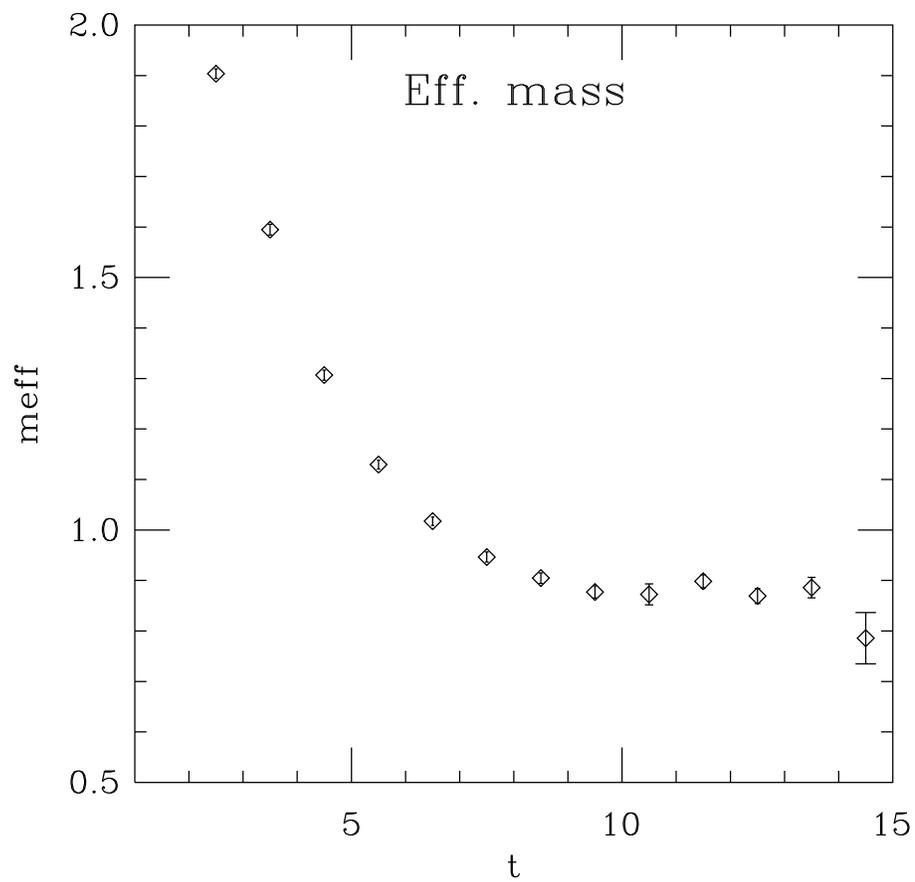}
\end{center}
   \caption{
Effective mass plot for proton correlators.
Unquenched data from 
UKQCD~\cite{Allton:2001sk}
with $\kappa$=0.1350 , $\beta$ = 5.2.
 }
\label{eq:protonCORR}
\end{figure}

In figure~\ref{eq:protonCORR} I show some data for a proton correlator.
The traditional way to plot a correlator is to use the effective 
mass plot.
\begin{equation}
m_{eff} = \log( \frac{c(t) } {c(t+1)} )
\label{eq:effectiveMASS}
\end{equation}
When only the first term in equation~\ref{eq:fitmodel} dominates,
then equation~\ref{eq:effectiveMASS} is a constant. There is a simple
generalisation of equation~\ref{eq:effectiveMASS} for 
periodic boundary conditions in time. 

In the past the strategy used to be fit far enough out
in time so that only one exponential contributed 
to equation~\ref{eq:fitmodel}. One disadvantage of doing this 
is that the noise increases at larger times, hence the 
errors on the final physical parameters become 
larger~\cite{Lepage:2001ym}.

Many of the most interesting questions in hadronic physics
involve the hadrons that are not the ground state with
a given set of quantum numbers, hence novel techniques 
that can extract masses of excited states
are very important.

Recently, there has been a new set of tools developed to extract physical
parameters from the correlator $c(t)$ in equation~\ref{eq:fitmodel}.
The new techniques are all based on the spectral representation of 
the correlator in equation~\ref{eq:timeSLICEDCORR}.
\begin{equation}
c(t) = \int_0^\infty \rho(s) e^{-st} ds
\end{equation}
The fit model in equation~\ref{eq:fitmodel} corresponds to a spectral
density of 
\begin{equation}
\rho(s) = a_0 \delta (s-m_0) + a_1 \delta (s-m_1)
\end{equation}
Spectral densities are rarely a sum of poles, except when
the particles can't decay such as in the $N_c \rightarrow \infty$
limit. Shifman reviews some of the physics in the spectral
density~\cite{Shifman:1999mk}.
 As the energy
increases it is more realistic to represent the spectral function as a 
continuous function.
Leinweber investigated a sum rule inspired approach to studying hadron
correlators~\cite{Leinweber:1995gt} on the lattice.  Allton and
Capitani~\cite{Allton:1998ax} generalised the sum rule analysis to
mesons.  The sum rule inspired spectral density is
\begin{equation}
\rho(s) = \frac{Z}{2 M}  \delta (s-M)
+
\theta( s - s_0) \rho^{OPE} (s)
\end{equation}
The operator product expansion is used to obtain
the spectral density. 
\begin{equation}
 \rho^{OPE} (s) = \sum_{n=1}^{n_0}
\frac{s^{n-1}}{(n-1)!} . C_n O_n(m_q, 
\langle :\overline{q}q: \rangle ) 
\label{cmn:eq:opespect}
\end{equation}
where $m_q$ is the quark mass and 
$:\overline{q}q:$ is the quark condensate.
The $C_n$ factors in equation ~\ref{cmn:eq:opespect} are
obtained by putting the OPE expression for the 
quark propagator in~\ref{eq:timeSLICEDCORR}.
The final functions that are fit to the lattice data involve
exponentials multiplied by polynomials in $1/t$ (where $t$ is the
time).
These kind of methods can also help to understand
some of the systematic
errors in sum rule calculations~\cite{Leinweber:1995gt}.
This hybrid lattice sum rule analysis is not in wide use
in the lattice community. See~\cite{Allton:2002mr} for recent developments.
There has been an interesting calculation of the mass of the 
charm quark using sum rule ideas and lattice QCD 
data~\cite{Bochkarev:1996ai}.

There is currently a lot of research into using the
maximum entropy method to extract masses from lattice
QCD calculations. Here I follow the description of the 
method used by CP-PACS~\cite{Yamazaki:2001er}. There are 
other variants
of the maximum entropy method
in use.
The maximum entropy approach~\cite{Nakahara:1999vy}
is based on Bayes theorem.
\begin{equation}
P(F\mid DH) \propto P(D \mid FH) P(F \mid H)
\label{eq:cmn:bayesTheorem}
\end{equation}
In equation~\ref{eq:cmn:bayesTheorem} $F$ is the spectral
function, $D$ is the data, and $H$ is the prior information
such as $f(w) > 0$, and 
$P(F\mid DH)$ is the conditional probability of 
getting $F$ given the data $D$ and priors $H$.
The replacement of the fit model in equation~\ref{eq:fitmodel}
is
\begin{equation}
c_f(t) =\int_0^\infty K(w,t) f(w) dw
\end{equation}
where the kernel $K$ is defined by
\begin{equation}
K(w,t) = e^{- w t} + e^{-w(T-t))}
\end{equation}
with $T$ the number of time slices in the time direction.

\begin{eqnarray}
P(D \mid F H ) & = & \frac{1}{Z_L} e^{-L} \label{cmn:fig:ChiSq} \\
L  & = & \frac{1}{2} \sum_{i,j}^{N_D} 
(c(t_i) -c_f(t_i)   ) D^{-1}_{i j} 
(c(t_j) -c_f(t_j)   )
\end{eqnarray}
where $Z_L$ is a known normalisation constant
and $N_D$ is the number of time slices to use.
The correlation matrix is defined by
\begin{equation}
D_{i j} = \frac{1}{N(N-1)}
          \sum_{n=1}^{N} ( c(t_i) -c^n(t_i)   ) ( c(t_j) -c^n(t_j) ) 
\end{equation}
\begin{equation}
c(t) = \frac{1}{N} \sum_{n=1}^{N} c^n(t_i)
\end{equation}
where N is the number of configurations and 
$c^n(t)$ is the correlator for the n-th configuration.
Equation~\ref{cmn:fig:ChiSq} is the standard 
expression for the likelihood used in the least square analysis.

The prior probability depends on a function $m$
and a parameter $\alpha$.
\begin{equation}
P(F \mid H m \alpha) = \frac{1}{Z_s(\alpha)} e^{\alpha S}
\end{equation}
with $Z_s(\alpha)$ a known constant.
The entropy $S$ is defined by:
\begin{equation}
S \rightarrow \sum_{i-1}^{N_w} [ f_i - m_i 
-f_i \log (\frac{f_i}{m_i})  ]
\end{equation}
The $\alpha$ parameter, 
that determines the weight between the entropy and $L$,
is obtained by
a statistical procedure. 
The physics is built into the $m$ function.
CP-PACS build some constraints from perturbation
theory into the function $m$.
The required function $f$ is obtained from the condition
\begin{equation}
\frac{\delta( \alpha S(f) -  L   )}{\delta f }\mid_{f=f_w} = 0 
\end{equation}
In principle the maximum entropy method offers the prospect of
computing the masses of excited states in a more stable way than
fitting multi-exponentials to data.  It will take some time to gain
more experience with these types of methods.  I review some of the
results from these techniques in sections~\ref{se:mesonexcite}
and~\ref{cmn:se:baryonexcite}.

A more pragmatic approach to constraining the fit parameters
was proposed by Lepage et al.~\cite{Lepage:2001ym}. The basic
idea is to constrain the parameters of the fit model
into physically motivated ranges.
The standard $\chi^2$ is modified to the $\chi^2_{aug}$.
\begin{equation}
\chi^2_{aug} = \chi^{2} + \chi^{prior}
\end{equation}
with the prior $\chi^2_{prior}$ defined by
\begin{equation}
\chi_{aug}^2 \equiv
\sum_{n} \frac{( a_n -\hat{a_n})^2 } { \hat{\sigma_{a_n}} }
+
\sum_{n} \frac{( m_n -\hat{m_n})^2 } { \hat{\sigma_{m_n}} }
\label{smn:eq:lepagechisq}
\end{equation}
The priors constrain the fit parameters to 
$m_n$ = $\hat{m_n} \pm \hat{\sigma_{m_n}}$
and $a_n$ = $\hat{a_n} \pm \hat{\sigma_{a_n}}$,
The derivation of  equation ~\ref{smn:eq:lepagechisq}
comes from Bayes theorem plus the assumption that the
distribution of the prior distribution is Gaussian
The idea is to ``teach'' the fitting algorithm what
the reasonable ranges for the parameters are.
Lepage et al.~\cite{Lepage:2001ym} do test the 
sensitivity of the final results
against the priors.

It is difficult to understand why certain things are
done in lattice calculations without an appreciation of the 
computational costs of lattice calculations.
The SESAM collaboration~\cite{Lippert:2002jm} estimated that the
number
of floating point operations ($N_{flop}$) needed for $n_{f}$ =2
full QCD calculations as:
\begin{equation}
N_{flop} \propto (\frac{L}{a})^{5}(\frac{1}{am_{pi}})^{2.8}
\label{eq:manyFLOPS}
\end{equation}
for a box size of $L$, lattice spacing $a$, and $N_{sample}$ is the
number of sample of the gauge fields in equation~\ref{eq:timeSLICEDCORR}.
A flop is a floating point operation
such as multiplication.
With  appropriate normalisation factors 
equation~\ref{eq:manyFLOPS} shows how ``big'' a computer
is required and how long the programs should run for.
Equation~\ref{eq:manyFLOPS} shows that it is easy
to reduce the statistical errors on the calculation as they go like
$\sim \frac{1}{\sqrt{N_{sample}}}$, but more expensive to change
the lattice spacing or quark mass. 

In some sense equation~\ref{eq:manyFLOPS} (or some variant of it) is the 
most important equation in numerical lattice QCD. To half the value
of the pion mass used in the calculations requires essentially 
a computer that is seven times faster.
Equation~\ref{eq:manyFLOPS} is not a hard physical limit.
Improved algorithms or techniques may be cheaper. 
There are disagreements between the various collaborations that do
dynamical calculations as to the exact cost of the simulations.  In
particular, a formulation of quarks on the lattice called improved
staggered fermions~\cite{Gottlieb:2001cf} 
has a much less pessimistic computational 
cost. The reason for this is not currently understood.

\section{SYSTEMATIC ERRORS}

The big selling point of lattice QCD calculations is that the
systematic errors can in principle be controlled.  To appreciate
lattice QCD calculations and to correctly use the results, the
inherent systematic errors must be understood.

For computational reasons, an individual lattice calculation is done
at a finite lattice spacing in a box of finite size with quarks that
are heavy relative
to the physical quarks.
The final results from several calculations are then
extrapolated to the continuum and infinite volume limit.  It is
clearly important that the functional forms of the extrapolations are
understood.  Most of the systematic errors in lattice QCD calculations
can be understood in terms of effective field
theories~\cite{Kronfeld:2002pi}.  A particularly good discussion of the
systematic errors from lattice QCD calculations in light hadron mass
calculations is the review by Gottlieb~\cite{Gottlieb:1997hy}.

The MILC collaboration make an amusing observation pertinent to error
analysis~\cite{Bernard:1993an}. In the heavy quark limit, the ratio of
the nucleon mass to rho mass is $3/2$, requiring one flop to
calculate.  This is within 30 \% of the physical number (1.22). This
suggests for a meaningful comparison at the $3 \sigma$ level, the
error bars should be at the 10\% level.

\subsection{UNQUENCHING} \label{cmn:se:unquench}

The high computational cost of the fermion determinant led
to development of quenched QCD, where the dynamics of the determinant 
is not included in equation~\ref{eq:reallyQCD}, hence the dynamics
of the sea quarks is omitted.  Until recently the majority
of lattice QCD calculations were done in quenched QCD.
I will call a lattice calculation unquenched,
when the dynamics of the sea quarks are 
included.

The integration over the quark fields in equation~\ref{eq:reallyQCD} can be
done exactly using Grassmann integration. The determinant is nonlocal
and the cause of most of the computational expense in
equation~\ref{eq:manyFLOPS}.  The determinant describes the dynamics of the
sea quarks.  Quenched QCD can be thought of as corresponding to using
infinitely heavy sea quarks.

\begin{figure}
\def\filename{loop.eps}
\begin{center}
\includegraphics[scale=0.6]{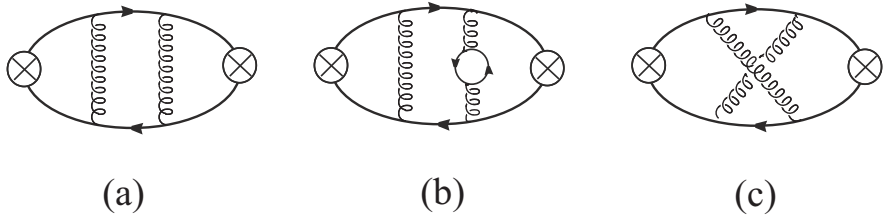}
\end{center}
\caption{
Different contributions to the pion correlator
from Chen~\cite{Chen:2002mj}}
\label{eq:quenchedLARGEn}
\end{figure}

Chen~\cite{Chen:2002mj} has made some interesting observations on the 
connection between quenched QCD and the large $N_c$ limit of QCD.
Figure~\ref{eq:quenchedLARGEn} shows some graphs of the pion 
two point function.
In unquenched QCD all three diagrams in figure~\ref{eq:quenchedLARGEn}
contribute to the two point function.
 In quenched QCD, only diagrams (a) and 
(c) contribute.
In the large $N_c$ (number of colours) limit only graphs of type (a)
in figure ~\ref{eq:quenchedLARGEn} contribute.
This argument suggests that in the large $N_c$ limit quenched
QCD and unquenched QCD should agree, hence for the real world $SU(3)$
case, quenched QCD and unquenched QCD should differ by $O(1/N_c)$ 
$\sim$ 30\%. Chen~\cite{Chen:2002mj} firms up this heuteristic argument 
by power counting factors of $N_c$ and discusses the effect of chiral
logs. This analysis suggests that the quenching error should be 
roughly 30\%, unless there is some cancellation such that the leading
$O(1/N_c)$ corrections cancel.

Quenched QCD is not a consistent theory, because
omitting the fermion loops causes 
problems with unitarity.
Bardeen~\cite{Bardeen:2001jm} et al. 
have shown that there is a problem with the 
non-singlet $0^{++}$ correlator in quenched QCD.
The problem can be understood using quenched chiral 
perturbation theory. The non-singlet $0^{++}$ propagator
contains an intermediate state of $\eta'-\pi$. The removal of fermion
loops in quenched QCD has a big effect on the $\eta'$ propagator.
The result is that a ghost state contributes to the scalar correlator,
that makes the expression in equation~\ref{eq:fitmodel} inappropriate to 
extract masses from the calculation.
Bardeen et al.~\cite{Bardeen:2001jm} predict that the ghost state will
make the $a_0$ mass increase as the quark mass is reduced below 
a certain point.
This behaviour was
observed by Weingarten and Lee~\cite{Lee:1999kv} 
for small box sizes (L $\le$ 1.6 fm).
The negative scalar correlator 
was also seen by DeGrand~\cite{DeGrand:2001tm} 
in a study of point to point correlators.
Damgaard et al.~\cite{Damgaard:2001js} also discuss
the scalar correlator in quenched QCD.

One major problem with quenched QCD is that it does not suppress zero
eigenvalues of the fermion operator~\cite{Bardeen:1998gv}.  A quark
propagator is the inverse of the fermion operator, so eigenvalues of
the fermion operator that are zero, or close to zero, cause problems
with the calculation of the quark propagator.  In unquenched QCD,
gauge configurations that produce zero eigenvalues in the fermion
operator are suppressed by the determinant in the measure.  Gauge
configurations in quenched QCD that produce an eigenvalue spectrum
that cause problems for the computation of the propagator are known as
``exceptional configurations''.  Zero modes of the fermion operator
can be caused by topology structures in the gauge configuration. 
The problem with exceptional configurations get worse as the 
quark mass is reduced.
The new class of actions,
described in the appendix~\ref{cmn:se:technicalDETAILS}
that have better chiral symmetry properties,
do not have problems with exceptional configurations

Please note that this section should not be taken as an apology for
quenched QCD.  As computers and algorithms get faster the parameters
of unquenched lattice QCD are getting ``closer'' to their physical
values. Hopefully quenched QCD calculations will fade away.

\subsection{LATTICE SPACING ERRORS} \label{se:LatticeSpacing}

Lattice QCD calculations produce results in
units of the lattice spacing. One experimental number must be 
used to calculate the lattice spacing from: 
\begin{equation}
a =
am_{latt}^{X} / m_{expt}^{X}
\end{equation}
As the lattice spacing goes to
zero any choice of $m_{expt}^{X}$ should produce the same lattice
spacing -- this is known scaling. Unfortunately, no calculations are 
in this regime yet. The recent unquenched calculations by the 
MILC 
collaboration~\cite{Bernard:2001av,Bernard:1999xx,Gray:2002vk} may be close.

Popular choices to set the scale are the mass of the rho,
mass splitting between the S and P wave mesons in charmonium,
and a quantity defined from the lattice potential called $r_0$.
The quantity $r_0$~\cite{Sommer:1994ce,Bali:2000gf} is 
defined in terms of the static potential $V(r)$ measured on the 
lattice.
\begin{equation}
r_{0}^{2} \frac{dV}{dr}\mid _{r_{0}} = 1.65
\end{equation}
Many potential~\cite{Sommer:1994ce}
models predict $r_{0}$ $\sim $ 0.5 fm.
The value of $r_{0}$  can not be measured experimentally,
but is ``easy'' to measure on the lattice.
The value of $r_{0}$ is a modern generalisation of the string tension.
Although it may seem a little strange to use $r_0$ to calculate the
lattice spacing, when it is not directly known from experiment.  There
are problems with all methods to set the lattice spacing.  For
example, to set the scale from the mass of the rho meson requires a
long extrapolation in light quark mass. Also it is not clear how to
deal with the decay width of the rho meson in Euclidean space.

The physics results from lattice calculations
should be independent of the lattice spacing.
A new lattice spacing is obtained by running at a different
value of the coupling in equation~\ref{eq:bareCOUPLING}.
In principle the dependence of quantities on
the coupling can be determined from renormalisation group
equations:
\begin{equation}
\beta(g_0) = a \frac{d}{da} g_0 (a)
\end{equation}
The renormalisation group equations can be solved 
to give the dependence of the lattice spacing on the 
coupling~\cite{Creutz:1983ev}.
\begin{equation}
a = \frac{1}{\Lambda} 
(g_0^2 \gamma_0)^{-\gamma_1 / (2 \gamma_0^2) }
e^{( -1/ (2 \gamma_0 g_0^2 )   ) }( 1 + O(g_0^2))
\label{eq:asymp}
\end{equation}
The $e^{ -1/ (2 \gamma_0 g_0^2 )} $ term in equation~\ref{eq:asymp}
prevents any weak coupling expansion converging for masses.

Equation~\ref{eq:asymp} is not often used in lattice QCD calculations.
The bare coupling $g_0$ does not produce very convergent series.
If quantities, such as Wilson loops
are computed in perturbation theory and from numerical
lattice calculations
the agreement between the two methods is very poor.
Typically couplings
defined in terms of more ``physical'' quantities, such as the
plaquette are used in lattice perturbative calculations~\cite{Lepage:1993xa}.

\begin{figure}
\def\filename{fig_mrho.ps}
\begin{center}
\includegraphics[scale=0.6]{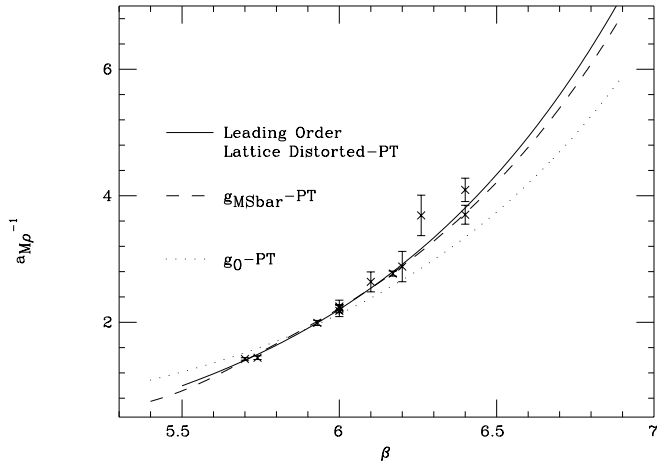}
\end{center}
   \caption{
The lattice spacing determined from the mass of the rho meson as
a function of the coupling ($\beta=\frac{6}{g^2}$)~\cite{Allton:1996kr} }
\label{eq:rhoMASS}
\end{figure}

Allton~\cite{Allton:1996kr}
tried to use variants of~\ref{eq:asymp} with 
some of the improved couplings. He also tried to model
the effect of the irrelevant operators. An example of 
Allton's result for the lattice spacing determined
from the $\rho$ mass as a function 
of the inverse coupling
is in figure~\ref{eq:rhoMASS}.

There are corrections to equation~\ref{eq:asymp} from lattice
spacing errors. For example the Wilson fermion action in
equation~\ref{eq:wilsonFERMION} differs from the continuum Dirac
action by errors that are $O(a)$. The lattice spacing corrections can
be written in terms of operators in a Lagrangian. These operators are
known as irrelevant.
Ratios of dimensional quantities are extrapolated to the
continuum limit using a simple polynomial.
\begin{equation}
\frac{a m_1}{ a m_2} = 
\frac{a m_1^{cont}}{ a m_2^{cont}} 
+ x a
+ O(a^2) 
\end{equation}
The improvement program discussed in appendix~\ref{cmn:se:technicalDETAILS}
designs
fermion actions to reduce the lattice spacing dependence
of ratios of dimensional quantities.
When computationally feasible,
calculations are done at (least) three lattice spacings and the results
are extrapolated to the continuum. This was the strategy
of the large scale GF11~\cite{Butler:1993ki} 
and CP-PACS calculations~\cite{Aoki:2002fd}.

It is very important to know the exact functional form of 
the lattice spacing dependence of the lattice results for a 
reliable extrapolation to the continuum. 
The quantum field theory nature of the 
field renormalises the correction polynomial. There are potentially
$O(a^n g^m)$ errors.
This is particularly a problem for states with a mass
that is comparable to the inverse lattice spacing.
There is physical argument that the Compton wavelength of a
hadron should be greater than the lattice spacing. This is a problem
for heavy quarks such as charm and bottom, that is usually solved
by the use of effective field
theories~\cite{Davies:1997hv,Davies:2002cx,Kronfeld:2002pi}. 
The masses of the excited light hadrons
are large relative to the inverse lattice spacing, so there may be
problems with the continuum extrapolations.

There have been a number of cases
where problems with continuum extrapolations have been 
found. For example Morningstar and Peardon~\cite{Morningstar:1999rf}  
found that the mass of the 
$0^{++}$ glueball had a very strange dependence 
on the lattice 
spacing.
Morningstar and Peardon~\cite{Morningstar:1999dh}
had to modify the gauge action to obtain results 
that allowed a controlled continuum extrapolation.
The ALPHA collaboration~\cite{Guagnelli:1999gu} 
discuss the problems of 
extrapolating the renormalisation constant associated with the
operator corresponding
to the average momentum of non-singlet parton densities
to the continuum limit, when the exact lattice spacing
dependence is not known.

In principle the formalism of lattice gauge theory does not put a
restriction on the size of the lattice spacing used.  The
computational costs of lattice calculations are much lower at larger
lattice spacings (see equation~\ref{eq:manyFLOPS}).  However, there
may be a minimum lattice spacing, set by the length scale of the
important physics, under which the lattice calculations become
unreliable~\cite{DeGrand:2000gq}.

There has been a lot of work to validate the instanton liquid model on
the lattice~\cite{Schafer:1998wv}. 
Instantons are semi-classical objects in the gauge fields.
The instanton liquid model models the gauge dynamics
with a collection of instantons of different sizes.
Some lattice studies claim to have
determined that there is a peak in the distribution of the size of the
instantons between 0.2 to 0.3 fm in quenched
QCD~\cite{Teper:1999wp,Hasenfratz:1998qk,Hasenfratz:1999ng}.
If the above estimates are correct, then lattice spacings of 
at least 0.2 fm would be required to correctly include the
dynamics of the instanton liquid on the lattice.
However, determining the instanton content of a gauge 
configuration is non-trivial, so estimates of size 
distributions are controversial. At least one group claims
to see evidence in gauge configurations against 
the instanton liquid model~\cite{Horvath:2001ir,Horvath:2002gk}.

There have been a few calculations of the hadron spectrum on lattices
with lattice spacings as coarse as 0.4
fm~\cite{Morningstar:1999rf,Alford:1995hw,Alford:1998yy,Alford:2000mm,DeGrand:1998pr}.
There were no problems reported in these coarse lattice calculations,
however only a few quantities were calculated.

\begin{figure}
\def\filename{r0_vs_k.ps}
\begin{center}
\includegraphics[scale=0.5]{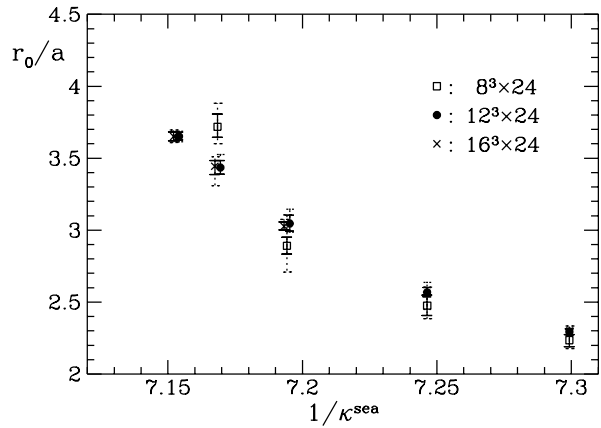}
\end{center}
   \caption{Lattice spacing dependence on the 
sea quark mass $1/\kappa_{sea}$
from ~\cite{Allton:1998gi}.
 }
\label{cmn:eq:r0withKappa}
\end{figure}

Another complication for unquenched calculations 
is that the lattice spacing depends on the 
value of the sea quark mass~\cite{Allton:1998gi}.
as well as the coupling. 
This is shown in figure~\ref{cmn:eq:r0withKappa}.
The dependence of the lattice spacing on the sea
quark mass is a complication, because for example the 
physical box size now depends on the quark mass.

Some groups prefer to tune the input parameters in their calculations 
so that the lattice 
spacing~\cite{Bernard:2001av,Irving:1998yu} is fixed
with varying sea quark mass.
Other groups prefer to work with a fixed bare 
coupling~\cite{Aoki:2000kp}.

\subsection{QUARK MASS DEPENDENCE}  \label{eq:massDEPEND}

The cost of lattice QCD calculations (see equation~\ref{eq:manyFLOPS})
forces the calculations to be done at unphysical large quark masses.
The results of lattice calculations are extrapolated to physical quark
masses using some functional form for the quark mass dependence. The
dependence on the light quark masses is motivated from effective field
theories such as chiral perturbation theory.  The modern view is
that the lattice QCD calculations do not need to be done at exactly
the masses of the physical up and down quarks, but the results can be
matched onto an effective theory~\cite{Sharpe:1998hh}.
In this section I briefly discuss the chiral
extrapolation fit models used to extrapolate hadron masses from lattice QCD
data to the physical points.  

The relationship between effective theories and lattice QCD is
symbiotic, because the results from lattice QCD calculations can also
test effective field theory methods.  The extrapolation of lattice QCD
data with light quark masses is currently a ``hot'' and controversial
topic in the lattice QCD community. The controversy is over the
different ways to improve the convergence of the effective theory.
There was a discussion panel on the different perspectives on using
effective field theories to analyse lattice QCD data, at the lattice
2002 conference~\cite{Bernard:2002yk}.  In section~\ref{se:params},
I show that the masses of the quarks used in lattice
calculations are getting lighter, that will help alleviate
many problems.

In the past chiral perturbation theory provided
functional forms that could used to analyse the quark mass
dependence of lattice data. In practise, the extrapolations
of lattice data were mostly done assuming a linear dependence 
on the quark mass~\cite{Butler:1993ki},
The  mass dependence from effective field theory
was essentially treated like a black box.
However, a number of high profile 
studies~\cite{Detmold:2001jb} have shown that a deeper understanding of the 
physics behind the expressions from chiral perturbation theory
is required.
The subject of effective
Lagrangians for hadronic physics is very large.
The reviews by Georgi~\cite{Georgi:1993qn}, Lepage~\cite{Lepage:1997cs} 
and Kaplan~\cite{Kaplan:1995uv} 
contain introductions to the physical ideas behind
effective 
field theory calculations.
Scherer~\cite{Scherer:2002tk} reviews chiral perturbation theory
for mesons and baryons.

The chiral perturbation theory for the lowest pseudoscalar particles
is the perhaps the most ``well defined'' theory. The effective field
theories for the vectors and baryon particles are perhaps more subtle
with concerns about convergence of the expansion.
In this section, I will
briefly describe effective field theories, starting with the
pseudoscalars and working up to nucleons via the vector particles.
I will try to explain some of the physical pictures
behind the functional dependence from effective field
theories, rather than provide an exhaustive list of 
expressions to fit to.

In Georgi's review~\cite{Georgi:1993qn} of the 
basic ideas behind effective theory,
he quotes the essential principles behind effective field theory as

\begin{itemize}

\item A local Lagrangian is used.

\item There are a finite numbers of parameters that describe
      the interaction each of dimension $(k-4)$.

\item The coefficients of the interaction terms of dimension 
      $(k-4)$ is less than or of the order of $(\frac{1}{M^{k}})$
      where $E < M$ for some mass independent $k$.

\end{itemize}

The above principles allow that physical processes to be 
calculated to an accuracy of $(\frac{E}{M})^k$ for a process
of energy $E$. The energy scale $M$ helps to organise the 
power counting. The accuracy of the results can be improved 
in a systematic way if
more parameters are included. This is one of the main 
appeals of the formalism.
As the energy of the process 
approaches $M$, then in the strict effective field 
theory paradigm, a new effective field theory should be
used with new degrees of freedom~\cite{Georgi:1993qn}.
For example, Fermi's theory of weak interactions breaks
down as the propagation of the $W$ becomes important.
For these process the electroweak theory can be used.

The effective field theory idea can be applied to hadronic physics.
Indeed historically, the origin of the effective field theory idea was
in hadronic physics.  A Lagrangian is written down in terms of hadron
fields, as hadrons are the more appropriate degrees of freedom at low
energies.  The Lagrangian is chosen to have the same symmetries as
QCD, hence in its domain of validity it should give the same physics
results as QCD. This is known as Weinberg's
theorem~\cite{Weinberg:1979kz}.

For ``large'' momentum scales the quark and gluon degrees
of freedom will become evident, so the theory based on the
chiral Lagrangian with hadron fields must break down. 
Unfortunately, it is not clear what the next effective
field theory is beyond the hadronic effective Lagrangian,
because it is hard to compute anything in low energy QCD.

The lowest order chiral Lagrangian for pseudoscalars is
\begin{equation}
{\cal L}_2 = \frac{F^2}{4} \mbox{Tr}
(\sum_{\mu} D_\mu V^\dagger D_\mu V - \chi^\dagger V - \chi V)
\label{ew:LowestChiral}
\end{equation}
\begin{equation}
V(x) = e^{  \phi(x) / F_0 }
\end{equation}
\begin{equation}
\phi(x) =
\left(
\begin{array}{ccc}
\pi^0 + \frac{1}{\sqrt{3}} \eta & \sqrt{2} \pi^{+} & \sqrt{2} K^{+} \\
\sqrt{2} \pi^{-} & -\pi^0 + \frac{1}{\sqrt{3}} \eta & \sqrt{2} K^{0}
\\
\sqrt{2} K^{-} &  \sqrt{2} \overline{K}^{0} & - \frac{2}{\sqrt{3}} \eta  
\end{array}
\right)
\end{equation}

The expansion parameter~\cite{Manohar:1984md}
 for the chiral 
Lagrangian of pseudoscalar mesons is
\begin{equation}
\lambda_{\chi} \sim m_\rho \sim 4 \pi F_{\pi} \sim 1 GeV
\label{eq:CHIRALscale}
\end{equation}
The power counting theorem of Weinberg~\cite{Weinberg:1979kz}
guarantees that the
tree level Lagrangian in equation~\ref{ew:LowestChiral}
will  generate the $E^4$
Lagrangian and non-analytic functions.

The next order terms in the Lagrangian 
for the pseudoscalars are
\begin{equation}
{\cal L}_4 = \sum_{i=1}^{10}  \alpha_i O_i
\label{cmn:eq:nextOrderL}
\end{equation}
The coefficients $\alpha_i$ are independent of the 
pion mass. They represent the high momentum behaviour.
The normalisation used in lattice calculations 
is connected to the one in Gasser and 
Leutwyler~\cite{Gasser:1985gg}
by $\alpha_i = 8 (4 \pi)^2 L_i $.
For example, one of the terms in the $O(p^4)$ Lagrangian is
\begin{equation}
O_1 = [ \mbox{Tr} ( D_\mu V D^\mu V^\dagger) ]^2
\end{equation}

It can be convenient to have quarks
with different masses in the sea, than those used in the valence
correlators. In lattice QCD jargon, this is known as 
``partial quenching''. These partially quenched theories 
can provide information about the real 
unquenched world~\cite{Bernard:1994sv,Sharpe:2000bc,Sharpe:1997by}.
The chiral Lagrangian 
predicts~\cite{Sharpe:1997by,Heitger:2000ay} 
the mass of the pion
to depend on the sea and valence quark masses
\begin{eqnarray}
M_{PS}^2  =  y_{12} (4 \pi F_0)^2
[
1  & + & 
\frac{1}{n_f} 
\left(
\frac{  
y_{11} (y_{SS} - y_{11}) \log y_{11}
-
y_{22} (y_{SS} - y_{22})  \log y_{22}
}
{y_{22} - y_{11} }
\right)
\nonumber
\\
&+ & 
y_{12} (2 \alpha_8 - \alpha_5 )
+ 
y_{SS} N_f (2 \alpha_6 - \alpha_4 )
]
\label{eq:pionModel}
\end{eqnarray}
The function $y_{ij}$ is related to the quark
mass via
\begin{equation}
y_{ij} = \frac{B_0 (m_i + m_j) }{(4 \pi F_0)^2}
\end{equation}
where $F_0$ is the pion decay constant.
The leading order term in equation~\ref{eq:pionModel}
represents the PCAC relation.

The terms like $\log y_{11}$ are known as ``chiral logs'' are caused
by one loop diagrams.  As the chiral logs are independent of the
Gasser-Leutwyler coefficients ($\alpha_4$, $\alpha_5$, $\alpha_6$,
$\alpha_8$), they are generic predictions of the chiral perturbation
theory.
A major goal of lattice QCD calculations is to detect the presence
of the chiral logs. This will give confidence that the lattice QCD
calculations are at masses where chiral perturbation theory is
applicable
(see~\cite{Wittig:2002ux} for a recent review of how close lattice 
calculations are to this goal).

There have been some attempts to use equation~\ref{eq:pionModel}
to determine the Gasser-Leutwyler coefficients from
lattice QCD~\cite{Heitger:2000ay,Bardeen:2000cz,Irving:2001vy,Nelson:2003tb}.
\begin{table}[tb]
\begin{center}
  \begin{tabular}{|c|c|c|} \hline
GL  coeff & continuum   & $n_f =2$~\cite{Irving:2002fx} \\ \hline
$\alpha_5(4 \pi F_\pi)$ & $0.5 \pm 0.6$ & 
$1.22^{+11}_{-0.16}(stat)^{+23}_{-26}$ \\
$\alpha_8(4 \pi F_\pi)$ & $0.76 \pm 0.4$ &
$0.79^{+5}_{-7}(stat)^{+21}_{-21}$ \\ \hline
\end{tabular}
\end{center}
  \caption{
where the mass of the sea quark is $m_s$ and
the masses of the valence quarks are $m_1$ and $m_2$.
Comparison of the results for the 
Gasser-Leutwyler coefficients from lattice
QCD (with two flavors)~\cite{Irving:2002fx},
to non-lattice estimates~\cite{Bijnens:1994qh}.. 
}
\label{tab:GLcoeff}
\end{table}
Table~\ref{tab:GLcoeff} contains  some results for the 
Gasser-Leutwyler coefficients
from a two flavour lattice QCD calculation at a fixed
lattice spacing of around 0.1 fm~\cite{Irving:2002fx},
compared to some non-lattice estimates~\cite{Bijnens:1994qh}.

It is claimed that resonance exchange is largely responsible for the 
values for  the values of the Gasser-Leutwyler 
coefficients~\cite{Ecker:1989te,Bijnens:1994qh}. For example, in this approach
the value of $\alpha_5$ is related to the properties of the scalar mesons.
\begin{equation}
\alpha_5^{scalar} = 8 (4 \pi)^2  \frac{c_d c_m}  {M_S^2}
\end{equation}
where $M_S$ is the mass of the non-singlet scalar and the 
parameters $c_d$ and $c_m$ are related to the decay of the
scalar into two mesons.

The $\alpha_5$ and $\alpha_8$  coefficients are important
to pin down the Kaplan-Manohar ambiguity~\cite{Kaplan:1986ru}
in the
masses chiral Lagrangian at order O($p^4$).
The latest results on estimating
Gasser-Leutwyler coefficients using lattice QCD
are reviewed by Wittig~\cite{Wittig:2002ux}.

Most lattice QCD calculations do not calculate at light enough quarks
so that the $log$ terms in equation~\ref{eq:pionModel}
are apparent. In practise many groups extrapolate the
squares of the light pseudoscalar meson as either
linear or quadratic functions of the light quark masses.
For example the CP-PACS collaboration~\cite{AliKhan:2001tx}
used fit 
models of the type 
\begin{equation}
m_{PS}^2 = 
  b^{PS}_S m_{S}
+ b^{PS}_V m_{V}
+ c^{PS}_S m_{S}^2
+ c^{PS}_V m_{V}^2
+ c^{PS}_{SV} m_{V} m_{S}
\end{equation}
for the mass of the pseudoscalar ($m^{PS}$)
where $m_{S}$ is the mass of the sea quark and 
$m_{V}$ is the mass of a valence quark.

In traditional chiral perturbation theory 
(see~\cite{Scherer:2002tk} for a review)
framework
$B$ is 
\begin{equation}
B_0 = - \frac{1}{3 F_0^2} \langle \overline{q} q \rangle
\end{equation}


It is possible that $B_0$ is quite 
small~\cite{Fuchs:1991cq,Knecht:1994zb},
say of the order of the pion decay constant.
In this scenario the higher order terms in quark
mass become important.
There is a formalism
called generalised chiral perturbation 
theory~\cite{Knecht:1994zb} that is general enough
to work with a small $B_0$.
The chiral condensate has been computed using quenched lattice 
QCD~\cite{Giusti:1998wy,Hernandez:1999cu}.
The generalised chiral perturbation theory predictions
have been compared to some lattice data by Ecker~\cite{Ecker:1998ai}. 
The next generation of experiments may be able to provide
evidence for the standard picture of a large $B_0$.

Morozumi, Sanda, and Soni~\cite{Morozumi:1992gc} used a linear
sigma model to study lattice QCD data. Their motivation was that the 
quark masses on lattice QCD calculations may be too large for
traditional chiral perturbation to be appropriate.


There is also a version of chiral perturbation theory developed for
quenched QCD, by Morel~\cite{Morel:1987xk}, 
Sharpe~\cite{Sharpe:1990me}, 
Bernard and Golterman~\cite{Bernard:1992mk}.
Quenched QCD is considered as QCD with scalar ghost quarks.
The determinant of the ghost quarks cancels the determinant of the
quarks. The relevant symmetry for this theory is 
$U(3 \mid 3) \times U(3 \mid 3)$. This is a graded symmetry,
because it
mixes fermions and bosons. The chiral Lagrangian is written
down in terms of a unitary field that transform under the 
graded symmetry group.

In this section, I follow the review by 
Golterman~\cite{Golterman:1997zc}.
The problems with quenched QCD can be seen by looking
at the Lagrangian for the $\eta'$ and $\overline{\eta}'$
(the ghost partner of the $\eta'$).

\begin{eqnarray}
{\cal L}_{\eta'} & =  &
\frac{1}{2} (\partial_\mu \eta')^2
-
\frac{1}{2} (\partial_\mu \overline{\eta}')^2
+ \nonumber \\
& & \frac{1}{2} m_\pi^2 ( (\eta')^2   - (\overline{\eta}')^2)
+
\frac{1}{2} \mu^2 (\eta'  - \overline{\eta}')^2 
+
\frac{1}{2} \alpha^2 (\partial_\mu(\eta'  - \overline{\eta}'))^2 
\label{cmn:eq:etaLagrangian}
\end{eqnarray}
The propagator for the $\eta'$ can be derived from 
equation~\ref{cmn:eq:etaLagrangian}.
\begin{equation}
S_{\eta'}(p) = 
\frac{1}{p^2 + m_\pi^2}
-
\frac{\mu^2 + \alpha p^2}{(p^2 + m_\pi^2)^2}
\label{cmn:eq:etaProp}
\end{equation}
The double pole in equation~\ref{cmn:eq:etaProp}
stops the $\eta'$ decoupling in quenched chiral perturbation
theory as $\mu$ gets large. This has a dramatic
effect on the dependence of the meson masses 
on the light quark mass. 
For example the dependence of the square of the pion 
mass on the light quark mass $m_q$ is
\begin{equation}
m_\pi^2 = A m_q (1 + \delta (\log(B m_q)) + C m_q )
\label{cmn:eq:PionGoCrazy}
\end{equation}
where $\delta$ is defined by
\begin{equation}
\delta = \frac{\mu^2}{24 \pi^2 f_\pi^2 }
\end{equation}
and $A$, $B$, and $C$ are functions of the parameters in the 
Lagrangian.
In standard chiral perturbation theory,
the $\delta$ term in equation~\ref{cmn:eq:PionGoCrazy}
is replaced by $\frac{m_\pi^2}{(4 \pi f_\pi^2) }$.
As the quark mass is reduced, equation~\ref{cmn:eq:PionGoCrazy}
predicts that the pion mass will diverge. 
Values of $\delta$ from 0.05 to 0.30 have been 
obtained~\cite{Wittig:2002ux} from lattice data.
It has been suggested~\cite{Wittig:2002ux} 
that the wide range in 
$\delta$ might be caused by finite volume effects in
some calculations.

The effective Lagrangians, encountered so far, assume
that the hadron masses are in the continuum limit.
In practise most lattice calculation use the 
quark mass dependence at a fixed lattice spacing.
The CP-PACS collaboration~\cite{Aoki:2002fd} have compared
doing a chiral extrapolation at finite lattice 
spacing and then extrapolating to the continuum, 
versus taking the continuum limit,
and the doing the chiral extrapolation.
In quenched QCD, the CP-PACS collaboration~\cite{Aoki:2002fd} 
found that the two methods only differed at the $1.5 \sigma$
level.

It is very expensive to generate unquenched gauge configurations with 
very different lattice spacings, so it would be very useful
to have a formalism that allowed chiral extrapolations at 
a fixed lattice spacing.
Rupak and Shoresh~\cite{Rupak:2002sm}
have developed a chiral perturbation theory formalism
that includes $O(a)$ lattice spacing errors for the Wilson
action. This type of formalism had already been 
used to
study the phase of lattice QCD with Wilson 
fermions~\cite{Sharpe:1998xm}.
An additional parameter $\rho$  is introduced
\begin{equation}
\rho \equiv 2 W_0 a c_{SW}
\end{equation}
where $W_0$ is an unknown parameter and 
$c_{SW}$ is defined in the appendix.
This makes the expansion a double expansion in
$\frac{p^2}{\Lambda_\chi^2}$ and $\frac{\rho}{\Lambda_\chi}$.
\begin{equation}
{\cal L}_2 = 
\frac{F^2}{4} \mbox{tr} 
(\sum_{\mu} D_\mu V^\dagger D_\mu V )
-
\frac{F^2}{4} \mbox{tr} 
(
(\chi+\rho) V
+
V
(\chi^\dagger + \rho^\dagger) 
)
\label{cmn:eq:OaLag}
\end{equation}
The Lagrangian in equation~\ref{cmn:eq:OaLag}
is starting to be used to
analyse
the results from lattice QCD 
calculations~\cite{Farchioni:2003bx}.


For vector mesons it less clear how to write down
an appropriate Lagrangian than for the pseudoscalar
mesons. There are a variety of different Lagrangians for 
vector 
mesons~\cite{Meissner:1988ge,Ecker:1989yg,Bijnens:1997ni,Jenkins:1995vb},
most of which are equivalent.
A relativistic effective Lagrangian for the vector 
mesons~\cite{Bijnens:1997ni} is 
\begin{equation}
{\cal L} = 
-\frac{1}{4}
\mbox{Tr} ( D_{\mu} V_{\nu} - D_{\nu} V_{\mu} )^2
+\frac{1}{2}
M_V^2  \mbox{Tr}(V_{\nu} V^{\nu})
\label{eq:relatvisitic}
\end{equation}
where $V_{\nu}$ contains the vector mesons.
As discussed earlier in this section, a crucial ingredient of the
effective field theory formalism is that a power counting scheme can
be set up. The large mass of the vector 
mesons complicates the power counting, hence other formalisms
have been developed.
Jenkins et al.~\cite{Jenkins:1995vb} wrote down a heavy meson effective
field theory for vector mesons.

In the heavy meson formalism the velocity $v$
with $v^2 = 1$ is introduced. 
Only the residual momentum ($p$) of vector mesons 
enters the effective theory.
\begin{equation}
k_V = M_V v + p
\end{equation}
In the large $N_c$ limit the meson fields live
inside $N_{\mu}$.
\begin{equation}
N_{\mu} = 
\left(
\begin{array}{ccc}
\rho^0_{\mu} + \frac{1}{\sqrt{2}} \omega_{\mu} 
&  \rho^{+}_{\mu} 
&  K^{\star +}_{\mu} \\
 \rho^{-}_{\mu} 
& -\frac{1}{\sqrt{2}}\rho^0_{\mu} + \frac{1}{\sqrt{2}} \omega_{\mu}^0 
&  K^{\star 0}_{\mu}
\\
K^{\star -} 
& \overline{K}^{\star 0}_{\mu} 
& \phi_{\mu}
\end{array}
\right)
\end{equation}

The Lagrangian for the heavy vector mesons
(in the large $N_c$ and massless limit )
is
\begin{equation}
{\cal L} = 
-i \mbox{Tr} N_{\mu}^\dagger (v . {\cal D}) N^{\mu}
-i g_2 \mbox{Tr} \{ N_{\mu}^\dagger , N_{\nu} \} A_{\lambda}  v_{\sigma}
\epsilon^{\mu\nu\lambda\sigma}
\label{eq:stativVector}
\end{equation}
The connection between the Lagrangian in 
equation~\ref{eq:relatvisitic} and the Lagrangian in 
equation~\ref{eq:stativVector} is discussed by the 
Bijnens et al.~\cite{Bijnens:1997ni}.

At one loop the correction to the $\rho$
mass~\cite{Jenkins:1995vb}
is
\begin{equation}
\Delta m_{\rho} = - \frac{1}{12 \pi^2 f^2}
(
g_2^2 (\frac{2}{3} m_\pi^3 + 2 m_k^3 +  \frac{2}{3} m_\eta^3  )   
+
g_1^2 (m_\pi^3)
)
\label{eq:loop}
\end{equation}
where $g_1$ and $g_2$ are parameters (related to meson decay)
in the heavy vector Lagrangian and $f$ is the pion decay
constant.
The next order in the expansion of the masses of vector mesons
is in the paper by Bijnens et al.~\cite{Bijnens:1997ni}.
The equivalent expression in quenched and partially quenched chiral perturbation 
theory has been computed by Booth et al.~\cite{Booth:1997hk} 
and Chow and Ray~\cite{Chow:1998dw}.
The heavy vector formalism suggests that 
for degenerate unquenched quarks the mass of the vector mass
should depend on the quark mass $m_q$ like
\begin{equation}
M^{Vec} = M^{Vec}_0 + M^{Vec}_1 m_q + M^{Vec}_2 m_q^{3/2} + O(m_q^2)
\label{eq:RHOfitMODEL}
\end{equation}

In practise it has been found to be difficult to detect the 
presence of the $m_q^{3/2}$ term in equation~\ref{eq:RHOfitMODEL}
from recent lattice calculations.
The mass of the light vector particle from lattice QCD calculations 
is usually extrapolated to the physical
point using a function that is linear or quadratic in the quark
mass. For example the CP-PACS collaboration~\cite{AliKhan:2001tx} 
used the fit model
\begin{equation}
M^{Vec} = A^{Vec}
 + B_{S}^{Vec} m_{S}
 + B_{V}^{Vec} m_{V} 
 + C_{S}^{Vec} m_{S}^2
 + C_{V}^{Vec} m_{V}^2
 + C_{V}^{Vec} m_{V} m_{S}
\label{eq:MVpragmatic}
\end{equation}
to extrapolate the mass of the vector meson ($M_{vec}$)  in terms of 
the sea ($m_{S}$)
and valence ($m_{V}$) quark masses in their 
unquenched calculations.
CP-PACS~\cite{AliKhan:2001tx}  
also investigated the inclusion of terms from the 
one loop calculation of the correction to the rho mass 
in equation~\ref{eq:loop}.

There is an added complication for the functional dependence of the
mass of the $\rho$ meson on the light quark mass, because in principle
the rho can decay into two pions (see section~\ref{eq:NotGoodAt}) .
This decay threshold complicates the chiral extrapolation model. The
first person to do an analysis of this problem 
for the $\rho$ meson in
lattice QCD was
DeGrand~\cite{DeGrand:1991ip}.
There was further work done by Leinbweber amd 
Cohen~\cite{Leinweber:1994yw}.
The effect of decay thresholds on hadron masses is also
a problem for the quark model~\cite{Geiger:1990yc,Isgur:1999cd}.

The Adelaide group~\cite{Leinweber:2001ac} have studied the issue of
the effect of the rho decay on the chiral extrapolation model of the
rho meson mass in more detail.  The physical motivation behind 
Adelaide group's program in the extrapolation of hadron masses
in the light quark mass  
has been reviewed by Thomas~\cite{Thomas:2002sj}.

\begin{figure}
\def\filename{Rho_SE.eps}
\begin{center}
\includegraphics{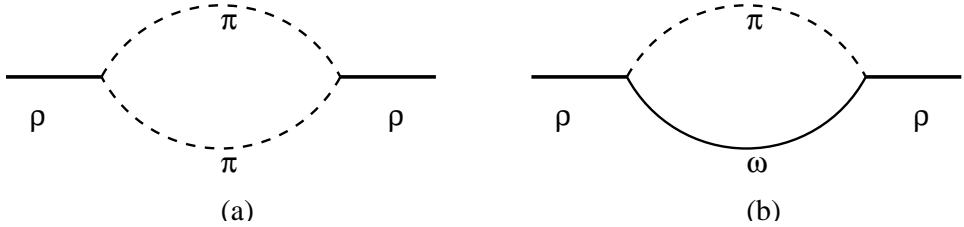}
\end{center}
   \caption{
Self energy contributions to the mass of the rho meson.
 }
\label{cmn:OneLoopRho}
\end{figure}

The mass of the rho is shifted by $\pi-\pi$ 
and $\pi-\omega$ intermediate states.
The effect of the two meson intermediates states can be found
by computing the Feynman diagrams in figure~\ref{cmn:OneLoopRho}
from an effective field Lagrangian.
The self energies from 
the $\pi-\pi$  ($\Sigma^\rho_{\pi\pi} $) and 
the $\pi-\omega$  ($\Sigma^\rho_{\pi\omega} $) intermediate states.
renormalize the mass of the $\rho$ meson.
\begin{equation}
m_\rho = \sqrt{ m_0^2 + \Sigma^\rho_{\pi\pi} + \Sigma^\rho_{\pi\omega} }
\end{equation}

To explain the idea, I will consider the self energy corrections from 
$\pi-\omega$ intermediate states in more detail.
\begin{equation}
\Sigma^\rho_{\pi\omega}  = - 
\frac{g_{\omega\rho\pi}^2 \mu_\rho}{12 \pi^2}
\int_0^\infty 
\frac{dk k^4 u^2_{\omega\pi}(k) }
{w_{\pi}(k)^2}
\label{eq:rhoGRAPH}
\end{equation}
where 
\begin{equation}
w_{\pi}(k)^2 = k^2 + m_\pi^2
\end{equation}
and $\mu_\rho$ is the physical mass of the $\rho$ meson.
The integral for the $\Sigma^\rho_{\pi\pi} $ is similar,
but the algebra is more complicated.

As the momentum increases the effective field 
theory description of the physics in terms of 
meson field breaks down. Adelaide prefer to parameterise
the breakdown of the effective field theory by
introducing a form factor $u_{\omega\pi}(k)$ 
at the interaction between the 
two pseudoscalars mesons and the vector meson.
A dipole form factor is used in equation~\ref{eq:rhoGRAPH}.
\begin{equation}
u_{\omega\pi}(k) = 
(
\frac
{ \Lambda_{\pi\omega}^2 - \mu_\pi^2 }
{ \Lambda_{\pi\omega}^2 + k^2 }
)^2
\end{equation}
The $\Lambda_{\pi\omega}$ parameter is a (energy) scale 
associated with the finite extent of the hadrons.
This is a fit parameter that the Adelaide group~\cite{Leinweber:2001ac}
determine from lattice data. They obtain~\cite{Leinweber:2001ac}  
$\Lambda_{\pi\omega}$ 
= 630 MeV.

It is instructive to look at the self energy $\Sigma^\rho_{\pi\omega}$ 
with a 
sharp cut off
\begin{equation}
\Sigma^\rho_{\pi\omega}  = - 
\frac{g_{\omega \rho \pi} \mu_\rho } {12 \pi^2}
(m_{\pi}^3 \arctan(\frac{\Lambda}{m_\pi})  + \Lambda^3 - \Lambda m_\pi^2   ) 
\label{eq:ossieCutOff}
\end{equation}
There is strong dependence on $\Lambda$ in equation~\ref{eq:ossieCutOff}.
The $\Sigma^\rho_{\pi\omega}$ term contains the $m_\pi^3$ term of
chiral perturbation theory.

In the Adelaide approach~\cite{Leinweber:2001ac} 
the fit model used to 
extrapolate the mass of the rho meson 
in terms of the 
quark mass is
in equation~\ref{eq:OssieRhoFit}.
\begin{equation}
m_{\rho} = c_0 + c_1 m_\pi^2
+ \frac{ 
\Sigma^\rho_{\pi \omega} (\Lambda_{\pi \omega} , m_\pi) +    
\Sigma^\rho_{\pi \pi} (\Lambda_{\pi \pi} , m_\pi) 
}{2(c_0 + c_1 m_\pi^2)}
\label{eq:OssieRhoFit}
\end{equation}
This is a replacement for the chiral extrapolation model 
in equation~\ref{eq:RHOfitMODEL}.

The Adelaide group~\cite{Leinweber:2001ac}  
note that the coefficient of the 
$m_{\pi}^3$ term in equation~\ref{eq:loop} is known,
hence this is a constraint on the fits from the lattice 
data. However, the actual fits to the mass of the rho particle
from lattice QCD, do not reproduce the known 
coefficient in equation~\ref{eq:loop}. When the Adelaide
group~\cite{Leinweber:2001ac}  
fit the expression in equation~\ref{eq:OssieRhoFit}
to the mass of the rho at the coarse lattice spacings
from calculations by UKQCD~\cite{Allton:1998gi}
 and CP-PACS, the correct
coefficient is obtained.

It is interesting to compare the 
Adelaide group's~\cite{Leinweber:2001ac}  
approach to the chiral extrapolation of the mass of the rho meson
to a more ``traditional'' effective field
theory calculation. Equation~\ref{eq:rhoGRAPH} looks similar to
a perturbative calculation with a cut off. In an effective 
field theory calculation terms with powers of $\Lambda$ would be
absorbed into the counter terms. In a field theory with a cut off,
the actual cut off should have no effect on the dynamics in the
effective theory. A strong dependence on the cut off would signify the 
breakdown of the effective theory. The lecture notes by Lepage 
discuss the connection between a cut off field theory and 
renormalisation in 
nuclear physics~\cite{Lepage:1997cs}.

In his review on effective field theories Georgi~\cite{Georgi:1993qn} 
quotes Sidney Coleman
as asking ``what's wrong with form factors?'' Georgi's answer is 
``nothing''. My translation  of Georgi's more detailed 
answer to Coleman's question is in the
next paragraph.

Both the Adelaide group's approach 
and effective field theory agree
for low momentum scales, hence both formalisms can reproduce the
non-analytic correction in equation~\ref{eq:loop} to the mass of the
rho, however the two formalisms differ in the treatment of the large
momentum behaviour. In the effective field theory paradigm the large
momentum behaviour is parameterised by a local Lagrangian with terms
that are ordered with a power counting scheme. In the Adelaide
group's approach the long distance physics is parameterised (presumably
with Coleman's blessing) by a form factor.  In an ideal world, the use
of an effective field theory is clearly superior to the use of a form
factors as the accuracy of the approximation is controlled.
For the rho and nucleon it is not obvious how to set up a
power counting scheme in the energy (although people are trying).
Also an effective field theory based on hadrons 
will no longer describe the physics at large momentum when
the quark and gluon degrees of freedom become important,
hence the use of a form factor may be a more pragmatic
way to control the long distance behaviour of hadronic 
graphs. It may be easier to introduce decay thresholds
in a form factor based approach.
The hard part of a form factor based approach 
is controlling the errors from the approximate nature
of the form factor. The Adelaide group~\cite{Leinweber:1999ig}
 do check the sensitivity of 
their final results by using different form factors.
For the light pseudoscalars, the standard effective field
theory formalism is clearly superior.

There is a ``tradition'' of not using 
``strict''
effective field
theory techniques for the $\rho$ meson.  For example 
DeGrand~\cite{DeGrand:1991ip} used a twice
subtracted dispersion relation to regulate the graph
in figure~\ref{cmn:OneLoopRho}.


Nucleons have also been incorporated into the Chiral Lagrangian 
approach (see~\cite{Scherer:2002tk} for a review).
\begin{equation}
\Psi = 
\left(
\begin{array}{c}
p \\
n
\end{array}
\right)
\end{equation}

\begin{equation}
{\cal L }_{\pi N} = 
\overline{\Psi} 
iD\hspace{-.6em}/
- \hat{m_N} 
+ \frac{\hat{g_A}}{2} \gamma_\mu \gamma_5 u_\mu
\Psi
\label{eq:ReltBaryon}
\end{equation}
where $\hat{m_N}$ and $\hat{g_A}$ are parameters 
in the chiral limit. 
\begin{eqnarray}
u^2        & = & U \\
u_\mu      & = & i u^\dagger \partial_\mu U u^\dagger \\ 
\Gamma_\mu & = & \frac{1}{2} [u^\dagger , \partial_\mu u   ] \\
D_\mu      & = & \partial_\mu + \Gamma_\mu
\end{eqnarray}
$U$ is the $SU(2)$ matrix containing the pion fields.

There are a number of complications with baryon chiral perturbation
based on the relativistic action  over meson perturbation
theory.  As discussed earlier the power counting is the key
principle in effective theory as it allows an estimate
of the errors from the neglected terms.
However, the nucleon mass in baryon chiral perturbation theory is the
same order as $4 \pi F_0$. This complicates the power counting.
Also the expansion is linear in the small momentum.

Most modern baryon chiral perturbation
theory calculations are done using ideas motivated 
from heavy quark effective field 
theory~\cite{Jenkins:1991es,Bernard:1992qa}.
The four momentum
is factored into a velocity dependent part and a small 
residual momentum part.
\begin{equation}
p_\mu = m v_\mu + k_\mu
\end{equation}

The baryon field is split into ``large''
and ``small'' fields.
\begin{eqnarray}
{\cal N}_v & \equiv & e^{i m v.x} P_{v+} \Psi \\
{\cal H}_v & \equiv & e^{i m v.x} P_{v-} \Psi \\
\end{eqnarray}
where the projection operator is defined by
\begin{equation}
P_{v+/-} = \frac{1 + v_\nu \gamma^\nu}{2}
\end{equation}

The leading order Lagrangian for heavy baryon chiral perturbation
theory (HBChPT) is
\begin{equation}
{\cal L }_{\pi N} = 
\overline{{\cal N}}_v ( i v.D + \frac{g_A}{2} S_v .u  ) {\cal N}
\label{eq:HeavtBaryon}
\end{equation}
where
\begin{equation}
S_v^{\mu} = \frac{i}{2} \gamma_5 \sigma^{\mu \nu} v_{\nu}
\end{equation}
In principle there are $\frac{1}{M_Q}$ 
corrections to equation~\ref{eq:HeavtBaryon}.
See the review article by~\cite{Scherer:2002tk} for a detailed
comparison of the relativistic and heavy Lagrangians.

The final extrapolation formula for the nucleon mass 
$M_N$ as a function of the quark mass $m_q$
is:
\begin{equation}
M^N = M^N_0 + m_q B + m_q^{3/2} C + .... 
\end{equation}
The coefficient $C$ is negative and is a prediction 
of the formalism.

The convergence of baryonic chiral perturbation theory is very poor
even in the continuum.  On the lattice the pion masses are even
larger, hence there are additional concerns about the convergence of
the predictions. 
Using chiral perturbation theory, 
Borasoy and Meissner~\cite{Borasoy:1997bx}
computed the nucleon mass (and other quantities) using
heavy baryonic chiral perturbation theory,
including
all quark mass terms up to and including the quadratic order.
The result for the nucleon mass is in equation~\ref{eq:MNconverge} for 
each order of the quark mass
\begin{equation}
M^N = \hat{m} ( 1 + 0.34 - 0.35 + 0.24 )
\label{eq:MNconverge}
\end{equation}
where $\hat{m}$ = 767 MeV. 
Note this is not a lattice calculation.
Although, when all the terms are summed up,
the correction is small to the nucleon mass, there are clearly
problems with the convergence of the series.  The corrections to the
masses of the $\Lambda$, $\Sigma$ and $\Xi$ baryons were 
also sizable.

Donoghue et al.~\cite{Donoghue:1998bs} blame the poor convergence of
baryoninc chiral perturbation theory on the use of dimensional
regularisation. Donoghue et al.~\cite{Donoghue:1998bs} argue
that the distances below the size of the baryon in
the effective field theory description breaks down, however 
the dimensionally regulated graphs include all length scales.
The incorrect physics from the graphs is compensated by
the counter terms in the Lagrangian. Unfortunately, this
``compensation'' makes the expansion poorly convergent.
Lepage~\cite{Lepage:1997cs} also gives an example
(from~\cite{Kaplan:1996xu}) from nuclear physics where using a cut off
gives a better representation of the physics than using minimal
subtraction with dimensional regularisation.

The graph in equation~\ref{eq:DonDRreg} occurs at one
loop for the baryon masses.
\begin{equation}
\int \frac{d^4 k }{(2\pi)^4}
\frac{k_i k_j}{ (k_0 - i \epsilon) (k^2 - m^2 + i \epsilon) }
=
-i \delta_{i j } 
\frac{ I(m)  }
{24 \pi}
\label{eq:DonDRreg}
\end{equation}
In dimensional regularisation~\cite{Jenkins:1992ts}
\begin{equation}
I_{dim-reg}(m) = m^3
\label{eq:DRfinalresult}
\end{equation}
The graph in equation~\ref{eq:DonDRreg} contributes the 
lowest non-analytic term in the octet masses
($M_N \propto m_\pi^3$).
Donoghue et al.~\cite{Donoghue:1998bs} point out that it
is a bit suspicious that the
result in equation~\ref{eq:DRfinalresult}
is finite when the integral in equation~\ref{eq:DonDRreg}
is cubicly divergent.

Consider now equation~\ref{eq:DonDRreg} regulated with a 
dipole regulator.
\begin{equation}
\int \frac{d^4 k }{(2\pi)^4}
\frac{k_i k_j}{ (k_0 - i \epsilon) (k^2 - m^2 + i \epsilon) }
(\frac{\Lambda^2}{\Lambda^2 - k^2  })
=
-i \delta_{i j } 
\frac{ I_{\Lambda}(m)  }
{24 \pi}
\label{eq:DonCutOff}
\end{equation}

\begin{equation}
I_{\Lambda}(m) = 
\frac{1}{2}
\Lambda^4
\frac
{2 m + \Lambda}
{m + \Lambda}^2
\end{equation}
In the limit $m < \Lambda$  limit
\begin{equation}
I_{\Lambda}(m) 
\longrightarrow
\frac{1}{2} \Lambda^3
-\frac{1}{2} \Lambda m^2
+ m^3
+ ...
\end{equation}
Hence for small $m$ the result from dimensional regularisation
is reproduced.

Up to this point the treatment of the integrals looks
very similar to the approach originally advocated by the 
Adelaide group~\cite{Leinweber:1999ig,Thomas:1999ae}.
However, Donoghue at al.~\cite{Donoghue:1998bs} treat $\Lambda$
as a cut off. Strong dependence on  $\Lambda$ is
removed via renormalisation. 
\begin{equation}
M_0^r  = M_0 - d \Lambda^3
\end{equation}
where $d$ is a function of the other renormalised 
parameters in the Lagrangian (such as the pion decay constant).
In the original work by the Adelaide 
group~\cite{Leinweber:1999ig,Thomas:1999ae}.
the parameter $\Lambda$ was a physical number that
could be extracted from the lattice data.
In the formalism of Donoghue at al.~\cite{Donoghue:1998bs},
the physical results should not depend 
on $\Lambda$, although a weak dependence on 
$\Lambda$ may remain because the calculations are
only done to one loop.
The results for the mass of the nucleon as a function
of the order of the expansion are
\begin{equation}
M^N = 1.143 - 0.237 + 0.034 = 0.940 \mbox{GeV}
\label{eq:MITconverge}
\end{equation}
with the cut off $\Lambda = 400$ MeV.
See~\cite{Meissner:1998um} for a brief
critique of formalism of Donoghue at al.~\cite{Donoghue:1998bs}.

The Adelaide group~\cite{Leinweber:1999ig,Thomas:1999ae} consider the 
one loop pion self energy to the nucleon and delta propagators.
The method is essentially the same as the one applied to the 
chiral extrapolation of the rho mass.
The Adelaide group~\cite{Leinweber:1999ig,Thomas:1999ae}
fit model for the nucleon mass is
\begin{equation}
M_N = \alpha_N + \beta_N m_\pi^2
+ \sigma_{NN}( m_\pi, \Lambda)
+ \sigma_{N\Delta}( m_\pi, \Lambda)
\label{eq:OssieMODEL}
\end{equation}
where $\sigma$ come from one loop graphs.
Equation~\ref{eq:OssieMODEL} is a three parameter fit model:
$\alpha$, $\beta$, and $\Lambda$.

The Adelaide group have applied the formalism of 
Donoghue at al.~\cite{Donoghue:1998bs}
to the analysis of lattice QCD 
data~\cite{Young:2002ib,Young:2002ib}
from CP-PACS. They have compared it with
their previous formalism~\cite{Young:2002ib,Young:2002ib}.

Lewis and collaborators have studied the lattice
regularisation of chiral perturbation 
theory~\cite{Lewis:2000cc,Borasoy:2002hz}.
This type of calculation can not be used to quantify the 
lattice spacing dependence of lattice 
results~\cite{Borasoy:2002hz}, as the lattice spacing dependence
of the two theories could very different.
However, it is interesting to explore different regularisation
schemes.
There is renewed interest in the relativistic 
baryon Lagrangian, because a method~\cite{Becher:1999he}.
called
``infrared regularisation''  allows
a power counting scheme to be introduced
(see~\cite{Scherer:2002tk} for a review).

All the above analysis of effective Lagrangians relied on 
perturbation theory to study the theory.
Hoch and Horgan~\cite{Hoch:1992ke} 
used a numerical 
lattice calculation
to study
the $SU(2) \times SU(2)$ non-linear model, for pions
and the nucleon.
The unitary matrix $V$
\begin{equation}
V = \frac{1}{F} (\sigma . 1 + i \pi_a \tau_a \gamma_5)
\end{equation}
\begin{equation}
\sigma^2 + \pi^2 = F^2 
\end{equation}
\begin{eqnarray}
S(V) & =  & -\frac{1}{4} F^2  \sum_x
\mbox{tr}
(\sum_{j=1}^{4}        
V(x)V(x+\mu_j)^\dagger + V(x+\mu_j)V^\dagger(x) ) \\
& &
-\frac{1}{4} F m_0^2  \sum_x
\mbox{tr}
(
U(x) + U(x)^\dagger
)
\nonumber
\end{eqnarray}
The numerical calculation was done with a small volume $8^4$,
and no finite size study was done.
The comparison of the lattice results 
with the perturbative results
 is complicated by the 
effect of the unknown parameters in the 
next order Lagrangian (equation ~\ref{cmn:eq:nextOrderL}
for example).
Hoch and Horgan~\cite{Hoch:1992ke} 
found that the lattice calculation disagreed 
with the predictions of one loop perturbation theory
for log divergent quantities.

A more conservative (some might say cowardly) approach to chiral
extrapolations is to only interpolate the appropriate hadron masses to
the mass of the strange quark, in an attempt to try to minimise the
dependence of any results on uncontrolled extrapolation to the light
quark masses.  One formalism~\cite{Allton:1997yv} for doing this is
called the ``method of planes''. Similar methods have been used by
other groups (see for example~\cite{Lacock:1995tq,Maiani:1986yj}).
Obviously, this type of technique is not useful to get the nucleon
mass.  In unquenched QCD, the sea quark masses should be extrapolated
to their physical values, so there is no way to avoid a chiral
extrapolation even for heavy hadrons.

It is traditional to plot the hadron masses before any
chiral extrapolations have been done, so as not to contaminate
the raw lattice data from the computer with any theory.
In an``Edinburgh plot''
the ratio of the nucleon to rho mass is plotted against
the pion to rho mass~\cite{Bowler:1985hr}. If there were no
systematic errors, such as lattice spacing dependence, then
the data should fall on a universal curve. It is also common
to use $1/r_0$~\cite{Sommer:1994ce} 
(see section \ref{se:LatticeSpacing}) as a 
replacement for the mass of the $\rho$.
There is also an APE plot that plots
the ratio of the nucleon to rho mass  against
the square of the pion to rho mass~\cite{Bacilieri:1989zq}. 
This parameterisation is meant to have a smoother mass dependence than the Edinburgh
plot.

\subsection{FINITE SIZE EFFECTS.}

The physical size of the lattice represents an obvious and an important
systematic error. One simple way to estimate the size of 
a hadron is to consider its charge radius. For example, the 
proton's charge radius is quoted as 0.870 (8) fm in 
the particle data table~\cite{Hagiwara:2002fs}.
The sizes of all recent unquenched lattice QCD calculations are
all above 1.6 fm (see section~\ref{se:params}). The fact that all
the physical lattice sizes were much bigger than the charge radius 
does not rule out finite size effects. In this section I will
discuss some of the mechanisms thought to be behind finite size 
effects in lattice data.

A nice physical explanation of the origin of 
finite size effects for hadron masses has
been presented by Fukugita et al.~\cite{Fukugita:1992hw}.
Consider a hadron in a box with length $L$ and periodic
boundary conditions. 
Most lattice QCD calculations use periodic boundary conditions in
space. The path integral formalism requires that the 
boundary conditions in time are anti-periodic~\cite{Sakita:1985ad},
although many groups use periodic boundary condition in time as well
as space.
The finite size of the box will 
mean that a hadron will interact with periodic images a distance
$L$ away. The origin of finite size effects is closely related to
nuclear forces.
The MILC collaboration have described a qualitative 
model for finite size effects based 
on nuclear density~\cite{Bernard:1993an}.
The nucleon is considered as part of 
nucleus comprising of the periodic images of the 
nucleon.

The self energy of the hadron ($\delta E$) will be 
\begin{equation}
\delta E = \sum_{\underline{n}} V (\underline{n} L ) 
\label{cmn:eq:ELdepend}
\end{equation}
where $V(x)$ is the potential between two hadrons
a distance $x$  apart.
Various approximations to $V(x)$ give different functional forms
for the dependence of the hadron mass on the volume.

In the large $L$ limit the potential is approximated by
one particle exchange ($e^{-m r} / r $). The interaction energy goes
like $V(0) + 6 V(L)$, so the dependence of the hadron mass will be
$\sim e^{-m_\pi L }$.
This argument can be made rigorous~\cite{Luscher:1982uv,Luscher:1983ma}.

It is useful to consider equation~\ref{cmn:eq:ELdepend} in momentum space
using the Poisson resummation formulae.
\begin{equation}
\delta E = \frac{1}{L^3} \sum_{\underline{n}} \hat{V} ( \underline{n} \frac{2  \pi}{L} )
\label{cmn:eq:ELdependMom}
\end{equation}

A more general expression for the potential 
between two hadrons can be derived if the spatial
size of the hadron is modelled with a form factor  
$F(k)$.
\begin{equation}
\hat{V} (k)  = \frac{F(k)^2 } {k^2 + m^2} 
\end{equation}
Momentum is quantised on the lattice in quanta 
of $ \frac{2 \pi a  } {L}$. In physical units
the value of a quantum of momentum is around 1 GeV,
hence the $\underline{k}= \underline{0}$ term in 
equation~\ref{cmn:eq:ELdependMom} should dominate the sum.
Therefore this model predicts that the masses should depend on 
the box size like:
\begin{equation}
M = M_{\infty} + c L^{-3}
\label{cmn:eq:FinitePrag}
\end{equation}
This model is physically plausible but not a rigorous 
consequence of QCD.
Fukugita et al.~\cite{Fukugita:1992hw} noted their data
could also be fit to the functional form.
\begin{equation}
m^2 = m_{\infty}^2 + c \frac{1} { L^3}
\end{equation}
rather then equation~\ref{cmn:eq:FinitePrag}.
Unfortunately only the theory for the
regime of point particle interacting at large distances
is really 
rigorous~\cite{Luscher:1982uv,Luscher:1983ma}..

Chiral perturbation theory can be used to compute
finite volume corrections. For example the ALPHA/UKQCD collaboration
have used~\cite{Garden:1999fg}
a chiral perturbation theory calculation
by ~\cite{Gasser:1987vb}  Gasser and Leutwyler to
estimate the dependence of the pion mass $m_{PS} (L)$
on the box size $L$.
\begin{equation}
\frac{m_{PS} (L) } {m_{PS} (\infty)}
= 1 + \frac{1}{N_f}
\frac{m_{PS}^2}{F_{PS}^2} g( m_{PS}  L)
\label{cmn:eq:pionWithVolume}
\end{equation}

\begin{equation}
g(z) = \frac{1}{8 \pi^2 z^2} \int_{0}^{\infty} 
\frac{dx}{x^2} e^{-z^2 x - 1 /(4 x)}
\label{eq:Ffs}
\end{equation}

Garden et al.~\cite{Garden:1999fg} used
equations~\ref{cmn:eq:pionWithVolume} 
and~\ref{eq:Ffs}
to show that the error in $m_{PS}$ was 0.1 \% for $m_{PS} L >
4.3$. This an example of the rule thumb that finite size effects in
hadron masses
become a concern for $m_{PS} L < 4 $. 
Colangelo et al.~\cite{Colangelo:2002hy}
are extending equation~\ref{cmn:eq:pionWithVolume} to the next
order.
Ali Khan et al.~\cite{AliKhan:2002hz}
have started to use chiral perturbation theory to study the finite
size effects on the nucleon.

It would be useful if some insight could be gained from finite size
effects in quenched QCD that could be applied to unquenched QCD where
the volumes are smaller.  Unfortunately, there are theoretical
arguments~\cite{Antonellig:1995ea} that show that finite size effects
in the unquenched QCD will be larger than in the quenched QCD.  A
propagator for a meson can be formally be written in terms of gauge
invariant paths.  Conceptually this can be understood using the
hopping parameter expansion. The quark propagators are expanded in
terms of the $\kappa$ parameter. The hopping parameter
expansion~\cite{Montvay:1987wh} was used in early numerical lattice
QCD calculations, but was found not to be very convergent for light
quarks.
\begin{equation}
c(t) = 
\sum_{C} \kappa_{val}^{L(C)} \langle W(C) \rangle
+ 
\sum_{C'} \kappa_{val}^{L(C')} \sigma_{val} \langle P(C') \rangle
\label{eq;finiteSIZE}
\end{equation}
where $W(C)$ are closed Wilson loops inside the lattice of 
length $L(C)$. $P(C')$ are Polykov lines that wrap around
the lattice in the space direction. The $\sigma_{val}$
parameter is 1 for periodic boundary conditions and 
$-1$ for anti-periodic boundary conditions for Polykov
lines that wrap around the lattice an odd number of times.

In quenched QCD $\langle P(C') \rangle = 0$ because of a $Z_3$
symmetry of the pure gauge action, so only the first term in
equation~\ref{eq;finiteSIZE} contributes to the correlator $c(t)$. 
The centre of SU(3), the elements that commute with 
all the elements of SU(3), is the $Z(3)$ 
group~\cite{Rothe:1997kp}.
The Wilson gauge action is invariant under the gauge links
being multiplied by a member of the centre of the group
on each time plane.
In unquenched QCD, the
$Z_3$ symmetry is broken by the quark action, so both Wilson loop and
Polyakov lines contribute to the correlators.
There is no clear connection between the arguments based 
on equation~\ref{eq;finiteSIZE}
and the nuclear force and chiral perturbation formalism for finite
size effects.

In their large scale quenched QCD
spectroscopy calculations the CP-PACS collaboration kept the box to be
3 fm~\cite{Aoki:2002fd}, so did not apply any corrections for the 
finite size of the lattice.
Using the finite size estimate from previous
calculations CP-PACS~\cite{Aoki:2002fd} estimated that the finite size
effects were of the order 0.6 \%. The GF11 group~\cite{Butler:1994em} 
did a finite size
study at a coarse lattice spacing ($a$ = 0.17 fm). The results for
ratios of hadron masses at finer lattice spacings were then
extrapolated to the infinite volume limit using the estimated
mass shift at the coarse lattice spacing. Gottlieb~\cite{Gottlieb:1997hy}
gives a detailed critique of the method used by the GF11 group 
to extrapolate their mass ratios to the infinite volume.

If a formalism could be developed that would
predict the dependence of hadron masses on the box size,
then this would help make the calculations 
cheaper (see equation~\ref{eq:manyFLOPS}). 
The additional savings in computer time
could be spent on reducing the size of the light quark masses used
in the calculations~\cite{Orth:2001mk}.
Gottlieb~\cite{Gottlieb:1997hy} suggested, the only way to control
finite size effects is to keep increasing the box size until the
masses no longer change.

The finite box size is not always a bad thing.  The size of the box
has been used to advantage in lattice QCD calculations 
(see~\cite{Luscher:1998pe} 
for a review).  A definition of the coupling is
chosen that is proportional to $1/L$ (where $L$ is the length of one
side of the lattice).  A recursive scheme is setup that studies the
change in coupling as the length of the lattice side $L$ is halved.
The Femtouniverse, introduced by Bjorken~\cite{Bjorken:1979hv}, is a
useful regime to study QCD in.  There is a chiral perturbative
expansion based on the limit $m_{\pi} << \frac{1}{L}$
(see~\cite{Damgaard:2001js} for a modern application).  See Van
Baal~\cite{vanBaal:2000zc} for a discussion of the usefulness of the
finite volume on QCD.

It seems possible to run calculations in a
big enough box to do realistic calculations. So the prospects are good
for a first principles lattice calculation of hadron masses without
resort to approximations in simulations of that kind required in
condensed matter systems in simulations of macroscopic size systems.

\section{AN ANALYSIS Of SOME LATTICE DATA} \label{cmn:se:typical}

To consolidate the previous material, I will work over
a simple analysis of some lattice QCD data.
I think 
it is helpful to understand the steps in lattice QCD 
calculation, if the ideal case where
QCD could be solved analytically is considered first. 
All the masses of the 
hadrons would be known as a function of the parameters
of QCD: quark masses ($m_i$) and coupling $g$.
\begin{equation}
M_H = f_H(m_u,m_d,m_s,m_c,m_b,g)
\end{equation}
As the masses and coupling are not determined by QCD, 
the equation for all the hadrons $H$ would have to be
solved to get the parameters. The solution would be 
checked for consistency that a single set of parameters
could reproduce the entire hadron spectrum.
For a calculation of a form factor, the 
master function $f_H$ would also depend on the momentum.

The formalism of lattice QCD is in some sense is a discrete
approximation to the function $f_H$. The result from 
calculation would be a table of numbers:
\begin{equation}
M_H = f_H^{latt}(m_u,m_d,m_s,m_c,m_b,g,L,a)
\end{equation}
An individual lattice calculation would also 
depend on the lattice spacing and lattice 
volume. Lattice calculations have to be done 
at a number of different lattice spacings
and volumes to extrapolate the dependence of 
$f_H^{latt}$ on $L$ and $a$.

By doing calculations with a number of different parameters the
results can be combined to produce physical results in much the same
way that could be done if the exact solution was known.  In particular
the lattice spacing and lattice volume have be extrapolated away to
get access to the function $f_H$.

To understand the procedure in more detail, I will work through a naive
analysis of some lattice QCD data from the UKQCD
collaboration~\cite{Bowler:1999ae}.  Table~\ref{tab:sampleLATTICEdata}
contains the results for the mass of the $\pi$ and $\rho$ particles in
lattice units from a quenched QCD~\cite{Bowler:1999ae}.  The lattice
volume was $24^3\;48$, $\beta$ = 6.2, and the ensemble size
was 216. The clover action using the ALPHA coefficients was used.  To
use the data in table~\ref{tab:sampleLATTICEdata}, the
secret language of the lattice QCD cabal must be converted to the
working jargon of the continuum physicist.

The ensemble size of 216 means that  216 snapshots of the 
vacuum (the value of $N$ in 
equation~\ref{eq:algorEXPLAIN}) were used to compute 
estimates of the $\rho$ and $\pi$ correlators.
A supercomputer was used to compute the correlators
for each gauge configuration 
from quark propagators (see equation~\ref{eq:algorEXPLAIN}).
The masses were calculated by fitting the correlator
to a fit model of the form in equation~\ref{eq:fitmodel},
using a $\chi^2$ minimiser such as MINUIT~\cite{James:1975dr}.
The error bars in the table in equation~\ref{tab:sampleLATTICEdata} 
come from a statistical procedure called 
the ``bootstrap'' method~\cite{Allton:1992sg}.
\begin{table}[tb]
\begin{center}
  \begin{tabular}{|c|c|c|} \hline
$\kappa$ & $a m_{\pi}$ & $a m_{\rho}$ \\ \hline
0.13460  & $0.2803^{+15}_{-10}$ &  $0.3887^{+32}_{-28}$ \\
0.13510  & $0.2149^{+19}_{-14}$ &  $0.3531^{+55}_{-51}$ \\
0.13530  & $0.1836^{+23}_{-18}$ &  $0.3414^{+72}_{-82}$  \\ \hline
\end{tabular}
\end{center}
  \caption{
Lattice QCD data for the 
masses of the pion and rho from lattice QCD calculations
from UKQCD~\cite{Bowler:1999ae}
}
\label{tab:sampleLATTICEdata}
\end{table}

\begin{figure}
\def\filename{ukqcd_data.ps}
\begin{center}
\includegraphics[angle=-90,scale=0.5]{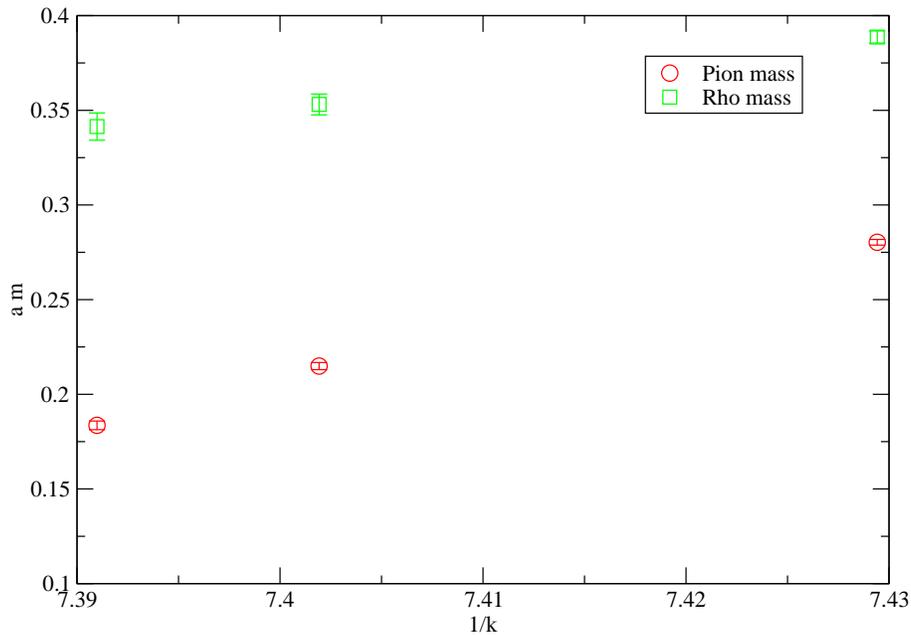}
\end{center}
   \caption{
Rho and pion masses 
in lattice units
as a function of inverse kappa value.
The data set had $\beta=6.2, and a volume of 24^3 48$.
The fermion operator was the non-perturbatively 
improved clover operator.
 }
\label{cmn:eq:rawLATTdata}
\end{figure}
A table of numbers of hadron masses (or even a graph) is not too
useful. A better way to encapsulate the hadron masses as a function of
quark mass is to use them to tune effective Lagrangians, as discussed
in section~\ref{eq:massDEPEND}.
To plot the data in 
table~\ref{tab:sampleLATTICEdata}
in a more physical form, I convert from
the $\kappa$ value to the quark mass
\begin{equation}
m_q = \frac{1}{2} ( 
\frac{1}{\kappa} 
    -
\frac{1}{\kappa_{crit}} 
)
\label{cmn:eq:kappaCRIT}
\end{equation}
There are additional $O(a)$ corrections to
equation~\ref{cmn:eq:kappaCRIT}~\cite{Luscher:1998pe}.  The
$\kappa_{crit}$ parameter is required because clover fermions break
chiral symmetry. The value of $\kappa_{crit}$ is the chosen to give a
zero pion mass.  Equation~\ref{cmn:eq:kappaCRIT} is the basis of the
computation of the masses of quarks from lattice QCD. However,
perturbative factors are required to convert the quark mass to a
standard scheme and scale.  This perturbative ``matching'' can be
involved, so the value of the quark mass is rarely used to indicate
how light a lattice calculation is.

The simplest
thing to do is to use a fit model in equation~\ref{eq:pionFITmass}.
\begin{equation}
m_{PS}^2 = S_{PS} m_q
\label{eq:pionFITmass}
\end{equation}
There are classes of fermion actions (see 
appendix~\ref{cmn:se:technicalDETAILS}), 
such as staggered 
or Ginsparg-Wilson actions, that do not have an additive 
renormalisation.

A simple fit to the data in table~\ref{tab:sampleLATTICEdata} with the 
fit model in equations~\ref{eq:pionFITmass} 
and~\ref{cmn:eq:kappaCRIT} gives  $\kappa_{crit} = 0.135828(6)$,
to be compared to $\kappa_{crit} = 0.135818^{+18}_{-17}$ from the
explicit analysis from UKQCD that included a $O(a)$ correction term.

To simplify the analysis, I will assume that 
the physical mass of the light quark is zero. 
I fit the  data for the mass of the rho in
table~\ref{tab:sampleLATTICEdata}
to the model in equation~\ref{eq:vectorFITmass}.
\begin{equation}
m_{V} = A_{V} + S_{V} m_{q}
\label{eq:vectorFITmass}
\end{equation}
The result for the $A_V$ parameter is $0.304(2)$.
If the mass of the light quark is assumed to be zero, then
$a m_\rho = 0.304$, hence the lattice spacing is 
2530 MeV (using $m_\rho = 770$ MeV).
This can be compared against the $a^{-1}$ = 2963 MeV
from $r_0$~\cite{Guagnelli:1998ud}.

As the masses of quarks get lighter more sophisticated fit models
based on the ideas in section~\ref{eq:massDEPEND} can be used.
Although the basic ideas behind the analysis outlined in this section
are correct, there are many improvements that can be made,
particularly if the rho and pion correlators for each configuration
are available.

\section{PARAMETER VALUES OF LATTICE QCD CALCULATIONS} \label{se:params}

The results from unquenched lattice QCD calculations, with the lattice
spacing and finite size effects accounted for, are the results from
QCD at the physical parameters of the calculation. Hence a key issue
in unquenched calculations is how close the parameters are to the
physical parameters. For example, as discussed in
section~\ref{eq:massDEPEND}, ideally the masses of the quarks must be
light enough to match the lattice results to chiral perturbation
theory. The parameters used in a lattice calculation are usually
dictated by the amount of computer time available, or what gauge
configurations are publicly available.

In this section, I will describe the current state of the art in the
parameters used in lattice QCD calculations of the hadron spectrum.
It is not entirely obvious which parameters to use to characterise a
lattice calculation. The obvious choice of using quark masses is
complicated by the need for renormalisation and running.  To show how
light the quarks are in a calculation, I plot the ratio of the
pseudoscalar mass to the vector mass for as a function of lattice
spacing. 
The danger of this type of plot is that it says nothing
about finite size effects. 
I usually just show the ratio for the lightest quark mass, as this is
the most computationally most expensive point. The error bars on the
ratio gives some indication on the statistical sample size. I have
always used a lattice spacing defined by $r_0 = 0.49 fm$ (see
section~\ref{se:LatticeSpacing}).

\begin{figure}
\def\filename{simul_params_quenched.ps}
\begin{center}
\includegraphics[angle=-90,scale=0.5]{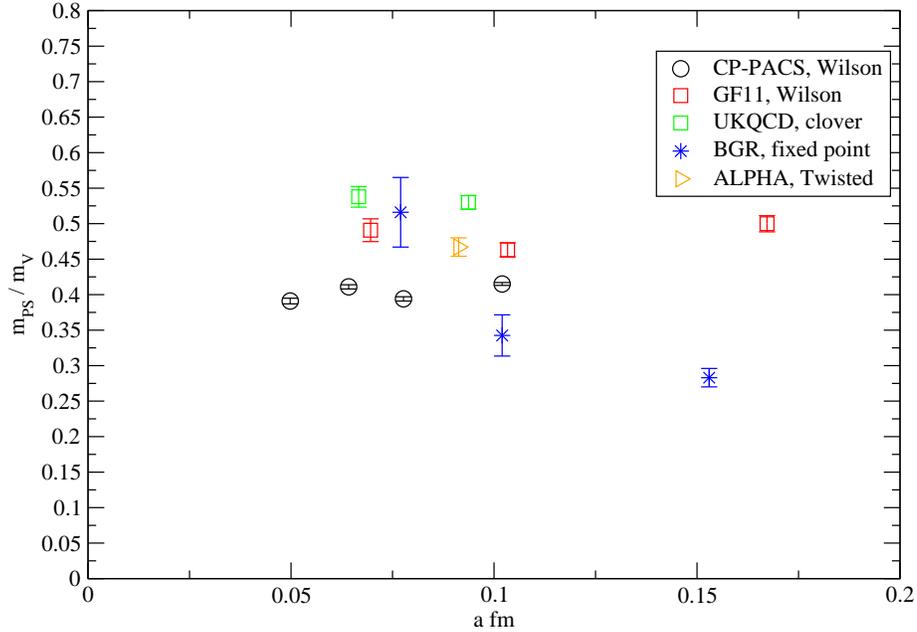}
\end{center}
   \caption{
Smallest ratio of pseudoscalar mass to vector mass as
a function of lattice spacing for a number of quenched 
QCD calculations. The labels have the collaboration name
followed by the type of fermion operator.
 }
\label{eq:quenchedMASSanda}
\end{figure}

In figure~\ref{eq:quenchedMASSanda}, I plot the smallest pseudoscalar
mass to vector mass ratio as a function of lattice spacing for some
recent large quenched calculations from the following collaborations:
GF11~\cite{Butler:1993ki,Butler:1994em},
CP-PACS~\cite{Aoki:1999yr}, UKQCD~\cite{Bowler:1999ae}
BGR~\cite{Gattringer:2003qx}, and ALPHA~\cite{DellaMorte:2001tu}.
The improvements in parameters
between the GF11 calculation~\cite{Butler:1993ki,Butler:1994em}, at
the start of 1990's and the CP-PACS calculation~\cite{Aoki:1999yr}
at the end of 1990's
can clearly be seen in the figure.

Both the main benchmark calculations by CP-PACS~\cite{Aoki:1999yr} and
GF11~\cite{Butler:1993ki,Butler:1994em} used both the Wilson fermion
and gauge actions. As I discuss in section~\ref{cmn:se:technicalDETAILS},
lattice QCD actions with better properties have
been developed.
For example, the point in figure~\ref{eq:quenchedMASSanda} from UKQCD
used the clover fermion action. The clover action is designed to have
smaller lattice spacing errors compared to the Wilson action.
Unfortunately, the calculations by UKQCD~\cite{Bowler:1999ae} and
QCDSF~\cite{Gockeler:1998fn} that used the non-perturbative improved
clover action reported problems with exceptional configurations, hence
for those lattice spacings the light quark masses can not be reduced
further at these lattice spacings. There is a formalism called
twisted mass QCD~\cite{DellaMorte:2001tu} that is a natural
extension of the clover action, that can be used to explore the light
quark mass regime~\cite{McNeile:2001yk}.

To reach lighter quark masses,
it seems likely, that new fermion actions such as the
overlap-Dirac, Domain wall, fixed point~\cite{Hasenfratz:2002rp}, 
and twisted mass QCD~\cite{DellaMorte:2001tu} will be
required to study the light quark mass region of quenched QCD, and
hence improve on the results of the CP-PACS calculation~\cite{Aoki:1999yr}.

The first results from these types of calculations are shown in the
figure~\ref{eq:quenchedMASSanda}.  Unfortunately, the fermion
operators that obey the Ginsparg-Wilson relation are computationally
more expensive than Wilson like actions. At the moment the error bars
are too large to be competitive with those from actions that use
Wilson type fermions. However the control of the systematic errors
in the calculations that use Ginsparg-Wilson operators 
are rapidly improving~\cite{Gattringer:2003qx}.


Table~\ref{tb:dynamicalPARAMS} shows the parameters of some recent
large scale unquenched calculations. I have also included the mass of
the lightest pion in the calculations.  Although the choice of the
lattice spacing in a specific calculation can have large
uncertainties, I feel that the mass of the pion in physical units
gives a more immediate measure of the ``lightness'' of the quarks in a
calculation.  For the table I just used the lattice spacing quoted by
the collaboration.
The review article by 
Kaneko~\cite{Kaneko:2001ux} contains a more thorough 
survey of the parameters of some recent unquenched calculations.
\begin{table}[tb]
\begin{center}\begin{tabular}{|c|c|c|c|c|c|}
\hline 
Collaboration &
$n_{f}$ &
a fm &
L fm &
$\frac{M_{PS}}{M_{V}}$ &
$m_{PS}$ MeV\\
\hline
\hline 
MILC~\cite{Bernard:1999xx}&
2+1&
0.09&
2.5&
0.4&
340\\
\hline 
CP-PACS~\cite{AliKhan:2000mw}&
2&
0.11&
2.5&
0.6&
900\\
\hline 
UKQCD~\cite{Allton:2001sk}&
2&
0.1&
1.6&
0.58&
600\\
\hline 
SESAM~\cite{Glassner:1996xi,Bali:2000vr}&
2&
0.074&
1.8&
0.57&
530\\
\hline
\end{tabular}
\end{center}
\caption{Typical parameters in recent unquenched 
lattice QCD calculations.}
\label{tb:dynamicalPARAMS}
\end{table}

\begin{figure}
\def\filename{simul_params.ps}
\begin{center}
\includegraphics[scale=0.6]{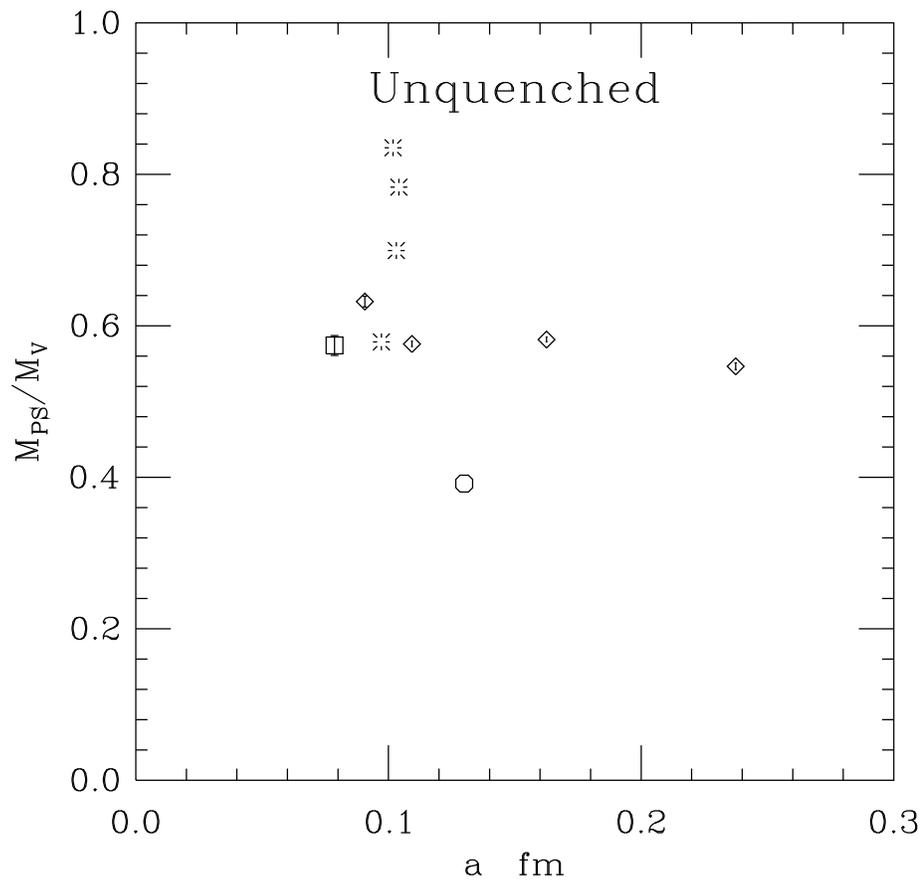}
\end{center}
   \caption{
Smallest ratio of pseudoscalar mass to vector mass as
a function of lattice spacing for a number of full
QCD calculations.
The data is from: 
CP-PACS~\cite{AliKhan:2000mw} (diamonds),
UKQCD~\cite{Allton:2001sk} (bursts), 
MILC~\cite{Bernard:1999xx} (octagon), 
and 
SESAM~\cite{Glassner:1996xi} (squares).
 }
\label{eq:fullQCDMASSanda}
\end{figure}

In figure~\ref{eq:fullQCDMASSanda} the ratio of the of masses
of the lightest pseudoscalar to vector for some recent full QCD
calculations is plotted. The effect of the cost formula for full
QCD~\ref{eq:manyFLOPS} can be clearly seen, as the finest lattice
spacing used is around 0.1 fm. Only the CP-PACS calculation had
a number of different lattice spacings. 
The non-perturbative 
improved clover action was constrained to lie around 0.1 fm,
because the computation of the clover coefficient was
too expensive at coarser lattice spacings~\cite{Jansen:1998mx}.

The most interesting point is from  the calculations done
by the 
MILC collaboration~\cite{Bernard:2001av}.
The MILC collaboration are currently running 
with a lattice spacing of 0.09 fm.

\section{THE MASSES OF LIGHT MESONS} \label{se:lightMESON}

Light mesons have a number of important uses in lattice QCD
calculations. In calculations that use Wilson like fermions, the mass
of pion is used to calculate the additive mass renormalisation
($\kappa_{crit}$ in equation~\ref{cmn:eq:kappaCRIT}). The $\rho$ is
sometimes used to set the lattice spacing.  The mass of one of the
mesons: kaon, $K^{\star}$, or $\phi$ is used to calculate the strange
quark mass. After the light quark masses are calculated any remaining
masses are used as a consistency check.  The masses of the quarks are
used in any further calculations, such as the computation of matrix
elements.

The interpolating operators for mesons to be used in equation~\ref{eq:timeSLICEDCORR}
are in table~\ref{tab:mesonOPERS}.
\begin{table}[tb]
\begin{center}
  \begin{tabular}{|c|c|c|} \hline
Operator  & $J^{PC}$   &  Lightest particle\\ \hline
$\overline{\psi}_1\gamma_5 \psi_2$ & $0^{-+}$ & $\pi$  \\ \hline
$\overline{\psi}_1\gamma_4 \gamma_5 \psi_2$ & $0^{-+}$ & $\pi$  \\ \hline
$\overline{\psi}_1\gamma_i \psi_2$ & $1^{--}$ & $\rho$  \\ \hline
$\overline{\psi}_1\gamma_4 \gamma_i \psi_2$ & $1^{--}$ & $\rho$  \\
\hline
$\overline{\psi}_1\gamma_i \gamma_j \psi_2$ & $1^{+-}$ & $b_1$  \\
\hline
$\overline{\psi}_1\gamma_i \gamma_5 \psi_2$ & $1^{++}$ & $a_1$  \\ \hline
$\overline{\psi}_1 \psi_2$ & $0^{++}$ & $a_0$  \\ \hline
  \end{tabular}
\end{center}
  \caption{
Interpolating operators for light mesons. The 1 and 2 subscripts 
label flavour
and 
show that the mesons are non-singlet.
}
\label{tab:mesonOPERS}
\end{table}
The $J^{PC}$ quantum numbers of the meson operators can be derived
using the standard representation of the parity $P$ and charge
conjugation operators from the Dirac theory.  The meson interpolating
operators are usually extended in space 
using one of the prescriptions in
section~\ref{cmn:sec:basicLGT}.  Most calculations concentrate on the
S-wave mesons as the signal to noise ratio is better for these
mesons, than for P-wave states. I discuss P-wave states in
section~\ref{se:mesonexcite}.

The QCD field strength tensor has also been used
with the fermion bilinears in table~\ref{tab:mesonOPERS}.
The QCD field strength tensor ($F_{jk}$) has specific $J^{PC}$
quantum numbers that can be used to obtain fermion
bilinear operators with different $J^{PC}$ quantum
numbers. For example the MILC collaboration~\cite{Bernard:1997ib}
used 
an interpolating operator of the form:
\begin{equation}
\epsilon_{ijk}\overline{\psi}_1  \gamma_i F_{jk} \psi_2
\label{cmn:eq:glueRHO}
\end{equation}
with $J^{PC}=0^{-+}$. 
The MILC collaboration~\cite{Bernard:1997ib} 
obtained
the same mass for the pion  using the operator
in equation~\ref{cmn:eq:glueRHO} as the 
pion operator in table~\ref{tab:mesonOPERS}.
The operator in equation~\ref{cmn:eq:glueRHO} is
more ``gluey'' than a $\overline{\psi} \gamma_5 \psi$
operator, so might be expected to couple to hybrid
(quark-antiquark pair with excited glue) pions. 
However, the pion is still the lightest state that
couples to the operator in equation~\ref{cmn:eq:glueRHO},
hence is the state extracted from the fits.

\begin{figure}
\def\filename{meson_Ldep.ps}
\begin{center}
\includegraphics[angle=-90,scale=0.5]{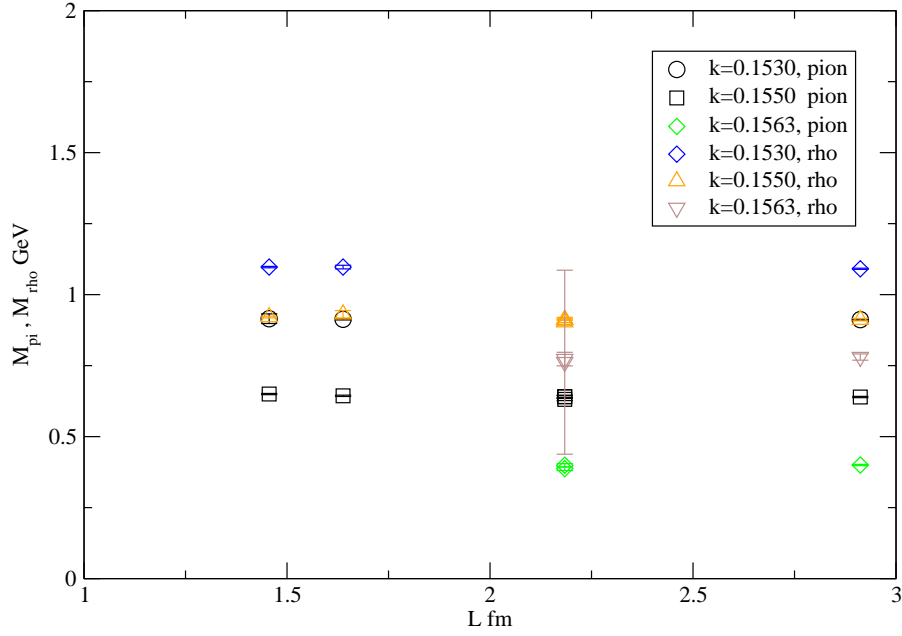}
\end{center}
   \caption{
Dependence of the mass of the $\pi$ and $\rho$ 
mesons on the box size
in quenched QCD for Wilson fermions at $\beta=6.0$.
The data was taken from the compendium of World data
in~\cite{Gockeler:1998fn}.
 }
\label{cmn:Lquenchmeson}
\end{figure}

In figure~\ref{cmn:Lquenchmeson}, I plot the 
mass of the $\pi$ and $\rho$ mesons as a function
of the physical box size for Wilson fermions at 
a fixed lattice spacing ($\beta=6.0,a \sim 0.1 fm$).
The data seems to show that a linear box size of 3 fm is 
big enough for ``small'' finite size errors. The errors on the masses
of the mesons
with the lighter quark masses are too big too draw 
any conclusions.

To get a more quantitative estimate of the volume dependence
UKQCD~\cite{Bowler:1999ae} studied finite size effects in quenched QCD
at $\beta=6.0$ with 
non-perturbatively improved clover fermions. Two volumes were 
used with sides 1.5 fm and 3 fm. For the pseudoscalar channel 
there was a $2 \sigma$ difference between the mass on the two volumes.
There were no statistically significant difference between
the masses for the vector particle between the two lattice
volumes.

\begin{figure}
\def\filename{cont_quench.ps}
\begin{center}
\includegraphics[angle=-90,scale=0.5]{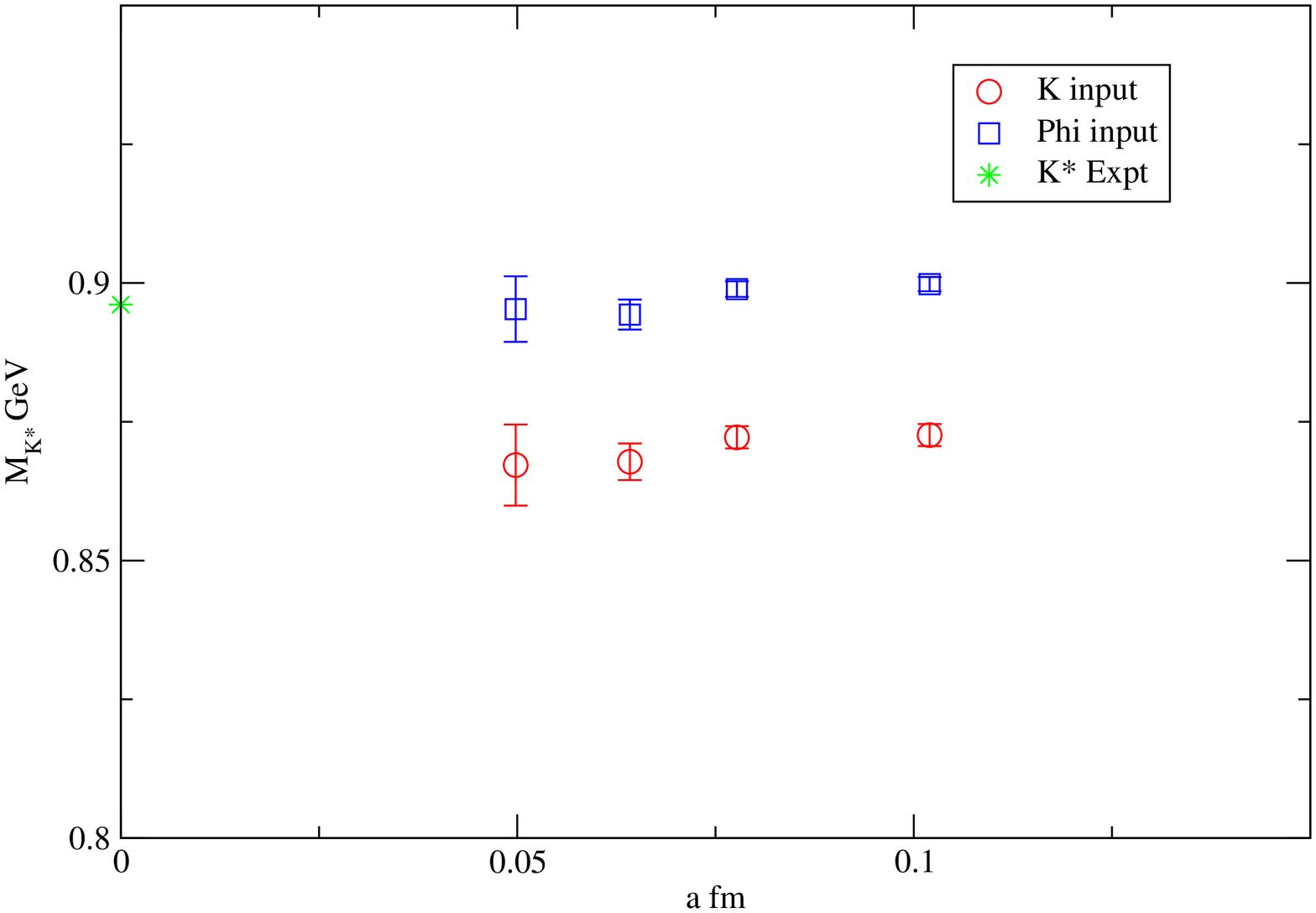}
\end{center}
   \caption{
Lattice spacing dependence of the $K^\star$ meson in quenched
QCD from the CP-PACS collaboration~\cite{Aoki:2002fd}.
 }
\label{cmn:aquenchDepend}
\end{figure}

To give some idea of the size of lattice spacing errors
in the masses of the light mesons, I plot the mass of the 
$K^\star$ meson as a function of the lattice spacing in 
figure~\ref{cmn:aquenchDepend} from 
the CP-PACS collaboration~\cite{Aoki:2002fd}.
The figure shows the difference between using the 
kaon or $\phi$ meson to determine the strange quark mass.
The clear difference between the mass of the $K^{\star}$
when the strange quark mass is determined in two different ways
is caused by the quenched QCD not being the theory of nature.
The results for meson masses in the continuum limit
are in table~\cite{Aoki:2002fd}.
\begin{table}[tb]
\begin{center}
  \begin{tabular}{|c|c|c|c|} \hline
Mass          & Result ($m_K$ ) GeV  &  Result ($m_\phi$ ) GeV & Expt GeV  \\ \hline
$m_K$         &     -                &  0.546(06)              &  0.496  \\  
$m_{K^\star}$ &     0.846(07)        &  0.891(05)              &  0.892  \\  
$m_\phi$      &     0.970(06)        &  -                      &  1.020  \\ \hline
  \end{tabular}
\end{center}
  \caption{
Masses of light S-wave mesons in quenched QCD 
from CP-PACS~\cite{Aoki:2002fd}.
The different analyses depend on which meson is used to
determine the strange quark mass.
}
\label{cmn:tab:cppacsMesonMass}
\end{table}

The physical summary of the results in
table~\ref{cmn:tab:cppacsMesonMass}
is that the hyperfine splitting is
too low from quenched lattice QCD. 
To isolate the reduction of the 
hyperfine splitting, 
  Michael and
Lacock~\cite{Lacock:1995tq,Maiani:1986yj} 
introduced 
the J parameter.
\begin{equation}
J = M_{V} \frac{d M_V} {d M_P^2 }\mid_{M_V = 1.8 M_P}
\label{cmn:eq:Jdefn}
\end{equation}
where $M_V$ and $M_P$ are the vector and pseuodscalar masses
respectively. The condition $M_V = 1.8 M_P$ corresponds to
the experimental ratio of $K^{\star}$ and $K$ masses.
This mass ratio was chosen so that an extrapolation 
to quark masses below strange was not required.
Some theoretical problems with the value of $J$
defined at a light quark reference point are discussed
by Leinweber et al.~\cite{Leinweber:2001ac}.

The J parameter has been chosen to be independent of the 
lattice spacing and a explicit definition of the 
quark mass. One experimental estimate for $J$
is obtained from
\begin{equation}
J  =  m_{K^{\star}} \frac{m_{K^{\star}} - m_{\rho} } {m_{K}^2 -
  m_{\pi}^2} 
   =  0.48 
\end{equation}
Including the uncertainty from using $\phi - K^\star$
difference rather than the $K^\star - \rho$ mass difference,
Michael and Lacock~\cite{Lacock:1995tq} estimate the 
experimental value of J to be $0.48(2)$.

In quenched QCD, Michael and Lacock~\cite{Lacock:1995tq}
obtained $J=0.37(2)(4)$ in disagreement with the 
experimental estimate. Using their bigger data set,
the CP-PACS collaboration~\cite{Aoki:2002fd}
obtained J = 0.346(23) from a continuum extrapolation
in quenched QCD.

The clear disagreement of the  $J$ parameter from
quenched QCD with experiment, in principle makes $J$ 
a good quantity to measure the effect of the 
sea quarks. The actual definition of J in 
equation~\ref{cmn:eq:Jdefn} for unquenched QCD is not trivial
because, as discussed in section~\ref{se:LatticeSpacing}, 
the lattice spacing does
depend on the mass of the sea quarks. The issues in defining 
$J$ for unquenched QCD are discussed by the 
UKQCD collaboration~\cite{Allton:2001sk}.

\begin{figure}
\def\filename{Jplot.ps}
\begin{center}
\includegraphics[scale=0.6]{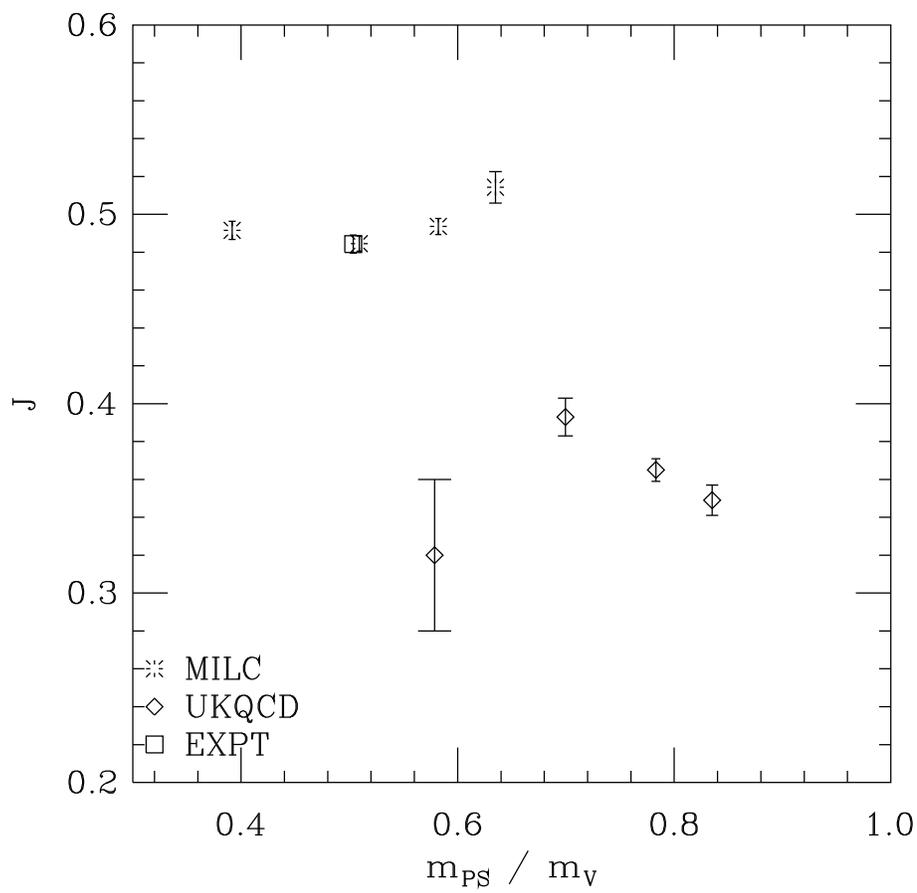}
\end{center}
   \caption{
Comparison of the J parameter 
from UKQCD~\cite{Allton:2001sk}  (diamonds)
and 
MILC~\cite{Bernard:2001av} (bursts).
The experimental point is the square.
 }
\label{cmn:fig:JPLOT}
\end{figure}

Figure~\ref{cmn:fig:JPLOT} compares the value of 
$J$ from the unquenched lattice QCD from 
UKQCD~\cite{Allton:2001sk} and MILC~\cite{Bernard:2001av}.
Both the calculations done by the 
UKQCD and MILC collaborations were at a fixed
lattice spacing. The calculation by MILC was done at
a lighter sea quark mass then UKQCD, that presumably 
explains why the value of J from MILC agrees
better with the experimental value. The MILC
result needs to be confirmed by a study of the lattice
spacing dependence.
Kaneko~\cite{Kaneko:2001ux} has recently reviewed 
the status of calculations of $J$ from unquenched
calculations.

The effect of unquenching on the hyperfine splitting
in light mesons has an important effect on 
the value of the strange quark mass extracted from lattice
data.
For example, the CP-PACS collaboration~\cite{AliKhan:2000mw} have 
used $n_f =2$ lattice QCD to calculate the strange quark mass.
CP-PACS's results are in table~\ref{tab:cppacsstrange}. 
The results show a sizable reduction in the mass of the 
strange quark between quenched and two flavour QCD.
Lubicz~\cite{Lubicz:2000ch} 
and Wittig~\cite{Wittig:2002ux}
review the determination of the masses of quarks from 
lattice QCD.
\begin{table}[tb]
\begin{center}
  \begin{tabular}{|c|c|c|} \hline
$n_f$  & input    &  $m_s$ MeV \\ \hline
2      &  $\phi$  &  $90^{+5}_{-11}$   \\
2      &  $K$     &  $88^{+4}_{-6}$   \\  \hline
0      &  $\phi$  & $132^{+4}_{-6}$   \\
0      &  $K$     & $110^{+3}_{-4}$   \\  \hline
  \end{tabular}
\end{center}
  \caption{
Mass of the strange quark from 
CP-PACS~\cite{AliKhan:2000mw}
in the $\overline{MS}$ scheme at a scale of 2 GeV.
}
\label{tab:cppacsstrange}
\end{table}

\subsection{P-WAVE MESONS AND HIGHER EXCITATIONS} \label{se:mesonexcite}

The calculation of the masses of ground state mesons and baryons is
essentially done to validate lattice QCD methods and to compute quark
masses and the strong coupling. For excited S-wave and
P-wave states, the issues are to determine the quark and
glue content of the state (as well as to check that the masses agree
with experiment). The signal to noise ratio is worse for P-wave mesons
than for S-wave mesons~\cite{DeGrand:1992yx}, so the calculations are
harder.  These type of calculations tend to be more exploratory, so
usually no attempt is made to take the continuum or infinite volume
limit. As the underlying lattice techniques mature, there will be
attempts quantify all the systematic errors.

The interpolating operators in table~\ref{tab:mesonOPERS} can be used
to create the P-wave $a_0$, $b_1$, and $a_1$ states. However, in the
quark model these states have zero wave function at the origin.  So
non-local interpolating operators with a node at the origin have
been tried~\cite{Martinelli:1983be,Hasenfratz:1982bw,Patel:1983ht}.
Reasonable results can be obtained using the 
operators in table~\ref{tab:mesonOPERS}.
Spin 2 states are not accessible from local interpolating operators,
so must be created using non-local operators.

A more general approach~\cite{Lacock:1996vy}
is to consider a generic 
non-local meson  operator at a specific time
$t$ (as is required for an interpolating operator for a time
correlation function).
\begin{equation}
O(\underline{r})
=
\overline{\psi}(\underline{x},t) 
\Gamma
\prod_{\underline{x},\underline{x}+ \underline{r}} U
\psi(\underline{x}+\underline{r},t)
\label{cmn:eq:interOPER}
\end{equation}
where the quark and anti-quark are separated 
by a distance $\underline{r}$
and $\Gamma$ ia an arbitrary gamma matrix.
The 
$\prod_{\underline{x},\underline{x}+ \underline{r}} U$ is a
set of gauge links that connects the quark to the anti-quark
in a gauge invariant way. In general
$\prod_{\underline{x},\underline{x}+ \underline{r}} U$ 
is not unique and will effect the transformation properties
of $O(\underline{r})$ under the cubic group.
Operators are designed to transform under specific representations 
of the cubic group.

\begin{table}[tb]
\begin{center}
\begin{tabular}{|c|c|c|} \hline
Continuum representation & cubic rep. \\ \hline
J=0 & $A_1$ \\
J=1 & $T_1$ \\
J=2 & $E$,$T_2$ \\
J=3 & $A_2$,$T_1$,$T_2$ \\
J=4 & $A_1$,$E$ , $T_1$,$T_2$ \\ \hline
\end{tabular}
\end{center}
\caption{
Representation of $SU(2)$ in terms of 
representations of the cubic group.
}
\label{cmn:tab:cubicREP}
\end{table}
The connection between the representations of the cubic group 
and the $SU(2)$ rotation group are in table~\ref{cmn:tab:cubicREP}.
The dimensions of the representations: $A_1$, $A_2$, $E$, $T_1$,
$T_2$ are 1,1,2,3, and 3 respectively. Hence the dimensions of the
representations match between the cubic group and
$SU(2)$ rotation group with representation of 
$2J+1$ in table~\ref{cmn:tab:cubicREP}.
 
If the gauge configurations are fixed to Coulomb gauge then non-local
interpolating operators based on spherical harmonics can be used.
This approach was studied by DeGrand and
Hecht~\cite{DeGrand:1992ng,DeGrand:1992yx}. As hadron masses are gauge
invariant quantities, gauge fixing at an intermediate stage should not
effect the final results.
Meyer and Teper discuss how to construct higher spin
glueball operators~\cite{Meyer:2002mk}.

There are number of interesting puzzles with the phenomenology of
``P-wave'' mesons.  For example, there are speculations that the
$a_0(980)$ particle is potentially not well described by a
$\overline{q}q$ state in the quark model. Its mass is very close to
the threshold for two kaon decay. There are many models that treat
this state as a $\overline{K}K$ molecules~\cite{Weinstein:1983gd} or
$\overline{q}\overline{q}qq$ state. As lattice QCD calculations use
the non-singlet $\overline{q}q$ operator to create this state, it may
not couple strongly to a molecular $\overline{qq}qq$ state. Therefore I
would expect that the lightest state in the $\overline{q}q$ channel to
be the $a0$(1450). However, this speculation should, and will be~\cite{Hart:2002sp}
tested in unquenched lattice calculations.

As discussed in section~\ref{cmn:se:unquench}, the interpretation of the
$a_0$ state is complicated by a quenched chiral
artifact~\cite{Bardeen:2001jm} in quenched QCD.  The largest
systematic study of the $a_0$ in quenched QCD was done by Lee and
Weingarten~\cite{Lee:1999kv}. Unfortunately, they could only find the
mass of the $a_0$ mass at the mass of the strange quark, possibly
because of problems with the artifact in this channel.  
Alford and Jaffe~\cite{Alford:2000mm}
review some of the earlier results for the 
mass of the $a_0$ from quenched QCD.
In a lattice
calculation that used domain wall fermions and modelled the quenched
chiral artifact, Prelovsek and Orginos~\cite{Prelovsek:2002qs}
obtained the lightest state in the $a_0$ channel to be 1.04(7) GeV.
At one coarse lattice spacing, Bardeen et al.~\cite{Bardeen:2001jm}
obtained a value of 1.34(9) GeV for the lightest state in the $a_0$
channel, also with an analysis that was aware of the quenched chiral
artifact.  The existing quenched lattice QCD data does not determine
the mass of the lightest non-singlet $0^{++}$ state

In a two flavour unquenched calculation the UKQCD
collaboration~\cite{Hart:2002sp} quote the preliminary result for the
mass of the $a_0$ to be 1.0(2) GeV at one lattice spacing. The MILC
collaboration~\cite{Bernard:2001av} have computed the $a_0$ mass with
2+1 flavours of sea quarks. The lightest mass of the $a_0$ in MILC's
calculation is 0.81 GeV~\cite{Bernard:2001av}.  The MILC
collaboration~\cite{Bernard:2001av} also claimed to see evidence for
the open decay channels of the $a_0$ state.  Hence, more work is
required in unquenched QCD to determine the mass of the lightest
non-singlet $0^{++}$ state.

The only values of the masses of the $a_1$ and $b_1$ mesons from
unquenched QCD, I could find were from the MILC
collaboration~\cite{Bernard:2001av}.  MILC obtained masses of the
$a_1$ and $b_1$ of 1.23(2) and 1.30(2) GeV compared to the
experimental values of 1.23(4) GeV and 1.2295(32) GeV
respectively~\cite{Hagiwara:2002fs}. There doesn't seem to be any
interesting experimental issues with the $a_1$ and $b_1$ mesons. As
noted by MILC~\cite{Bernard:2001av}, according to the quark model the
masses of the $a_1$ and $b_1$ mesons should be similar to the mass of
the $a_0$. Hence, any large splitting between the masses of $a_0$ and
$a_1$ and $b_1$ is indicative of dynamics beyond the simple quark
model.

The most surprising aspect of hadron spectroscopy are
the existence of Regge trajectories.
Empirically, the square of the mass of a hadron with mass $m(l)$
is linearly related to the spin by:
\begin{equation}
l = \alpha' M^2(l) + \alpha(0)
\label{eq:regge}
\end{equation}
Although the Regge trajectories can be explained 
by models (see~\cite{Brisudova:1998wq} for a review),
in particular the string model, the existence
of Regge trajectories  has not been shown to be
a rigorous consequence of QCD.
Equation~\ref{eq:regge} is a very useful tool in classifying
baryon states. Historically some of the spin assignments of baryons
were guessed from equation~\ref{eq:regge}~\cite{Hey:1983aj}.
Equation~\ref{eq:regge} is also useful in scattering experiments
(see~\cite{Forshaw:1997dc} for a discussion).

The simple relation in equation~\ref{eq:regge} has recently been
challenged by a number of authors. The improved precision of the
experimental data on hadron masses has allowed fits to the spectrum
that seem to show some nonlinearity in the relation between the square
of the meson masses and their spin~\cite{Tang:2000tb}.  Brisudova et
al.~\cite{Brisudova:1999ut} discuss various hadronic models that can
reproduce linear and nonlinear Regge trajectories. The nonlinearities
in the Regge trajectories were related to flux tube
breaking~\cite{Brisudova:1998wq}, hence the study of Regge
trajectories is complimentary to the study of string
breaking~\cite{Bernard:2001tz}.  Brisudova et
al.~\cite{Brisudova:1998wq} make the prediction that are no light
quark quarkonia beyond 3.2 GeV due to the termination of the Regge
trajectories.  If this conjecture is correct, then this might simply
searches for hybrid and glueballs in the region from 3 to 5 GeV.

The problem for lattice QCD calculations is that it is hard to
construct interpolating operators with spin higher than 2 on the
lattice.  As higher orbital states have larger masses than ground
state mesons, they have a worse signal to noise ratio, so it is hard
to extract a signal from the calculations.

\begin{figure}
\def\filename{chew_frautschi.ps}
\begin{center}
\includegraphics[angle=-90,scale=0.5]{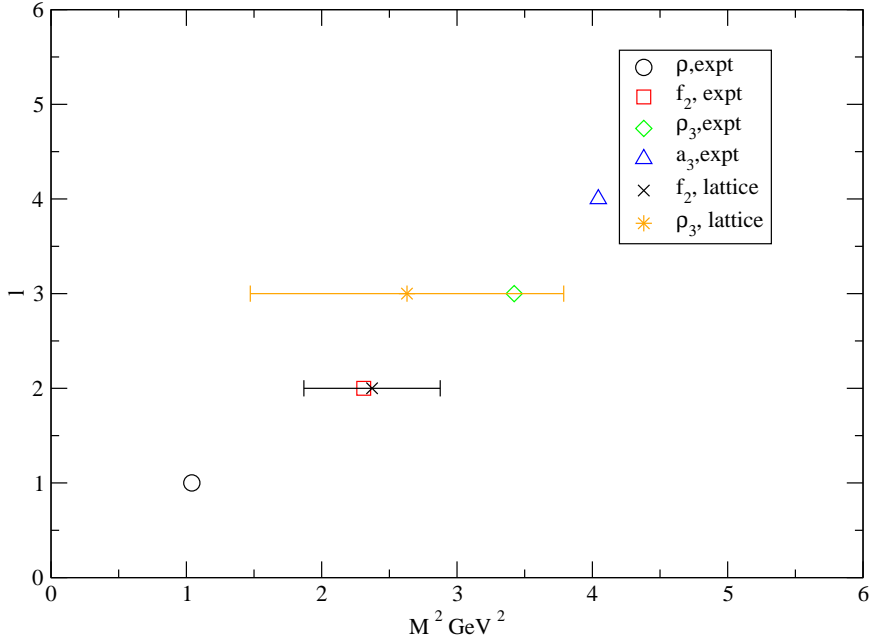}
\end{center}
   \caption{
Square of the mass of the meson mass versus its spin $l$.
The octagons are the experimental numbers and the bursts are the 
lattice QCD
data
from UKQCD~\cite{Lacock:1996vy}.  The quarks have the mass of the 
strange quark. There are only error bars on the lattice data.
 }
\label{eq:chewFIGURE}
\end{figure}

There have been some calculations of spin 3 masses from lattice QCD
calculations. The UKQCD collaboration~\cite{Lacock:1996vy} studied
spin 2 and spin 3 states in a quenched lattice QCD calculation, In
figure~\ref{eq:chewFIGURE}, I plot their results in a Chew-Frautschi
plot~\cite{Chew:1962eu} with experimental data from the particle data
table.

There have been a few attempts to study the spectroscopy of excited
mesons using lattice QCD. As discussed in
section~\ref{cmn:sec:basicLGT} the computation of the masses of
excited states requires a multi-exponential fit to the lattice
correlators that can be unstable.

The CP-PACS collaboration~\cite{Yamazaki:2001er} used the maximum
entropy method (briefly revived in section~\ref{cmn:sec:basicLGT}) to
study the excited spectrum of the rho and pion mesons. The
calculations were done in quenched QCD, using the same data set that
was used for their calculation of the ground state
masses~\cite{Aoki:1999yr}.

\begin{figure}
\def\filename{rho.eps}
\begin{center}
\includegraphics[scale=0.6]{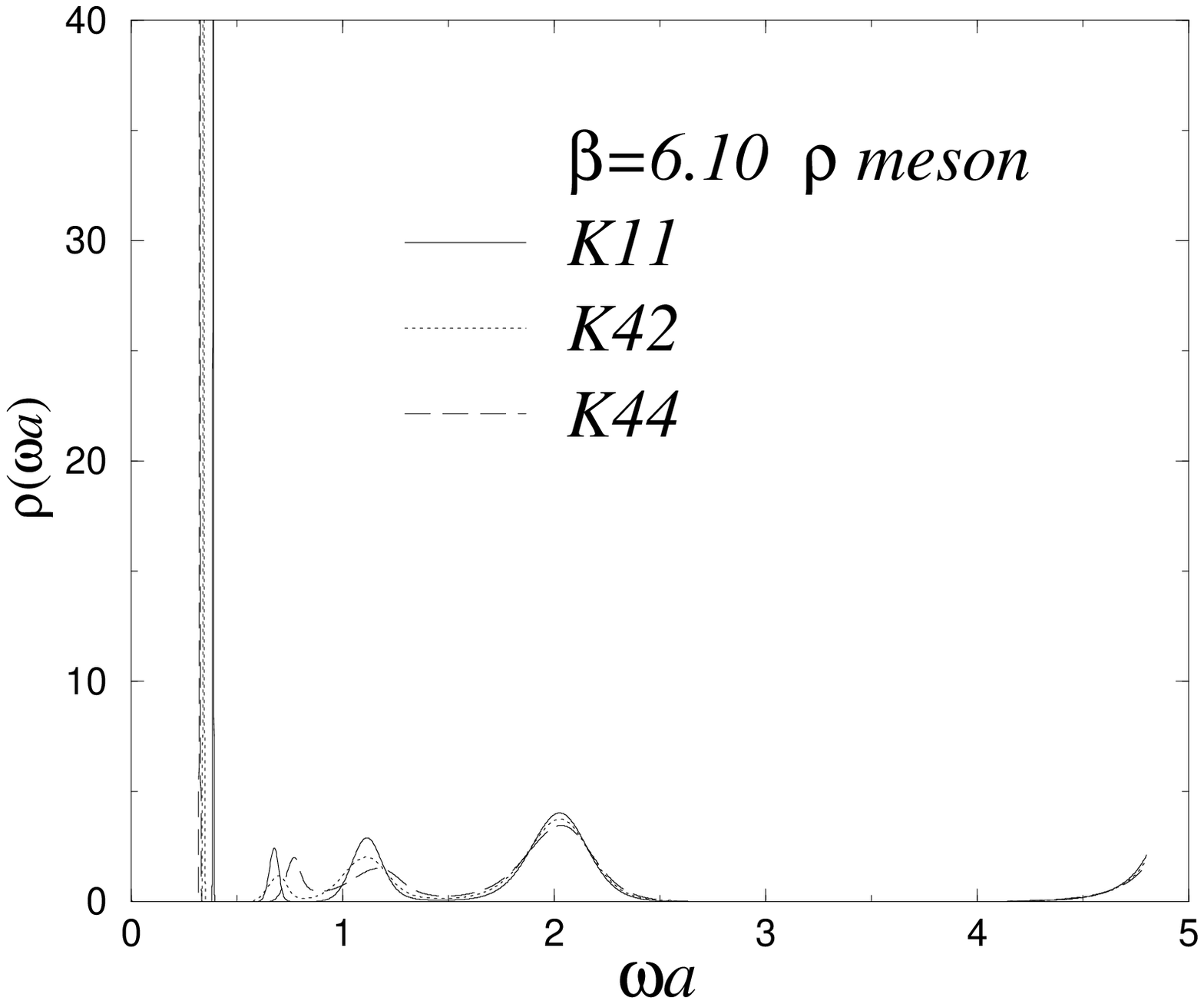}
\end{center}
   \caption{
Spectral function (from MEM) from the CP-PACS collaboration for the 
$\rho$ correlators. The lattice
spacing is $a^{-1}$ = 2.58 GeV. The K11, K42, and K44 keys
are the plots for mesons with different quark masses.
 }
\label{cmn:spectralCPACS}
\end{figure}

\begin{figure}
\def\filename{rhog.eps}
\begin{center}
\includegraphics[scale=0.6]{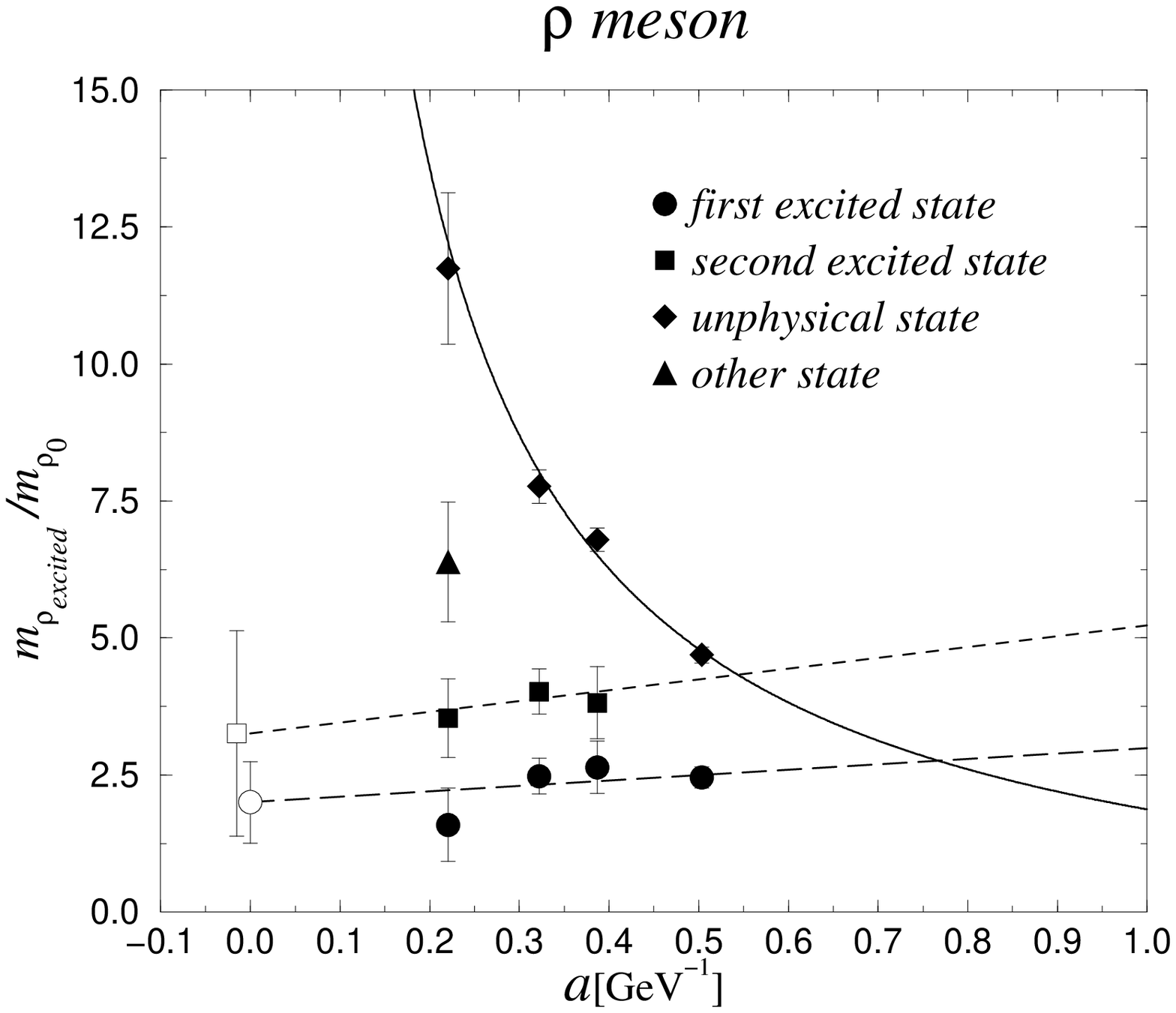}
\end{center}
   \caption{
Continuum extrapolation of the masses in the 
vector channel from CP-PACS.
 }
\label{cmn:excitedRHO}
\end{figure}

A spectral density from CP-PACS~\cite{Yamazaki:2001er}
is in figure~\ref{cmn:spectralCPACS}. The masses of the states
are obtained from the peaks in figure~\ref{cmn:spectralCPACS}.
Figure~\ref{cmn:excitedRHO} from CP-PACS~\cite{Yamazaki:2001er} 
shows the masses of the 
excited $\rho$ meson as a function of lattice spacing.
The diverging graph in figure~\ref{cmn:excitedRHO}
is thought to be a bound state
of fermion doublers.

The final results from CP-PACS~\cite{Yamazaki:2001er} were that the
first excited state of the pion had a mass of 660(590) MeV and the
mass of the first excited rho meon was 1540 (570) MeV from quenched QCD.
The errors also include an estimate of the error from taking the 
continuum limit.

Experimentally~\cite{Hagiwara:2002fs}, the first 
excited pion is the $\pi(1300)$ with the
mass of $1300 \pm 100$ MeV.  The excited states of the $\rho$ meson are
more interesting.  There are two states: $\rho(1459)$ with a mass of $1465
\pm 25$ and the $\rho(1700)$ with a mass of $1700 \pm 20$ MeV.
Donnachie~\cite{Donnachie:2001zg} reviews the evidence for a hybrid
state in the $1^{--}$ channel.

The errors on the excited states from CP-PACS are really too big for a
meaningful comparison with experiment. The results from
CP-PACS~\cite{Yamazaki:2001er} need confirmation from other, less
Bayesian based fitting methods, such as variational  methods.

\section{THE MASSES OF LIGHT BARYONS}

An important, but perhaps slightly boring, goal of lattice QCD is to
compute the mass of the nucleon with reliable errors from first
principles. The nucleon is the most important hadron for the real
world of the general public, but the nucleon's role in the esoteric
domain of particle physics is as ``background'' to the more
interesting stuff.  In this section, I will discuss the highlights of
some recent large scale quenched and unquenched QCD calculations.  As
the aim of lattice QCD calculations for these baryons is to validate
lattice QCD before the spectroscopy of more interesting hadrons is
attempted, the focus will be on the error analysis.

In principle the mass of the nucleon should be an ideal quantity to
compute on the lattice, because it is stable within QCD, hence there
are no concerns with the formalism due to decay widths.  The main
complication with getting an accurate value of the nucleon mass is
 the large chiral extrapolation required (see
section~\ref{eq:massDEPEND}) caused by the large quark masses used
in the calculations.  
In the majority of lattice QCD calculations,
electromagnetism is ignored, so I refer to the generic nucleon rather
than the neutron or proton.

The ``standard'' interpolating operators for the
nucleon~\cite{Montvay:1994cy,Hoek:1986sr} 
\begin{eqnarray}
N_{1}^{1/2+}  & = &  \epsilon_{ijk} ( u_i^T C \gamma_5 d_j) u_k
\nonumber \\
N_{2}^{1/2+}  & = &  \epsilon_{ijk} ( u_i^T C d_j) \gamma_5 u_k
\nonumber 
\end{eqnarray}
where $u$ and $d$ are operators that create 
the up and down quark respectively and $C$ is the 
charge conjugation matrix.
In the non-relativistic limit (keeping the upper
components) the $N_2$ operator vanishes, so it
doesn't couple strongly to the nucleon. Empirically
the $N_2$ operator has been found useful for $N^{\star}$ states.
Leinweber~\cite{Leinweber:1995nm}
discusses the connection between the interpolating
operators for the nucleon 
used on the lattice and those used in QCD sun rules.

The nucleon correlator 
is constructed, in the standard way, by 
creating a nucleon at the origin, 
and then destroying it at
time t later.
\begin{displaymath}
C_{\pm}(t)  =  \sum_{\underline{x}}
\langle 0 \mid
N(\underline{x},t) 
( 1 \pm \gamma_4) 
\overline{N}
(\underline{0},0) 
\mid   0 \rangle
\end{displaymath}
The specific representation of equation~\ref{eq:fitmodel}) is
slightly more subtle.
\begin{eqnarray}
C_{+} (t) & \rightarrow  & A^{+} e^{-m_{+} t }
+ A^{-} e^{-m_{-} (T-t) } \nonumber \\
C_{-} (t) & \rightarrow & A^{-} e^{-m_{-} t }
+ A^{+} e^{-m_{+} (T-t) } \nonumber 
\end{eqnarray}
where $m_{+}$ and $m_{-}$  are the masses of the 
lightest positive and negative parity states.

\begin{figure}
\def\filename{nucleon_qu_L.ps}
\begin{center}
\includegraphics[angle=-90,scale=0.5]{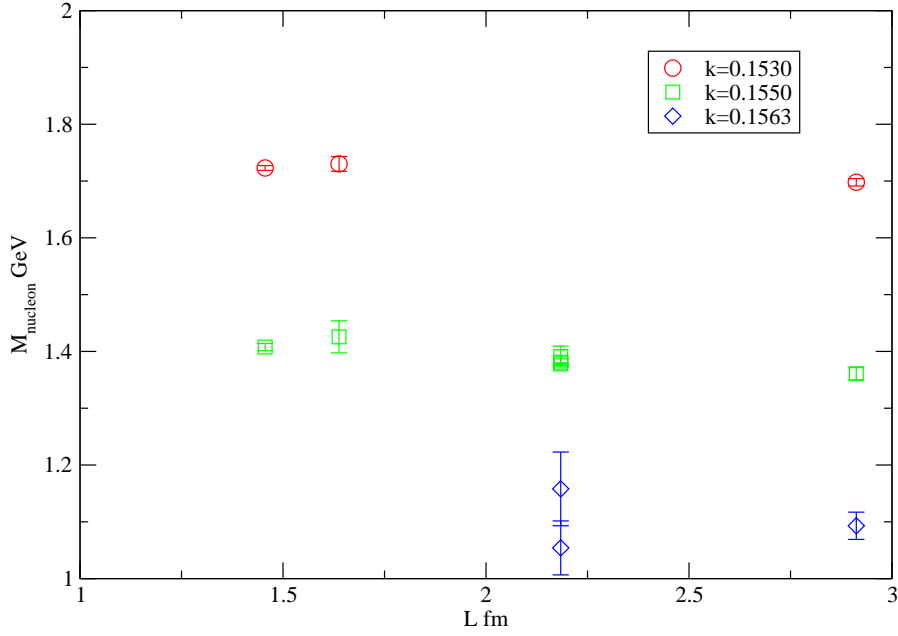}
\end{center}
   \caption{
Dependence of the mass of the nucleon on the box size
in quenched QCD for Wilson fermions at $\beta=6.0$.
The data was taken from the compendium of World data
in~\cite{Gockeler:1998fn}.
 }
\label{cmn:LquenchNucl}
\end{figure}

To illustrate the finite size effects on the mass of 
the nucleon in quenched QCD, I plot the nucleon mass 
as a function of box size for Wilson fermions at
$\beta=6.0$ in figure~\ref{cmn:LquenchNucl}. This data
set was chosen because there was data at a number
of different volumes. For all the data I used the 
lattice spacing of a= 0.091 fm~\cite{Guagnelli:1998ud}.
The dependence of the nucleon mass on the length of 
the lattice looks ``mild'' from figure~\ref{cmn:LquenchNucl}.
Unfortunately. the statistical errors on the lightest data are too large
to make any statement about finite size effects.

For many years, the quality of a lattice QCD calculations was judged by
the final value of ratio of the nucleon mass to rho mass.
I now discuss some recent quenched QCD calculations in
more detail.

There was a large scale calculation of the quenched QCD spectrum by 
the MILC collaboration~\cite{Bernard:1998db} that used staggered
fermions. The study included four different lattice spacings:
$a^{-1}$ from 0.63 to 2.38 GeV and also investigated finite
size  effects.

The MILC collaboration~\cite{Bernard:1998db}   
tested 12 different chiral extrapolation
models based on generic model for the dependence
of a hadron mass $M_H$ on the quark mass $m_q$.
\begin{equation}
M_H = M 
+ a m_q^{1/2}  
+ b m_q 
+ c m_q^{3/2} 
+ d m_q^{2}
+ e m_q^{2} \log m_q
\label{eq:cmn:fitMODEL}
\end{equation}
Only three of the fit parameters (a,b,c,d,e) were varied in 
a fit model. The fit model in equation~\ref{eq:cmn:fitMODEL}
was broad enough to include both full and quenched chiral perturbation
theory. It is important to test more than the quenched
chiral perturbation theory results, because it is not clear 
that the data is in the regime where the expressions are valid are valid.
The simple linear fit, where only $b$ was non-zero 
in equation~\ref{eq:cmn:fitMODEL}, was not consistent with the 
data~\cite{Bernard:1998db}.

The MILC collaboration found that the coefficient of the 
$m_q^{1/2}$ term from their fits was the wrong sign from 
the expectations from quenched chiral perturbation theory,
hence this term was not included in the their final fits.
For the $\rho$ extrapolation the coefficient of the 
$m_q^{3/2} $ term was an order of magnitude lower than
expected 
The analysis of the chiral extrapolations was complicated
by the flavour symmetry breaking terms of 
staggered fermions~\cite{Bernard:1998db}. This seems to be a 
perennial feature of the staggered 
fermion formalism~\cite{Bernard:2001yj}.

\begin{figure}
\def\filename{mNmrho_quench.eps}
\begin{center}
\includegraphics[angle=-90,scale=0.5]{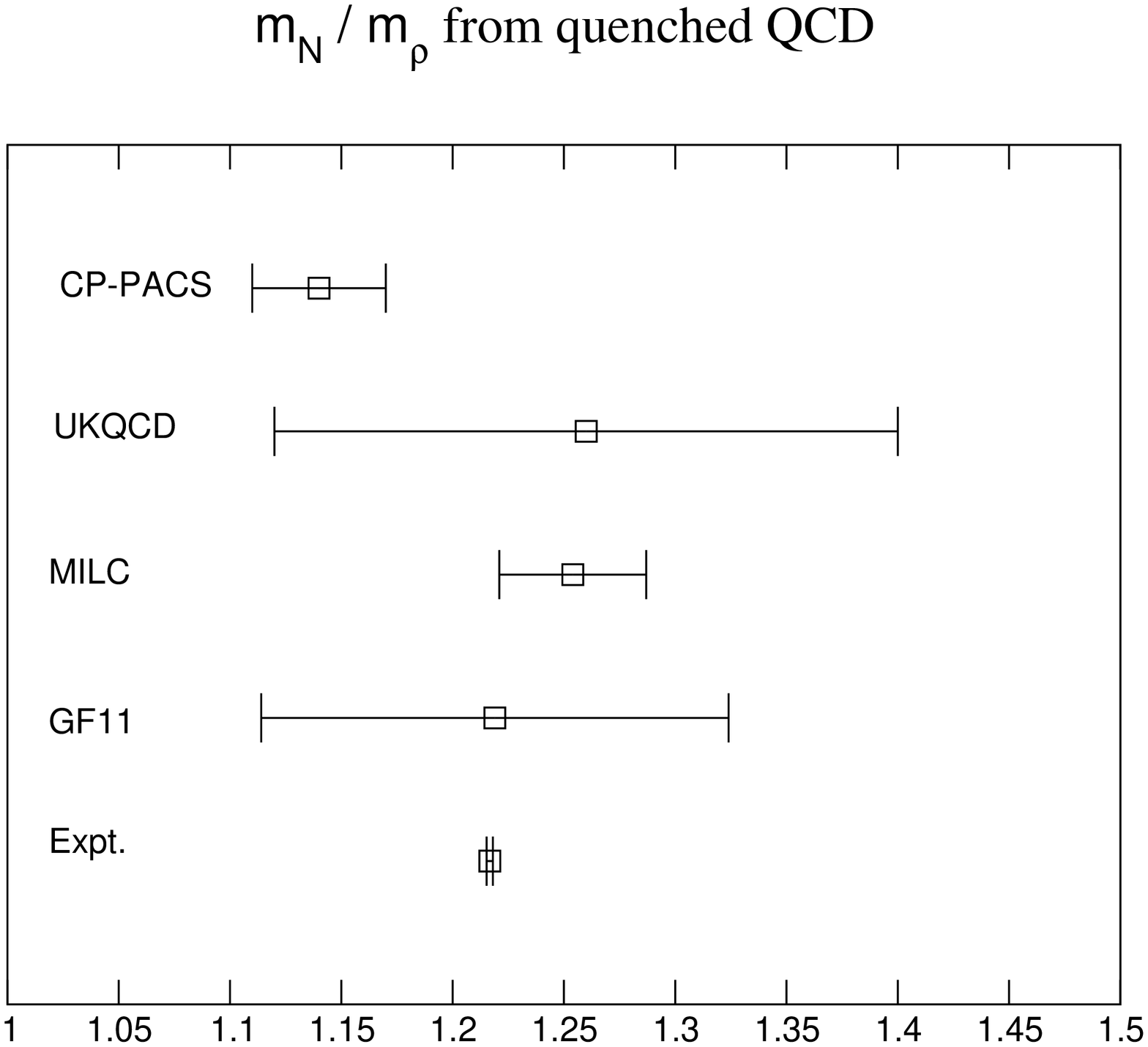}
\end{center}
   \caption{
The ratio of the nucleon mass to rho mass from 
several recent large scale quenched QCD calculations.
(~\cite{Butler:1994em,Aoki:2002fd,Bernard:1998db,Bowler:1999ae}.
).
 }
\label{cmn:mNmrhoquench}
\end{figure}

The result from the MILC collaboration~\cite{Bernard:1998db} for the
ratio of the nucleon to rho mass ratio was $m_N / m_\rho$ = $1.254 \pm
0.018 \pm 0.027 $, where the first error is statistical and the second
error is systematic (in the summary figure~\ref{cmn:mNmrhoquench}) I
have added the two errors in quadrature).

Kim and Ohta~\cite{Kim:1999ur} 
studied quenched QCD using staggered fermion
with a smaller lattice spacig ($a^{-1}$ = $3.63 \pm 0.06 $ GeV)
and lighter quarks than were
used by the MILC collaboration~\cite{Bernard:1998db}.
The spatial length was $2.59 \pm 0.05 fm$
and their lightest quark mass was 4.5 MeV.
They fitted  the same chiral extrapolation models
as MILC~\cite{Bernard:1998db} did, but also 
had problems unambiguously detecting the 
predictions from quenched chiral perturbation theory.
Kim and Ohta also studied some chiral extrapolation
formulae, suggested by~\cite{Mawhinney:1996qy}, 
that looked for finite volume effects masquerading
as quenched chiral logs.
Kim 
and Ohta's result~\cite{Kim:1999ur}  was 
$m_N/m_\rho$  = $1.24 \pm 0.04 (stat) \pm 0.02(sys)$


The CP-PACS collaboration~\cite{Aoki:2002fd}
found $m_N / m_{\rho}$ = 1.143 (33)(18). 
In the fits of the vector particle as a function
of the quark mass, CP-PACS excluded the 
$m_q^{3/2}$ term, but included the $m_q^{1/2}$ term
The coefficient of the $m_q^{1/2}$ term was found to be 
an order of magnitude less than the naive expectation.

In
figure~\ref{cmn:mNmrhoquench}, I plot the results for the ratio of the
nucleon and rho masses from some recent quenched lattice QCD
calculations~\cite{Butler:1994em,Aoki:2002fd,Bernard:1998db,Bowler:1999ae}.
I have selected calculations where an attempt was made to take the
continuum limit. 
For comparison, in the strong coupling limit $g \rightarrow \infty$ the hadron
spectrum can be computed analytically. In this limit the 
ratio of the nucleon and rho masses is
$\frac{m_N}{m_\rho} = $ 1.7~\cite{Smit:2002ug}

The agreement with experiment for the $m_N/m_\rho$ ratio from quenched
QCD is surprisingly good. 
Although agreement at the 
10 \% level may sound quite impressive, errors of this
magnitude are too large for QCD matrix elements required
in determining CKM matrix elements from 
experiment~\cite{Beneke:2002ks}.
The analysis of Booth et al.~\cite{Booth:1997hk} using the
non-analytical terms from quenched chiral perturbation theory estimated
that the value of $m_N/m_\rho$ from quenched QCD could be as low as
1.0. This situation is not seen in the lattice results, however none
of the quenched QCD calculations have detected the
predicted dependence
of the mass of the $\rho$ meson  on the quark mass,
To confirm the quark mass dependence predicted by
quenched chiral perturbation theory requires 
new calculations with lighter quarks.
The QCDSF collaboration~\cite{Gockeler:1999yj}
obtained a coefficient of the 
$m_q^{1/2}$ term from their fits that was roughly in 
agreement with expectations from quenched chiral perturbation
theory.

\begin{figure}
\def\filename{MassFinal.ps}
\begin{center}
\includegraphics[scale=0.8]{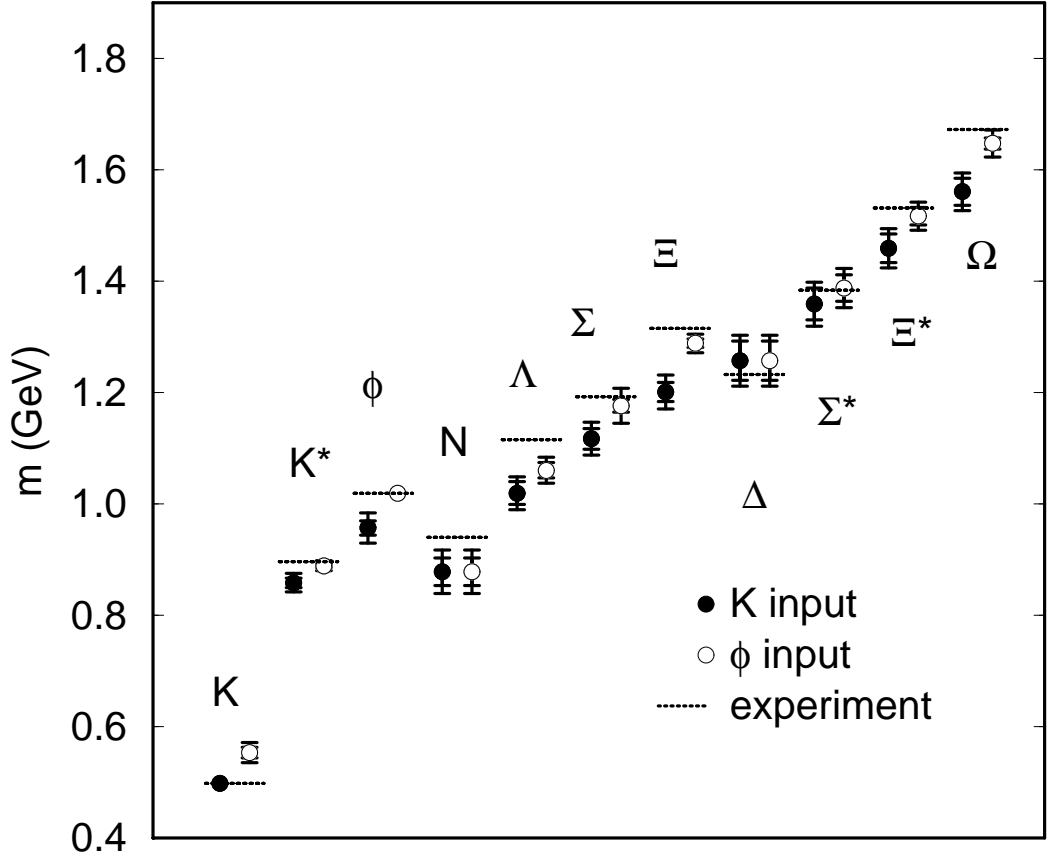}
\end{center}
   \caption{
Final results for the light hadron spectrum from CP-PACS in
quenched QCD~\cite{Aoki:2002fd}.
 }
\label{cmn:CP-PACSspectrum}
\end{figure}

A summary of some of the 
hadron spectrum from quenched QCD from 
the work by CP-PACS~\cite{Aoki:2002fd} is shown
in figure~\ref{cmn:CP-PACSspectrum}.
The ALPHA collaboration notice that the largest deviations
from experiment of the quenched lattice results 
from CP-PACS 
are for resonant hadrons~\cite{Garden:1999fg}
if the lattice spacing is determined from the
nucleon mass. 
The BGR collaboration also note a similar trend
in their data~\cite{Gattringer:2003qx}.
This trend could be modelled using
techniques similar to those used 
by Leinweber and Cohen~\cite{Leinweber:1994yw} to
study the $\rho$ meson.
It is difficult to see 
from figure~\ref{cmn:CP-PACSspectrum}
that the hadrons with large widths have
masses in quenched QCD that differ more
from experiment than those with smaller widths.

As discussed by Aoki~\cite{Aoki:2000kp} 
and the CP-PACS collaboration~\cite{Aoki:2002fd}
the value of the mass of the nucleon from
CP-PACS~\cite{Aoki:2002fd} and MILC~\cite{Bernard:1998db} 
differ at the $2.5 \sigma$ level. The two calculations 
used different fermion formulations, each with a different
set of potential theoretical problems, that should in principle
produce the same results in the continuum limit.

The main criticisms of the CP-PACS quenched study are on their
treatment of the quenched chiral perturbation theory.
Wittig reviews the lattice results for detecting the 
quenched chiral log~\cite{Wittig:2002ux}. 
The new calculations that use 
light fermion actions with better chiral symmetry properties
are disagreeing with the result from CP-PACS. These new calculations
are
done at a fixed (small volume) and large lattice spacings, so perhaps
there are systematic errors in their results~\cite{Wittig:2002ux}.
The CP-PACS and MILC collaborations  used the mass of the $\rho$  to determine
the lattice spacing. It would have been interesting to see the
dependence of the final results on using different quantities 
to determine the lattice spacing.

Although it may be possible to do a more systematic study of quenched
QCD at lighter quark masses than CP-PACS~\cite{Aoki:1999yr}, I am not
sure that it is worth it. As discussed in section~\ref{eq:massDEPEND}
the crucial non-analytic terms in quenched and unquenched chiral
perturbation theory are very different, so calculations with light
quarks in quenched QCD will have very little direct relevance for
unquenched calculations. As the masses of the quarks in unquenched
calculations decrease we should start to see the effect of the 
decay of the hadrons. The effect of particle decay on the 
mass spectrum can not be studied in quenched QCD.

\begin{figure}
\def\filename{mass_proton.ps}
\begin{center}
\includegraphics[angle=-90,scale=0.5]{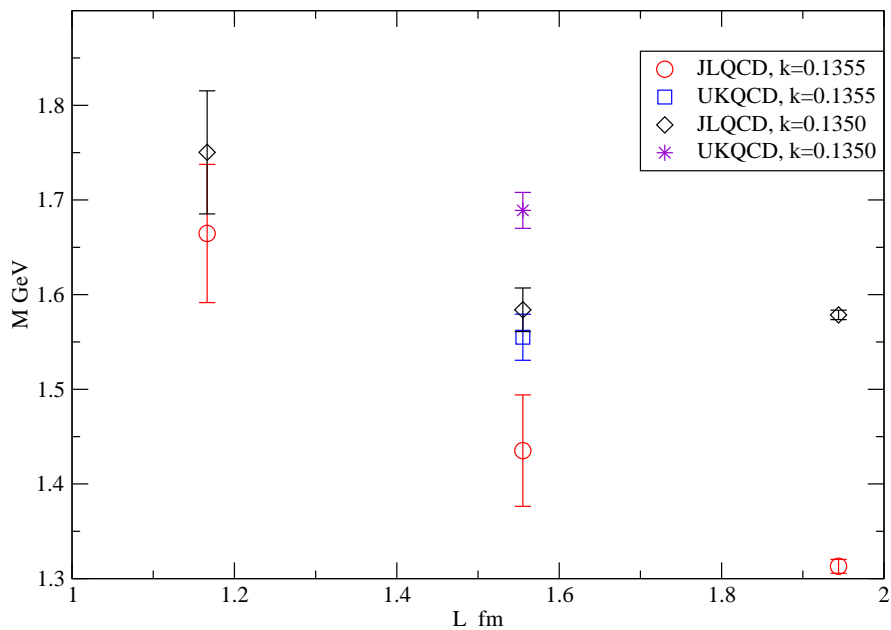}
\end{center}
   \caption{
Dependence of the mass of the nucleon on the box size
from 
UKQCD~\cite{Allton:1998gi} 
and JLQCD~\cite{Aoki:2002uc}.
These are two flavour unquenched calculations.
 }
\label{cmn:finiteVOLJLQCD}
\end{figure}

I will now describe the results from recent unquenched lattice QCD
calculations. As usual the systematic errors must first 
be discussed.
There have been two recent studies of finite size effects 
in the nucleon from unquenched QCD,
carried out by the JLQCD~\cite{Aoki:2002uc} and UKQCD~\cite{Allton:1998gi}. 
The results
are plotted in figure~\ref{cmn:finiteVOLJLQCD}.
One slight concern with the nice study of finite size effects
from JLQCD~\cite{Aoki:2002uc} is the large statistical errors
on the two smaller volumes. Also there is some disagreement between
the results from UKQCD and JLQCD on the $16^3$ volume.
As stressed by the MILC collaboration~\cite{Bernard:1993an}
a careful control of statistical errors is required to see definitive
evidence for the effect of the box size on the masses of hadrons.

\begin{figure}
\def\filename{cpacs_cont_limits.ps}
\begin{center}
\includegraphics[angle=-90,scale=0.5]{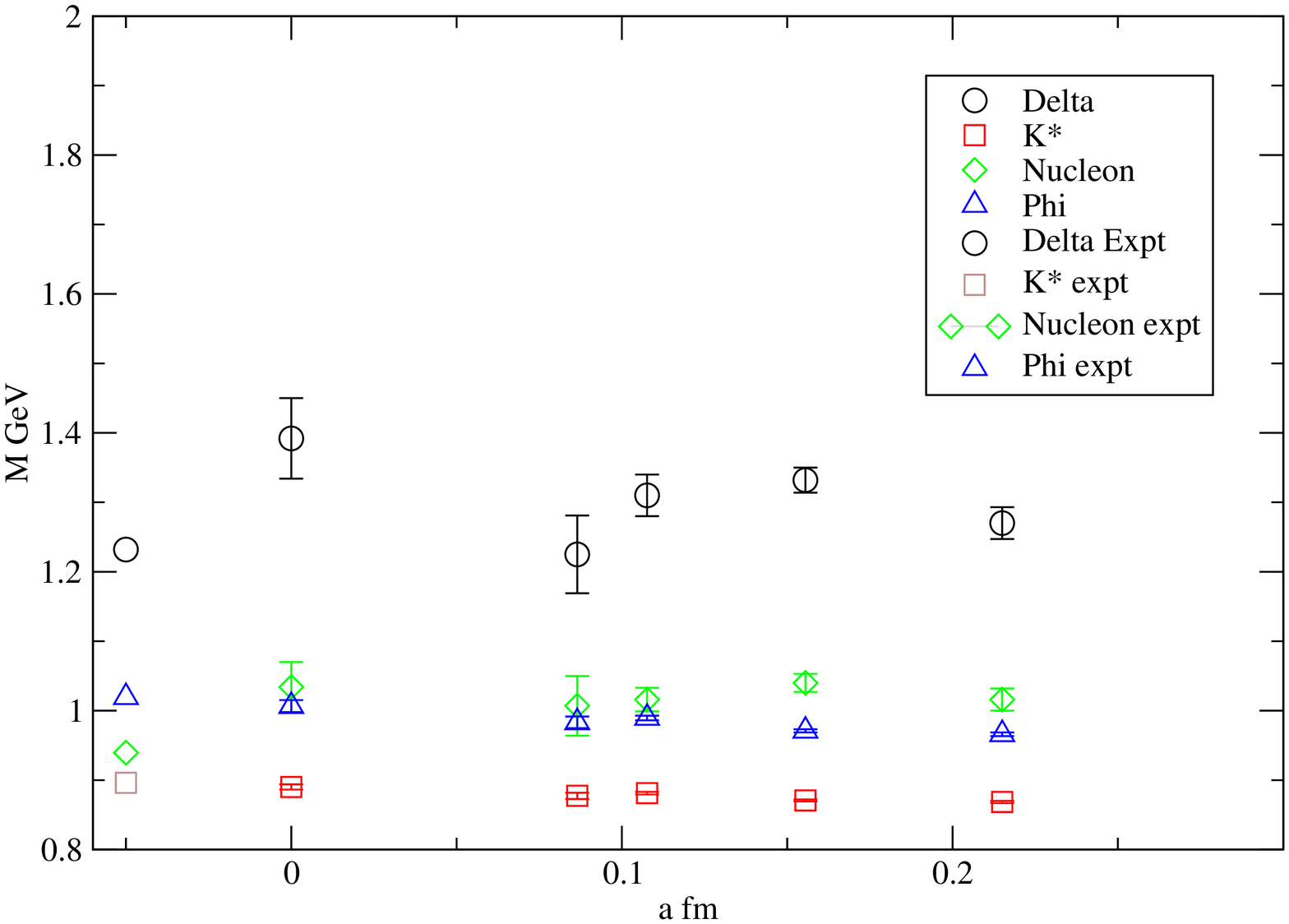}
\end{center}
   \caption{
Dependence of hadron masses on the lattice spacing.
from CP-PACS~\cite{Aoki:2002fd}
in 2 flavour unquenched QCD. Data from the finest lattice
spacing was not used in the continuum extrapolation.
 }
\label{cmn:cpPaCScont}
\end{figure}

In figure~\ref{cmn:cpPaCScont} I plot the dependence
of some hadron masses on the lattice spacing 
from the calculations by the CP-PACS 
collaboration~\cite{Aoki:2002fd}.
This should give some 
idea of the size of the lattice
spacing errors in unquenched QCD.

As discussed in section~\ref{se:lightMESON} the main 
``successes'' of unquenching have occurred in the meson sector.
A detailed comparison of the baryon spectrum with experiment is obscured 
by lattice spacing and finite size effects~\cite{AliKhan:2001tx,Aoki:2002uc}.
In a preliminary analysis the MILC collaboration~\cite{Bernard:1998ni}
 found that
$m_M/m_\rho$ from two flavour unquenched QCD in the continuum
limit was higher than the value in the quenched QCD, so the quenched
value agreed better with experiment than the unquenched result.
The MILC collaboration~\cite{Bernard:2001av} are now 
studying this issue using a better version 
of staggered fermions with 2+1 unquenched flavours.

In table~\ref{cmn:modelVlattice}, I compare the results from lattice
QCD with the old results from the Isgur-Karl quark
model~\cite{Isgur:1979be}.  Although the Isgur-Karl model agrees
better with experiment than the two lattice QCD calculations, because
the lattice QCD calculations are based on the QCD Lagrangian, the
hadron masses can be used to extract quark masses. This is not
possible from quark models where there is no way of relating the
constituent quark masses to the masses in the Lagrangian.  The more
interesting tests of the quark model occur for excited baryons. I
discuss the lattice results for these hadrons in
section~\ref{cmn:se:baryonexcite}.

In fact the agreement between the hadron masses from the 
quark model and experiment is actually too good. The quark
model calculation of Isgur and Karl~\cite{Isgur:1979be}
does not include the dynamics of hadron decay.
For example the $\Delta$ baryons have decay widths of
around 120 MeV~\cite{Hagiwara:2002fs}. It might have been 
expected that the errors in the quark model predictions
for the hadron masses would be of the order of the decay width.
Isgur and Geiger~\cite{Geiger:1990yc,Isgur:1999cd}
have developed a formalism to absorb some of 
the effects of quark-antiquark loops into the potential.
This issue is also reviewed by 
Capstick and Roberts~\cite{Capstick:2000qj}..

\begin{table}[tb]
  \begin{tabular}{|c|c|c|c|c|} \hline
Baryon & Expt.  & Isgur-Karl MeV &   CP-PACS (quenched) &  CP-PACS (unquenched) \\ \hline
$N$              &  940  & 940   & 878(25)   &  1034(36)  \\
$\Lambda$        &  1116 & 1110  & 1019(20) &  1160(32)  \\
$\Sigma$         &  1193 & 1190  & 1117(19) &  1202(30)  \\
$\Xi$            &  1315 & 1325  & 1201(17) &  1302(28)  \\
$\Delta$         &  1232 & 1240  & 1257(35) &  1392(58)  \\
$\Sigma^{\star}$ &  1384 & 1390  & 1359(29) &  1488(49)  \\
$\Xi^{\star}$    &  1532 & 1530  & 1459(26) &  1583(44)  \\
$\Omega$         &  1673 & 1675  & 1561(24) &  1680(41)  \\ \hline
\end{tabular}
  \caption{
Comparison of the hadron masses from unquenched and 
quenched lattice QCD  calculations with the
results from the Isgur-Karl quark model~\cite{Isgur:1979be}.
The unquenched data is taken from CP-PACS~\cite{AliKhan:2001tx}.
The quenched data is also from CP-PACS~\cite{Aoki:1999yr}.
For the lattice results the strange quark mass is
set by the kaon mass.
}
\label{cmn:modelVlattice}
\end{table}

\subsection{EXCITED BARYON STATES} \label{cmn:se:baryonexcite}

Recently there has been a lot of work on the spectroscopy of excited
nucleon states from lattice QCD. This research is mostly
motivated by the experimental program at the
Jefferson lab~\cite{Rossi:2003np,Burkert:2002nr,Burkert:2001nv}
The accurate spectroscopy of the $N^{\star}$ states will be an
accurate test of our understanding of the forces and effective degrees of
freedom in hadrons~\cite{Isgur:2000ad}. 
Realistically, lattice QCD calculations may only be able to
obtain one or two excited baryon states from
a specific channel. However this is enough to (potentially)
solve some very interesting and long standing  puzzles.

At the moment the identification of the Roper resonance 
from lattice QCD is controversial, so I will introduce 
some notation to prevent ambiguity. I call the nucleon the 
$N$ state, the first excited nucleon with positive parity
the $N'$ state, and the first excited nucleon with negative
parity the $N^{\star}$ state. In the particle data table, the 
$N'$  state would be the $N(1440)$ and the $N^{\star}$ state would be 
the $N(1535)$.

There is a potential  artifact associated with 
quenched QCD for baryon states~\cite{Labrenz:1996jy,Gockeler:2001db}.
The mass of the $\eta' N$ state is experimentally close to 
the mass of the $N(1440)$  and $N(1535)$ states. 
In quenched QCD the $\eta'$ is treated
incorrectly, so the intermediate $\eta' N$  state is also incorrect.
This is the analogue of the artifact in the scalar correlator
found by Bardeen at al.~\cite{Bardeen:2001jm} discussed in 
section~\ref{cmn:se:unquench}. 
Dong et al.~\cite{Dong:2003zf} claim to have seen the 
correlator, that would be positive definite in unquenched QCD,
for the $N^{\star}$ state go negative for pion masses below 248 MeV.
If this artifact causes the 
correlator to go negative this may ``confuse'' fitting techniques
such as the maximum entropy method that relies on a 
positive definite correlator.
There is no chiral artifact in unquenched QCD.
Work has started on studying the Roper resonance 
using 
unquenched QCD~\cite{Maynard:2002ys}.

In table~\ref{tab:NNstar} I have collected some  results for the
ratio of the $N^{\star}$ mass to the nucleon mass 
from some recent quenched lattice QCD calculations.
Only the calculation by 
Gockeler et al.~\cite{Gockeler:2001db}
took the continuum limit.
The experimental number corresponds to the mass of the 
$N(1535)$  divided by the nucleon mass about: 1.63. It is 
not clear what effect the 150 MeV width of 
the $N(1535)$ will be on the lattice result.

\begin{table}[tb]
  \begin{tabular}{|c|c|c|} \hline
Reference &  Comments & $M_{N^{\star}} / M_N$  \\ \hline
Blum et al.~\cite{Sasaki:2001nf} & Domain wall &  $1.49(9)$  \\
Gockeler et al.~\cite{Gockeler:2001db} & Clover & $1.50(3)$  \\
Broemmel et al.~\cite{Broemmel:2003jm} & Overlap & $1.77(7)$  \\
Dong et al.~\cite{Dong:2003zf} & Overlap & $1.67(12)$  \\ 
Nemoto et al.~\cite{Nemoto:2003ft} &  anisotropic clover &
$1.463(51)$ \\ \hline
\end{tabular}
  \caption{
Ratio of the mass of the parity partner of the nucleon
to the nucleon mass from the quenched QCD. 
}
\label{tab:NNstar}
\end{table}

The nature of the Roper resonance ($N(1440)$) is still a mystery.
There is an experimental signal for this state, but it is not clear
what the quark and glue composition of this hadron is. On the lattice 
three quark interpolating operators are used to study the Roper 
state. If the mass of the Roper is not reproduced, then this would be
evidence that additional  dynamics, beyond
three valence quarks, is important for this state.

The quark model has problems reproducing the 
experimental mass of the Roper resonance.
Using a simple harmonic oscillator potential 
to study the hadron spectrum, Isgur and Karl~\cite{Sasaki:2001nf,Isgur:1979wd} 
used a oscillator quantum of 250 MeV. In quark
model language, the $N^{\star}$ state would have 
one quantum above the ground state, and the 
$N'$ state would have two quantums above the ground state.
This predicted ordering is opposite to the experimental
masses of the $N(1440)$ and $N(1535)$.
Capstick and Roberts review the nature of the Roper resonance in
the context of potential models~\cite{Capstick:2000qj}.
Isgur also discusses the problems of the Roper resonance in
the quark model.~\cite{Isgur:2000ad}. The predictions of 
the quark model for the lowest excitations of the nucleon improve
if a more realistic potential is used, and the mixing 
between states is taken into account~\cite{Capstick:2000qj,Isgur:2000ad}.

There are predictions from flux tube and bag models that the lightest
hybrid baryon (three quarks with excited
glue)~\cite{Page:2002mt,Barnes:2000vn} is $J^P=1^+$ with a mass in the
region 1.5 to 1.9 GeV, hence close to the mass of the Roper resonance.

Sasaki,  Blum, and Ohta~\cite{Sasaki:2001nf} 
studied the first excited
state of the nucleon at a fixed lattice spacing of
0.1 fm and with a physical length of the lattice as
1.7 fm. Their calculations were done in quenched
QCD with quark mass in the range $M_{PS}/M_V$ 0.59 to 0.9.
The excited state masses were extracted using a variational
technique with two basis states that were different
interpolating operators for the nucleon. 
Sasaki et al~\cite{Sasaki:2001nf} could only
obtain a signal for the $N'$ state for pion masses
above 600 MeV. The mass of the $N'$ state was larger
than the $N^{\star}$ state.
If the variational technique was not biased by
truncation of the sum of excited states, then the
calculation of Sasaki~\cite{Sasaki:2001nf}  should be
able to resolve the ground and first excited states.
For the negative parity states, the masses obtained
from the ground and first excited state were 
degenerate. This could be interpreted as the 
variational technique not being able to resolve
two states. Experimentally the lightest $N^\star$ states
are the $N(1535)$ and $N(1650)$.

Melnitchouk et al.~\cite{Melnitchouk:2002eg} have also 
studied the spectrum the masses of the $N$, $N^{\star}$, and  $N'$
states. Their raw data for the splitting between the $N$ 
and $N^{\star}$ states is consistent 
with other groups~\cite{Sasaki:2001nf,Gockeler:2001db}.
The mass for the $N'$ state was much higher than the mass
of the experimental Roper 
resonance.

Dong at
al.~\cite{Dong:2003zf} claim
agreement between the mass of the 
$N'$ state from their calculation and 
the mass of the Roper resonance from experiment.
The calculation was done with the lattice
spacing of 0.2 fm and a physical box lengths of 2.4 and  3.2 fm.
The calculations used very light valence quarks (for
the lattice anyway). The lightest pion mass was 180 MeV.
The excited state masses were
extracted using the constrained curve fitting method developed
from the proposal by Lepage at al.~\cite{Lepage:2001ym}.
Dong at
al.~\cite{Dong:2003zf} reported that the mass of the 
$N'$ state started to decrease rapidly with pion masses 
below 400 MeV. The mass of the $N'$ state was less than that of the 
$N^{\star}$ state for pion mass of about 220 MeV. The mass quoted for the 
$N'$ state is 1462(157) MeV. The large error bars on this result means
that a 2 $\sigma$ statistical fluctuation would give 1776 MeV.

Sasaki~\cite{Sasaki:2003xc} has reported a study of the finite size effects
on the mass of the excited state of the nucleon.
The excited state of the nucleon was studied at the 
fixed lattice spacing of 0.09 fm. Three physical
box sizes were used: 1.5 fm, 2.2 fm and 3.0 fm.
The mass of the $N'$ state was extracted
using the maximum entropy method. The $N'$ and $N$
states  had finite size effects that increased
as the  light quark masses were reduced.
The final result for the $N'$ state looks as though
it would extrapolate 
to the experimental value from pion masses around
600 MeV. No dramatic decrease in the mass of the $N'$
is required.
The picture from the calculation of Sasaki~\cite{Sasaki:2003xc}
disagrees with that from Dong et al.~\cite{Dong:2003zf}.

Broemmel et al~\cite{Broemmel:2003jm} have tried to study the Roper
resonance using lattice QCD. They used an overlap-Dirac fermion
operator with a lattice spacing of 0.15 fm and two physical lattice
sizes of 1.8 and 2.4 fm. To study excited states they used a
variational technique based on three interpolating operators. They
could get a signal for the nucleon with pion masses as low as 220 MeV.
Unfortunately, they could extract a signal for excited nucleon states
with pion masses above about 550 MeV. This stopped them being able to
confirm the mass dependence of the $N'$ state claimed by Dong et
al.~\cite{Dong:2003zf}. This analysis did not include the effect of
the quenched chiral artifact in this channel.
Broemmel et al~\cite{Broemmel:2003jm} extracted both the ground and 
first excited state of the negative parity nucleon channel.

From the very interesting 
recent 
studies~\cite{Sasaki:2001nf,Dong:2003zf,Sasaki:2003xc,Broemmel:2003jm} of the 
Roper resonance on the lattice it is clear that 
careful attention will have to be paid to the 
systematic errors. 
Pion masses below 
200 MeV may be required. It would be good to have
variational calculations with wider sets of interpolating
operators as a check on the various Bayesian based fitting
techniques. For example it would be reasonable for 
``fuzzed``  nucleon operators~\cite{Lacock:1995qx} 
to couple  more to 
hybrid baryon states because they contain more 
glue.
It is a high priority that other groups
try to reproduce the results of
Dong at
al.~\cite{Dong:2003zf}.
and Sasaki~\cite{Sasaki:2003xc}.

One particular concern with getting the mass of the $N(1440)$ is that
its width is 380 MeV. The decay of $N(1440)$ will effect its mass.
This is also a difficulty for potential model calculations.  Capstick
and Roberts~\cite{Capstick:2000qj} estimate that ignoring the  width
of the $N(1440)$ causes an uncertainty of the order of 100 MeV on the
mass within the potential model framework.

Although I have focused on the excited states of the nucleon,
there is growing body of work on the parity partners of other 
baryons~\cite{Nemoto:2003ft,Melnitchouk:2002eg}.

\section{ELECTROMAGNETIC EFFECTS} \label{cmn:se:emEffects}

The majority of lattice QCD calculations do not incorporate the effect
of the electromagnetic fields in hadron mass calculations.  This is
reasonable, because the dominant interaction for hadron masses is the
strong force. I review the work done on including electromagnetic
fields in lattice QCD calculations.

The theory of QED has been 
studied by many groups using
lattice techniques. The formalism is similar to that
discussed in section~\ref{cmn:sec:basicLGT}, 
except that the gauge group
is U(1). However, there are some 
conceptual differences, because the U(1) gauge theory is 
not asymptotically free.
Issues relating to the non-asymptotically free
nature of QED have been studied in a non-perturbative 
way on the lattice~\cite{Gockeler:1997me,Kim:2001am}.
Also there are compact and non-compact versions
of the lattice $U(1)$ theory.
In non-compact QED the gauge fields $A_{\mu}^{em}$ take
to the range $-\infty$ to $\infty$.

To study electromagnetic effects on the hadron spectrum the 
electromagnetic fields have been quenched. The dynamics of 
the sea quarks have not been included in the generation
of the U(1) gauge fields.
A comparison between the use of background fields
in sum rules and lattice QCD has been made by
Burkardt et al.~\cite{Burkardt:1996vb}.

The most ``comprehensive'' 
study of the effect of electromagnetism on the 
masses of hadrons has been performed by 
Duncan et al.~\cite{Duncan:1996xy,Duncan:1997be}.
They used a non-compact version of QED
The gauge fields were generated using the action
\begin{equation}
S_{em} = \frac{1}{4 e^2} \sum_{x} \sum_{\nu \mu}
(D_{\mu} A_{\nu}(x) - D_{\nu} A_{\mu}(x) ) 
\end{equation}
where $e$ is the electromagnetic charge.
The $A_{\mu}(x)$ fields were subject to the linear 
Coulomb condition. The fields were promoted to compact fields
$U(x)_{\mu} = e^{\pm i q A_{\mu}(x)}$. This field coupled to a
quark field with charge $\pm q e $.

Currently, lattice QCD calculations are usually not accurate to 10
MeV, the order of magnitude of electromagnetic effects on the masses
of light hadrons.  To increase the mass splitting Duncan et
al.~\cite{Duncan:1996xy,Duncan:1997be} used large charges (2 to 6
times the physical values) and then matched onto chiral perturbation
theory that included the photon field.

The chiral extrapolation fit 
model used for pseudoscalars with electromagnetic fields
\begin{equation}
m_P^2 =   A(e_{q} ,e_{\overline{q}}  )
            +   m_q B(e_{q} ,e_{\overline{q}}  )
            +   m_{\overline{q}} B(e_{q} ,e_{\overline{q}}  )
\end{equation}
where $e_{q}$, $e_{\overline{q}}$
($m_q$, $m_{\overline{q}}$) are the (masses) of the quark and anti-quark

The calculations of Duncan et al.~\cite{Duncan:1996xy,Duncan:1997be}  
were done with one coarse 
lattice spacing $a^{-1} \sim 1.15$ GeV with a box size of
2 fm.
Some of the results for the mass splittings from Duncan et al. are in
table~\ref{cmn:tab:electromagEFFECTS}. I have included both the 
data and the corrected results, where
theoretical expressions 
were used 
to correct for finite volume effects.
\begin{table}[tb]
\begin{center}
  \begin{tabular}{|c|c|c|c|} \hline
Mass splitting  & Raw Lattice QCD (MeV)& Corrected MeV    &  Experiment (MeV)\\ \hline
$m_{\pi^{+}} - m_{\pi{-}}$   & 4.9(3) &  5.2(3)  & 4.594)  \\ \hline
$m_n$ - $m_p$  &  2.83(56)   & 1.55(56)  & 1.293  \\ \hline
$m_{\Sigma^0} -m_{\Sigma^+}$ & 3.43(39)  & 2.47(39) & 3.27 \\ \hline
$m_{\Sigma^-} -m_{\Sigma^0}$ & 4.04(36)  & 4.63(36) & 4.81 \\ \hline
$m_{\Xi^-} -m_{\Xi^0}$ & 4.72(24)  & 5.68(24) & 6.48 \\ \hline
  \end{tabular}
\end{center}
  \caption{
Electromagnetic mass splittings from 
Duncan at al.~\cite{Duncan:1996xy,Duncan:1997be} 
}
\label{cmn:tab:electromagEFFECTS}
\end{table}

There has some work on computing electromagnetic 
polarizabilities~\cite{Fiebig:1989en,Zhou:2002km,Christensen:2002wh}
from lattice QCD. 
The electric ($E$) and 
magnetic ($B$)
polarizabilities measure the interaction of a hadron with 
constant electromagnetic fields. Under the electromagnetic field
the mass of the hadron is shifted by $\delta m$.
\begin{equation}
\delta m = -\frac{1}{2} \alpha E^2 + 
           -\frac{1}{2} \alpha B^2 
\label{cmn:polarSHIFT}
\end{equation}
The $\alpha$ and $\beta$ quantities should be computable
from QCD.
Holstein~\cite{Holstein:2000yj} reviews the theory
and experiments behind the nucleon polarizabilities.
The experimental values for $\alpha$  and $\beta$
are extracted from Compton scattering experiments
(see ~\cite{MacGibbon:1995in} for example3).
A comparison of the results from lattice QCD
to models and experiment can be found in two
recent papers~\cite{Zhou:2002km,Christensen:2002wh}.

In contrast to work of Duncan et al.
the formalism used for electromagnetic 
polarizations~\cite{Fiebig:1989en} uses static
electromagnetic fields.
The $SU(3)$ gauge fields are modified by multiplying
\begin{equation}
U_{1}(x) \rightarrow  e^{ i \alpha q E x_4}   U_{1}(x)
\label{cmn:eq:emgauge}
\end{equation}
where $x_4$ is the Euclidean time variable and $E$ is 
the constant electric field. The phase factor in 
equation~\ref{cmn:eq:emgauge} can be linearised.
Smit and Vink have described how to put a constant magnetic field
on the lattice~\cite{Smit:1987fn}.

There are speculations that in very strong magnetic fields
($B\ge 5\times 10^{14}$ T),
the proton will become unstable to the decay to 
neutrons~\cite{Bander:1993ku}.
Magnetic fields of this intensity may be realised
in the universe~\cite{Grasso:2000wji}.
The original estimate~\cite{Bander:1993ku} 
of the instability of the proton
was done in the quark model.
In an attempt to remove some of the uncertainty in the 
hadronic calculation, Rubinstein et al.~\cite{Rubinstein:1995hc}
used lattice
QCD to study the dependence of the masses of the proton
and neutron on the magnetic field.

Some early
lattice calculations included the magnetic fields to look at the
magnetic moments of hadrons~\cite{Bernard:1982yu,Martinelli:1982cb}.
However, it is best to calculate magnetic
moments from form 
factors~\cite{Draper:1989bp,Draper:1990pi,Gadiyak:2001fe}, so 
electromagnetic fields are no longer used.

Electric fields were used in the first (unsuccessfully) attempts to
compute the electric dipole moment of the
neutron~\cite{Aoki:1989rx,Aoki:1990ix}.  The calculation of the
neutron electric dipole moment has recently been
reformulated~\cite{Guadagnoli:2002nm} in a way that does not require the 
use of electromagnetic fields.

\section{INSIGHT FROM LATTICE QCD CALCULATIONS}

The start of the book Numerical Methods for Scientists and Engineers
by Hamming has the immortal phrase: ``the purpose of computing is
insight not numbers.''  In the previous sections I have described how
lattice QCD is used to compute the masses of hadrons.  This may give
the impression that lattice QCD is essentially just a black box that
produces the masses of hadrons without any insight into the physical
mechanisms or relevant degrees of freedom.
In this section I hope to show that lattice QCD can
also help to explain the 
physical mechanisms behind the
hadron mass spectrum.
 A very good overview of the
type of insight wanted from hadronic physics, that contrasts the
hadron spectroscopy approach to the study of confinement with the
results from DIS type studies, is the paper by Capstick et
al.~\cite{Capstick:2000dk}.  Note however for the B physics
experimental program, high precision numbers with reliable error bars
are required to look for evidence for physics beyond the standard
model of particle physics~\cite{Yamada:2002wh}.

Isgur's motivation~\cite{Isgur:2000ad} for studying
the $N^\star$ particles is based on trying 
to understand the important degrees of freedom that describe
low energy QCD.  Increasingly, lattice QCD calculations are
being used to provide insight into the dynamics of QCD.
Some of Isgur's last 
papers~\cite{Isgur:1999ic,Isgur:2000ts,Horvath:2001ir,Horvath:2002gk} 
were devoted to using
lattice QCD to validate the quark model picture of hadronic
physics.

At first sight the model building approach to studying
QCD appears to give more insight into the dynamics of QCD.
It is usually quite easy to study the effect of adding new
interactions to the model.
There are many models for 
QCD interactions: quark model~\cite{Capstick:2000qj}, 
instanton
liquid model~\cite{Shuryak:2001as}, or bag model~\cite{Thomas:2001kw}.  
The main problem with the model based
approach to hadron spectroscopy is that is very difficult to judge
whether the assumptions in the models are valid.  The models are only
believed when they provide a reasonable description of ``most''
experimental data with a good $\chi^2$, but this does not 
prove that they are correct.
Different models of the QCD dynamics can be based on very different
physical pictures, but may give equally valid descriptions of
experimental data.  For example the physical assumptions behind the
bag model seem to be very different to the assumptions behind the
instanton liquid model~\cite{Shuryak:2001as}.
The major advantage of lattice QCD calculations is that the
theory can be mutilated in a controllable way, so the 
physical mechanism underlying a process can be studied.  

For example, the question about what mechanism in QCD causes a
linearly rising potential for heavy quarks at intermediate distances
is not something that can be answered by the quark model.
Greensite~\cite{Greensite:2003bk} reviews 
the work  on studying ``confinement''  using lattice QCD).

In this section I will focus on various attempts to explain
the value of the mass splitting between the nucleon and delta.
According to the quark model~\cite{Hagiwara:2002fs}, the masses of the nucleon and 
delta are split by a spin-spin hyperfine term:
\begin{equation}
H_{HF} = - \alpha_S M \sum_{i > j} 
( \vec{\sigma} \lambda_a)_i  ( \vec{\sigma} \lambda_a)_j
\label{eq:HFterm}
\end{equation}
where the sum runs over the constituent quarks and 
$\lambda_a$ is the set of SU(3) unitary spin matrices 
(a runs from 1 to 8).
In perturbative QCD a term of the form~\ref{eq:HFterm}
is naturally generated, but it is not clear whether the 
equation~\ref{eq:HFterm} is 
relevant for light relativistic quarks.

\begin{figure}
\def\filename{EDcont.ps}
\begin{center}
\includegraphics[scale=0.6]{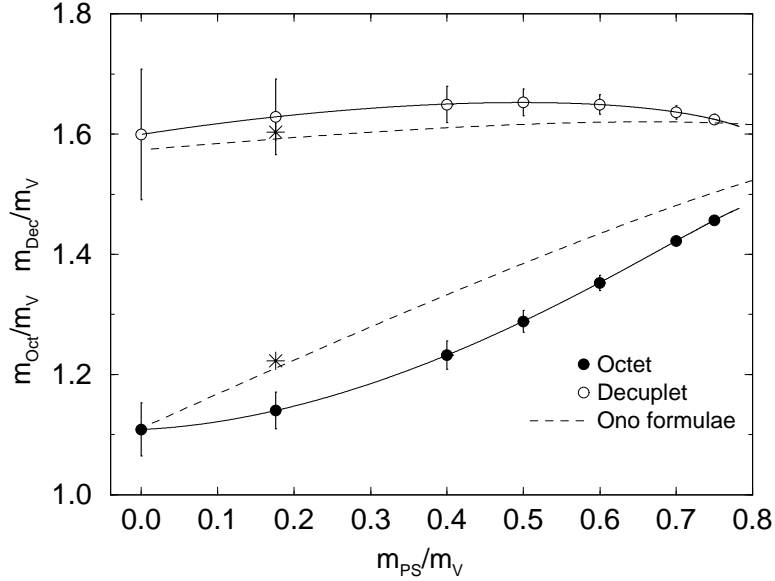}
\end{center}
   \caption{
Edinburgh plot from CP-PACS~\cite{Aoki:2002fd}
 from quenched QCD.
The lattice data is compared against the quark model of 
Ono~\cite{Ono:1978ss}.
 }
\label{cmn:cpPaCSEDplot}
\end{figure}

Lattice QCD calculations can produce more masses than
experiment, so the hyperfine interaction can be 
tested~\cite{Loft:1989ak,Iwasaki:1989qq}. 
A pictorial measure of the quark mass dependence of the 
masses of the octet and decuplet is given by
the ``Edinburgh'' plot from CP-PACS~\cite{Aoki:2002fd}
is in figure~\ref{cmn:cpPaCSEDplot}.
The continuous curve is from a quark model  by 
Ono~\cite{Ono:1978ss} that uses:
\begin{eqnarray}
M_{baryon} & =  & M_b + \sum_{i} m_i + \xi_b 
\sum_{i > j} \frac{ \vec{S_i} \vec{S_j}  }{ m_i m_j } \nonumber \\
M_{meson} & =  & M_m + \sum_{i} m_i + \xi_b 
\sum_{i > j} \frac{ \vec{S_i} \vec{S_j}  }{ m_i m_j } 
\label{cmn:eq:OnoModel}
\end{eqnarray}
The agreement between the lattice data and Ono's model
is reasonable in figure~\ref{cmn:eq:OnoModel}.
The lattice QCD data from CP-PACS~\cite{Aoki:2002fd}
in figure~\ref{cmn:cpPaCSEDplot}  is
almost precise 
enough
to show deviations from the model by 
Ono~\cite{Ono:1978ss} (equation~\ref{cmn:eq:OnoModel}).

In the instanton liquid model~\cite{Schafer:1998wv}, 
the vacuum is made up of a liquid of
interacting instantons.  There has been a lot of work on comparing
the instanton liquid model against lattice QCD.  The basic idea is to cool
the gauge configuration~\cite{Negele:1998ev}. This removes the
perturbative part of the gauge field and leaves the classical
configurations, that can be compared against the predictions of the
instanton liquid model.  The cooling procedure is essentially a way of
smoothing the perturbative noise from the configurations. This
perturbative noise presumably has something to do with the one gluon
exchange term. 

In the first study of this on the lattice~\cite{Chu:1994vi}, 
the masses of
the nucleon, $\rho$, $\Delta$ and $\pi$ were measured in the usual way.
The gauge configurations were then cooled and the 
simple hadron spectrum was measured again. The masses for the
$\rho$ $\pi$ and nucleon particles 
were
qualitatively the same before and after cooling. As the cooling
does not effect the instantons this suggested that the mass splittings
are largely due to instantons.
The mass splitting between the nucleon and $\Delta$ was
reduced by smoothing. Chu et al.~\cite{Chu:1994vi}
claimed that this was due to the problems with extracting the 
mass of the $\Delta$ from the lattice data. Later work on this 
issue~\cite{Kovacs:1999an,DeGrand:2000gq}
has not returned to the 
effect of instantons on the
mass splitting between the nucleon and $\Delta$.
Rosner and Shuryak~\cite{Shuryak:1989bf} have shown that
simple instanton interactions can give a reasonable 
representation of some baryon mass splittings.

In a series of papers, Glozman~\cite{Glozman:1999vd} and collaborators
have argued that a interaction based on Goldstone boson exchange
between constituent quarks gives a better description of the mass
spectroscopy of baryons, than interactions of the form in
equation~\ref{eq:HFterm}.

This is not the place for a detailed review of the case for and
against a interaction based on the exchange of Goldstone Bosons
(GBE). For a critique of GBE you can look at the paper by
Isgur~\cite{Isgur:1999jv}
and the review article by Capstick and 
Roberts~\cite{Capstick:2000qj}.  
Here I will just discuss the evidence for
GBE from lattice QCD.

\begin{figure}
\def\filename{valence_prop.eps}
\begin{center}
\includegraphics[scale=0.6]{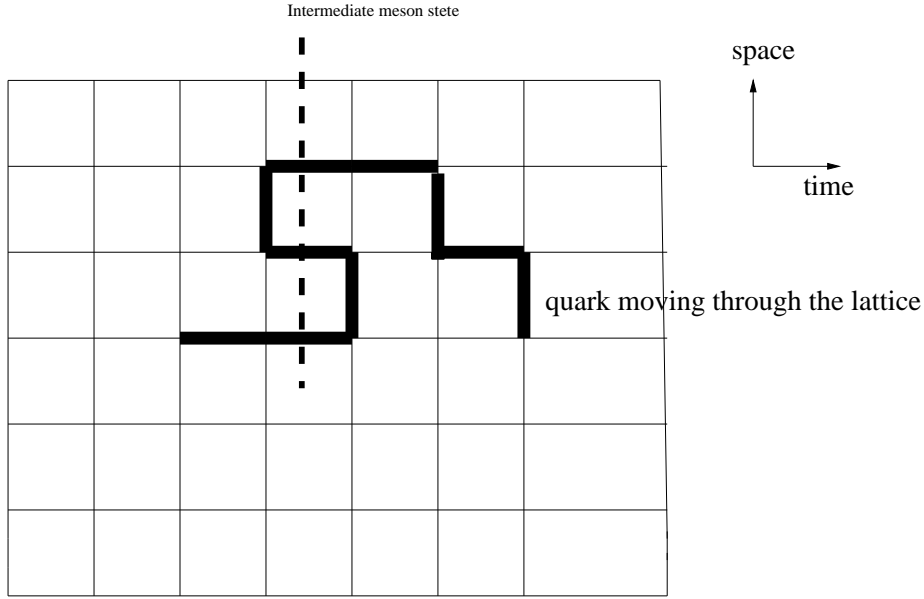}
\end{center}
   \caption{
Z graph for a quark.
 }
\label{cmn:fig:Zgraph}
\end{figure}

The Kentucky group~\cite{Liu:1998um} have introduced valence QCD, a mutilated
version of lattice QCD, that omits ``Z'' graphs from the formalism.
The aim was to study the foundation of the quark model.
In quenched QCD the sea quark loops are omitted, however there
are still higher Fock states from intermediate ``Z'' states.
Figure~\ref{cmn:fig:Zgraph} shows that a quark going backwards in time can be
interpreted as meson state.

The lattice version of valence QCD is the Wilson fermion action 
(equation~\ref{eq:wilsonFERMION})
with the backwards hopping terms removed.
\begin{eqnarray}
S_{f}^{W} = \sum_{x}
& ( &
-\kappa \sum_{i=1 }^{3}
\{
 \overline{\psi}_{x}(1-\gamma_{i})U_{i}(x)\psi_{x+i}
+
\overline{\psi}_{x+i}(\gamma_{i}+1) U_{i}^{\dagger}(x)\psi_{x}
\} \nonumber \\  \nonumber
& + &
\overline{\psi}_{x}\psi_{x}
 -\kappa \overline{\psi}_{x+\hat{t}}(\gamma_{4}+1)U_{\hat{t}}^\dagger(x)\psi_{x}
+
 \overline{\psi}_{x}(1 - \gamma_{\hat{t}}) \psi_{x}
)
\label{eq:wilsonFERMIONVALENCE}
\end{eqnarray}

The Kentucky group~\cite{Liu:1998um}  studied valence
QCD in a lattice calculation at $\beta=6.0$ with a 
volume of $16^3 \; 24$, and sample size of 100.
The Kentucky group~\cite{Liu:1998um} 
also used valence QCD to study form factors
and matrix elements, but I just will focus on their
results for the hadron spectrum.

\begin{figure}
\def\filename{qQCDvQCD.ps}
\begin{center}
\includegraphics[scale=0.8]{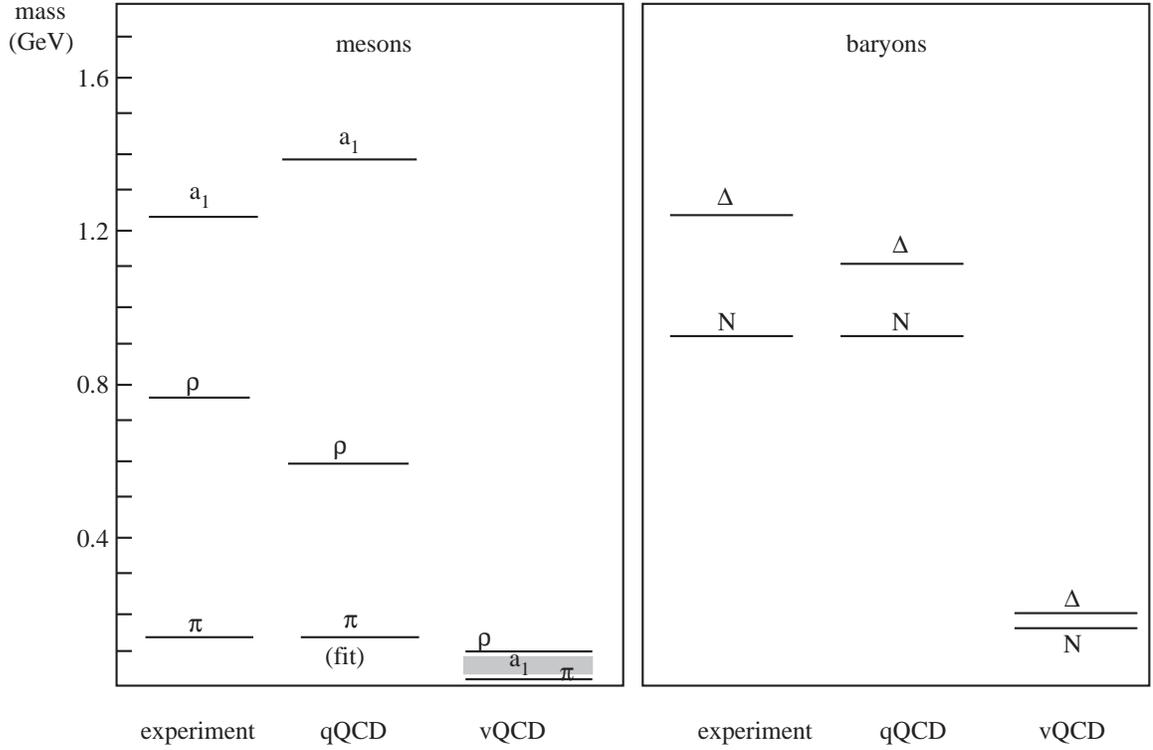}
\end{center}
   \caption{
Spectrum of valence (vQCD)
 and quenched (qQCD) QCD from~\cite{Isgur:1999ic}.
 }
\label{cmn:fig:valenceSPECTRUM}
\end{figure}

In figure~\ref{cmn:fig:valenceSPECTRUM}, 
I show a comparison of some hadron masses
(from the comment by Isgur~\cite{Isgur:1999ic})
The main conclusion from figure~\ref{cmn:fig:valenceSPECTRUM}
is that the hyperfine splittings seem to have been
reduced for light hadrons.

Valence QCD still has the physics of gluon exchange,
so the near degeneracy of the nucleon and $\Delta$,
suggests that the ``Z'' graphs are the important mechanism behind
the nucleon-delta mass splitting. 
For heavy hadrons the results from VQCD
are not suppressed relative to the those from 
quenched QCD. For example, the VQCD prediction for 
the mass splitting between $B^{\star}$ and $B$ is
$45.8 \pm 0.4$ MeV, compared to the experimental 
value of 46 MeV.
So the one gluon exchange is important for heavy quarks
(beyond charm), but less important for light quarks.
The VQCD calculations have much reduced hadron masses. For example.
the nucleon mass in VQCD is reduced by 700 MeV from quenched QCD.  The
reduction of the vector meson mass was 537 MeV. The Kentucky
group~\cite{Liu:1998um} attributed the reduction in hadron masses to
reduction of the dynamical or constituent quark mass due to the
omission of  Z graphs.

Isgur~\cite{Isgur:1999ic} criticised the conclusions from the 
study of lattice VQCD. In particular, Isgur notes that the 
one boson exchange interactions, based on Z graphs, operates between
two quarks,
so should not effect mesons made from a quark  and anti-quark.
So the near degeneracy between of the $\rho$ and $\pi$  masses 
in figure~\ref{cmn:fig:valenceSPECTRUM} would not be expected 
from the suppression of Z graphs.
Also Isgur~\cite{Isgur:1999ic}
 noted that the hadron spectrum from valence QCD was
radically different from experiment (and quenched QCD). This might
cause additional problems, if for example the wave-functions were
very different for valence QCD compared to the real world,
then this would effect the hyperfine splittings.
The Kentucky group did study matrix elements that are
related to the wave function of the states, but did
not see any problems~\cite{Liu:1999kq}.

The current lattice calculations are currently not able to determine
the correct mechanism 
for the mass splitting between the $\Delta$ and nucleon.
Much of the lattice work has concentrated on validating one 
particular model, rather than also falsifying competing models.
This type of physics is necessarily  qualitative.
For example, in the spirit of theory mutilation,
ideally only one piece of physics must be removed at one time.
It is
not clear what effect that valence QCD has on the instanton
structure of the vacuum. The modification of the Wilson fermion
operator in equation~\ref{eq:wilsonFERMIONVALENCE} will also effect
the zero mode structure that is of crucial importance to 
instanton inspired models. 
The operator in equation~\ref{eq:wilsonFERMIONVALENCE}
does not obey
\begin{equation}
M^{\dagger} = \gamma_5 M \gamma_5
\label{cmn:eq:hermLatt}
\end{equation}
for the Wilson operator $M$.
There is a generalisation of equation~\ref{cmn:eq:hermLatt} 
in valence QCD.
There has been some work on trying to ``disprove'' 
the instanton liquid model by studying lattice QCD gauge
configurations~\cite{Horvath:2001ir}. Edwards reviews the work 
by many groups on this~\cite{Edwards:2001ei}.

Lattice QCD simulations have been used to test other assumptions made
in models of the QCD dynamics.  For example, there are some models of
hadronic structure that are based on diquarks~\cite{Anselmino:1993vg}.  
A critical assumption
in diquark models is that two quarks actually do cluster to form a
diquark. This assumption has been tested in a lattice gauge theory
calculations by the Bielefeld group~\cite{Hess:1998sd}, where they
found no deeply bound diquark state in Landau gauge.
Leinweber~\cite{Leinweber:1993nr} has  claimed that 
lattice QCD data on the charge radii of hadrons
provides evidence against scalar diquark clustering.
As a test of the MIT bag
model~\cite{Lissia:1993gv} and Skyrme model~\cite{Chu:1994sy},
density operators from the models
were compared against results from quenched QCD.

The large $N_c$ limit of QCD provides much insight 
into QCD (see~\cite{Donoghue:1992dd} for a review).
Teper~\cite{Lucini:2001ej}
and collaborators have studied the glueball
masses in the large $N_c$  limit. Lattice QCD
calculations can be done with any gauge group.
Teper~\cite{Lucini:2001ej} et al. studied
the glueball masses for $N_c$ = 2,3,4,5.
This allowed them to estimate the size of the 
corrections to the $N_c  \rightarrow \infty$ 
limit. As the $O(1/N_c^2)$ corrections are small, it 
was important to control both the finite volume and 
lattice spacing errors.

\section{WHAT LATTICE QCD IS NOT GOOD AT} \label{eq:NotGoodAt}

In the previous sections I have implicitly assumed
that all the hadrons are stable. In the real world,
most hadrons are unstable to strong decays. For example the 
$\rho$ meson has a mass of 770 MeV and a decay width of 
150 MeV. Most lattice practitioners never ``worry'' about the 
$\rho$'s decay width (in public at least).
The determination of 
a hadron's
mass from
experiment is inextricably linked to the determination of the 
decay width. This is perhaps most dramatically demonstrated by the 
problem of missing baryon resonances~\cite{Koniuk:1980vw}. The quark
model predicted more excited baryons~\cite{Isgur:1979wd}
 than were actually seen in 
$N \pi$ reactions. 
It was claimed that the additional states were 
not seen because they coupled very weakly to  the $N \pi$ channel. 
The quark model
did predict that the missing baryons might be seen in $N \pi \pi$
reactions~\cite{Koniuk:1980vw}. 
There are experiments at the Jefferson lab
that are trying to detect
these ``missing resonances''~\cite{Burkert:2002nr}.

Lattice QCD calculations have to be done in Euclidean space for
convergence of the path integral~\cite{Glimm:1987ng}.  This implies
that the amplitudes and masses from lattice calculations are real.
This makes the study of resonances nontrivial, because decay widths are
inherently complex quantities. 
This is also a problem for calculations
with a finite chemical potential, although there has been some
progress in this area~\cite{Kogut:2002kk}.  

I do not discuss the very elegant
formalism of L\"{u}scher~\cite{Luscher:1991cf} for studying
unstable particles in a finite volume.
The formalism and results
from the scattering formalism are reviewed by Fiebig and Markum in this
volume~\cite{Fiebig:2002kg}. 
In this section, I would like to describe the possible implications for
mass determinations from standard correlation functions.

The momentum is quantised on a lattice of length $L$
and periodic boundary conditions. The momentum of mesons 
can only
take values:
\begin{equation}
p_n = \frac{2 \pi n }{a L}
\end{equation}
where $n$ is an integer between 0 and $L-1$.  For a typical lattice,
$L=16$ and $a^{-1}$ = 2.0 GeV, so the quantum of momentum is 0.79 GeV.
The quantisation of momentum makes the coupling of the state to
the scattering states different to that in the continuum. This 
``feature'' has been used to advantage by L\"{u}scher~\cite{Luscher:1991cf}
in his formalism.

The quantisation of momentum has important consequences for 
mesons that decay via $P$ wave
decays such as the $\rho$ meson. A $\rho$ meson at rest can only
decay to two pions with momentum $p$ and $-p$. The decay threshold is
$2 \sqrt{m_\pi^2 + (\frac{2\pi}{L})^2 }$. The quantisation of momentum on
the lattice does not effect the threshold for $S-$wave decays, 
An example of a S-wave decay is the decay of 
the $0^{++}$ meson to pairs of mesons. The 
quarks in current lattice calculations are almost light enough
for the strong decay of the flavour singlet $0^{++}$~\cite{Hart:2001fp}.

Hadron masses are extracted from lattice QCD calculations
using two point correlators (equation~\ref{eq:fitmodel}).
However the use of equation~\ref{eq:fitmodel} may 
not be appropriate for hadrons that can decay.
The most naive modification
of the lattice QCD formalism
caused by the introduction of decay
widths is the replacement 
\begin{equation}
m \rightarrow m + \frac{\Gamma i }{2}
\end{equation}
where $\Gamma$ is the decay width.
This modifies equation~\ref{eq:fitmodel} 
\begin{equation}
c(t) = a_0 e^{ -m_0 t } e^{ -\frac{\Gamma i t}{2} }    + \cdots
\label{eq:fitmodelMINK}
\end{equation}
that is used  to extract the masses from correlators.
There are a number of problems with equation~\ref{eq:fitmodelMINK},
so a more thorough derivation is required.

I review the work by Michael~\cite{Michael:1989mf} 
(see also the text book by Brown~\cite{Brown:1992db} )
on the effect of decays
on two point functions.
I will consider two scalar fields 
($\sigma,\pi$) interacting with the
interaction $\sigma \pi \pi$
The mass of the $\pi$ particle is $\mu$.
To study the implications of particle decay on the 
two point correlator, consider the renormalisation
of the propagator of the $\sigma$ particle in Euclidean space
\begin{equation}
P_B^{-1}(p) = p^2 + m^2
\end{equation}
The effect of the interaction of the $\pi$
particle with the $\sigma$ particle
renormalises the $\sigma$ propagator.
\begin{equation}
P^{-1}(p) = p^2 + m^2 -X(p^2)
\end{equation}
where $X$ is the self energy.

Masses are extracted from lattice QCD calculations
using the time sliced propagator:
\begin{eqnarray}
G(t) &= & \frac{1}{2 \pi} \int^{\infty}_{-\infty}
dp_0 e^{ i p_0 t } P(p_0,\underline{0}) \\
& = &
\frac{1}{\pi} 
\int_{2 \mu}^{\infty}
dE e^{-E t} \rho(E)
\end{eqnarray}
where 
\begin{equation}
\rho(E) 
=
\frac{Im X(E) }
{(m^2 - E^2 -X)^2}
\end{equation}

If the pole around $E \sim -M$ is neglected 
then the expression for the spectral density is
\begin{equation}
\rho(E) = \frac{1}{2 M} 
\frac{ \gamma(E)  }
{ (M-E)^2 + \gamma(E)^2  }
\end{equation}

In the limit 
$\gamma  \ll  (M - 2 \mu) , ( M - 2 \mu ) t \gg 1 $
\begin{equation}
G(t) = \frac{1}{2 M} 
e^{ -M t } \cos(\gamma t ) + 
\frac{\gamma e^{- 2 \mu t }}
{2 \pi M (M-2\mu)^2  t }
\label{cmn:eq:corrRESmodel}
\end{equation}

\begin{figure}
\def\filename{resonance_plot.ps}
\begin{center}
\includegraphics[angle=-90,scale=0.5]{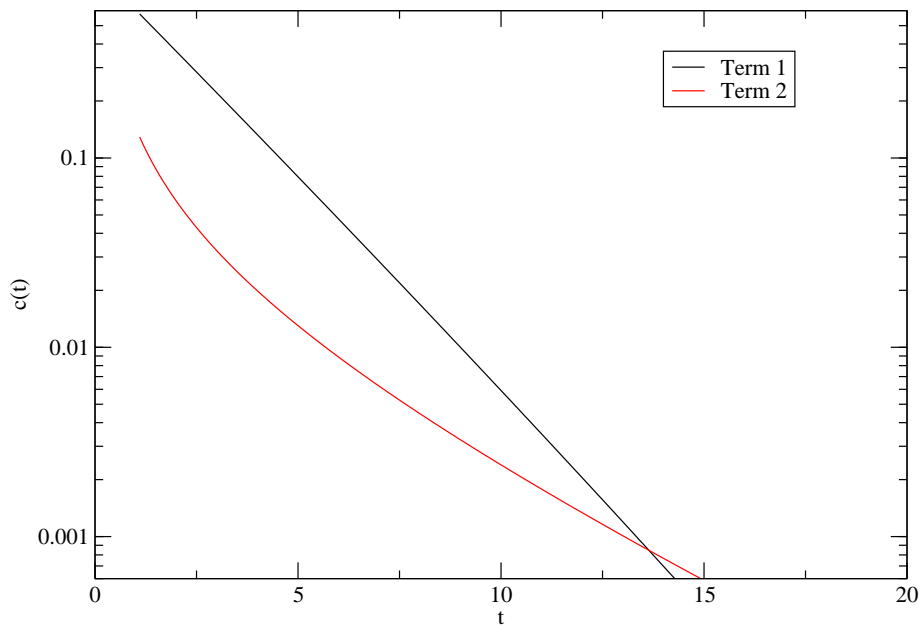}
\end{center}
   \caption{
Resonant correlators from equation~\ref{cmn:eq:corrRESmodel}.
 }
\label{cmn:plot:resCORR}
\end{figure}

In figure~\ref{cmn:plot:resCORR}, I plot 
separately the log of the first and 
second terms in equation~\ref{cmn:eq:corrRESmodel}
using the parameters: ($M=0.5,\gamma=0.05,\mu=0.1$).
The breakdown of a pure exponential decay can be
seen for the second term.

There is no evidence for the breakdown of the simple 
exponential fit model in current lattice
calculations (apart from the effects of chiral artifacts
in quenched QCD~\cite{Bardeen:2001jm}).
This effect may become apparent as the sea quark
masses are reduced to where particle decay is energetically
allowed.

The maximum entropy approach~\cite{Asakawa:2000tr,Yamazaki:2001er} to
extracting masses from correlators produces an estimate of the
spectral density, so in principle could be used to extract the decay
widths of particles. It is not clear to me whether a decay width
obtained from an analysis based on maximum entropy method would be
physical. Yamazaki and Ishizuka~\cite{Yamazaki:2002ir} have recently
compared the maximum entropy approach to studying unstable particles to
the method advocated by L\"{u}scher~\cite{Luscher:1991cf} in a
model theory. Yamazaki and Ishizuka claimed good agreement 
between the two methods.

Another way of looking for ``evidence'' of resonant behaviour is to
look for peculiarities in the quark mass dependence of the hadron
masses.  One of the first applications of this idea was by DeGrand who
studied the effect of $\rho$ decay on the quark mass dependence
of the rho mass~\cite{DeGrand:1991ip}.  The mixing of the $\rho$
correlator with $\pi\pi$ states at the nonzero momentum
makes the mass dependence of the $\rho$ correlator more
complex than described by the theory in section~\ref{eq:massDEPEND}.
The MILC collaboration~\cite{Bernard:2001av} have recently claimed to
see some evidence for the decay of the $a_0$ (non-singlet $0^{++}$) 
meson by comparing the
mass dependence of the $a_0$ particle 
in unquenched and quenched QCD.

The MILC collaboration~\cite{Bernard:1993an} 
tried to look for the evidence of
$\rho$ meson decay by studying the dispersion relation of the rho 
particle. The momentum can be injected into particle
correlators, so that the hadron masses can be computed 
at nonzero momentum. The signal to noise gets worse, as the 
momentum increases, so typically the dispersion relation is only known
for a few values of momentum.
On the lattice the dispersion relation is of the 
form~\cite{Bhattacharya:1996fz}
\begin{equation}
\sinh^2 E = \sinh^2 m + \sin^2 p
\label{eq:cmn:dispersionCONT}
\end{equation}
The ``improvement'' program described in the appendix,
aims to make the lattice dispersion relation closer to the 
continuum one. Some clever lattice theorists call this
computing ``the speed of light''~\cite{Alford:1998yy}.

The relationship in equation~\ref{eq:cmn:dispersionCONT} predicts
that the dependence of the 
hadron mass as a function of momentum. Consider an
interpolating operator for the $\rho$  meson:
$\overline{\psi}\gamma_{i}\psi$. If a unit
of momentum is injected in the $z$ direction, then the 
$\overline{\psi}\gamma_{z}\psi$ operator will couple to
the $\rho$ with one unit of moment, as well as two pions 
with unit momentum. The two states will mix and the masses will
be modified. 
However, the operators: $\overline{\psi}\gamma_{y}\psi$, 
$\overline{\psi}\gamma_{x}\psi$ do not couple to the 
two pion states, essentially because $\rho \pi\pi$ interaction is
zero for these kinematics. A signal for rho decay would be 
a different mass from the $\rho$'s polarised perpendicular
and parallel to the momentum of the state. The MILC collaboration
did not see this effect, perhaps because of artifacts with the 
type of fermions used. The UKQCD collaboration have recently claimed to see
evidence for $\rho$ decay via this mechanism~\cite{McNeile:2002fh}
that was consistent with the results from other methods.

In this review I have focused on just computing the masses 
of hadrons. 
The computation of decay widths from first principles is the holy
grail of hadron spectroscopy. 
There have been very few attempts at the calculation of the decay
widths from lattice QCD. The hadronic coupling constants are computed
from matrix elements, rather than trying to fit expressions
like equation~\ref{cmn:eq:corrRESmodel} 
to data.
Additional correlators over the standard two
point correlators are required to be calculated. 
There have been
some
attempts to compute the rho to two pion coupling
constant~\cite{Gottlieb:1984rh,Altmeyer:1995qx,Loft:1989sy,McNeile:2002fh}.
Perhaps the most famous calculation 
decay widths is the calculation of 
partial widths for the decay of $0^{++}$ glueball were calculated by the GF11
group~\cite{Sexton:1995kd}. A clarification of the formalism 
for studying decay widths and mixing in lattice QCD calculations is
described in~\cite{Michael:1989mf,McNeile:2000xx}.

\section{CONCLUSIONS}

The computation of the light hadron mass spectrum is essential 
to checking the validity of lattice QCD techniques.
For example, to reliably extract quark masses
requires a precise and consistent calculation
of the masses of the light mesons. A precise determination of the 
light hadron mass spectrum would unequivocally demonstrate that 
non-perturbative quantities can be extracted 
from a physically relevant quantum field theory.

The recent large scale quenched lattice QCD calculations of the
spectrum of the lightest hadrons 
from CP-PACS~\cite{Aoki:2002fd,Aoki:1999yr} 
are the benchmark for
future calculations. After nearly twenty years of work the deviations
of the predictions of quenched QCD from experiment have been
quantified. The next generation of large scale quenched QCD
calculations (if they are worth doing at all) will probably use the
new fermion operators, such as overlap-Dirac operator, to push to
lighter quark masses below $M_{PS}/M_V$ of 0.4.
The light quark mass region of quenched QCD is full of 
pathologies, such as quenched chiral logs, that will be 
fun to study theoretically, but of limited or no  relevance to 
experiment. Quenched QCD calculations will still be of some
value for many interesting quantities where the 
uncertainty is larger than the inherent error of quenched QCD.

The main challenges in unquenched hadron spectrum calculations is
determining the lattice spacing dependence and reducing the size of
the quark masses used in the calculations. The unquenched calculations
of the CP-PACS collaboration~\cite{AliKhan:2001tx} have shown that at
least three lattice spacings will be needed to obtain high
``quality'' results.  Lattice QCD calculations with 2+1 flavours of
quarks are starting to produce important 
results~\cite{Bernard:2001av}. The effective field theory 
community is increasingly
doing calculations specifically to analyse
data from lattice QCD calculations.

Motivated by the ``new nuclear physics'' experimental programs at
facilities, such as 
Jefferson lab~\cite{Rossi:2003np,Burkert:2002nr,Burkert:2001nv}, 
lattice calculations are
starting to be used to study interesting particles such as the $N^{\star}$s.
The use of more sophisticated interpolating operators and more
advanced statistical techniques (such as maximum entropy techniques)
may allow some information to be obtained on some of the lowest excited
states of hadrons. The determination of the masses 
of excited  states
from lattice QCD would be a big 
step forward for hadron spectroscopy, if this was indeed possible.

The theory of hadronic physics is the ultimate postmodern playground,
as it sometimes seems that the use of a particular model for hadronic
physics, from a mutual incompatible set of possibilities, 
is almost a
matter of personal preference~\cite{Pickering:1984tk}. 
One advantage of lattice QCD
calculations is that they provide qualitative information about
physical mechanisms, that is not directly accessible from experiment.
If as Shuryak~\cite{Shuryak:2001as} claims,
the physical picture
behind the bag model and the instanton liquid model are different,
then only one picture is correct, so one of them must be discarded.
I hope that qualitative lattice QCD calculations can help simplify
the theories behind hadronic spectroscopy by ruling out the underlying
pictures behind certain classes of models.  
The aim of simplifying the theory of hadronic physics is a stated aim
of the current experimental 
program~\cite{Isgur:2000ad,Capstick:2000dk}.

There are number of interesting ``challenges'' for hadron spectrum
calculations beyond the critical task of reducing the errors in
lattice QCD calculations:

\begin{itemize}

\item Can the resonant nature of the $\rho$ meson be determined 
      from lattice QCD?

\item 
  Can improved lattice calculations determine the structure of the
  Roper resonance (first excited state of the nucleon).  For over
  thirty five years, there have been many speculations on the nature
  of the Roper resonance.  Can lattice QCD close this issue?

\item Can the physical mechanism behind the hyperfine splittings
      in mesons and baryons be determined from lattice QCD?

\item Can lattice QCD calculations be used to simplify 
      hadronic physics by ruling out (or perhaps even
      validating) the bag model?

\end{itemize}

\section{ACKNOWLEDGEMENTS}

This work is supported by PPARC.  
I thank members of the UKQCD and MILC collaborations
for discussions.
I thank Chris Michael for reading the manuscript.

\appendix

\section{TECHNICAL DETAILS}  \label{cmn:se:technicalDETAILS}

To outsiders (physicists who live in continuum), the lattice 
gauge theory community most seem like a very inward looking bunch. 
A large fraction of research in lattice gauge is on improving
the methods used in lattice calculations. Improving the methods 
used in lattice calculations leads to smaller and more believable error bars
and hence is a good thing! 
Currently, the biggest improvements in the 
methodology of lattice QCD calculations are coming from ``better``
lattice representations of the continuum Dirac operator.
The ``new'' lattice representations of the Dirac operator
have either reduced lattice spacing dependence or a 
better chiral symmetry.

The importance of the dependence of the results on the lattice
spacing has been stressed through out this review. As the results
of lattice calculations are extrapolated to the continuum,
the calculations would be more precise if the 
lattice spacing dependence of quantities was weak.
The computational cost of 
reducing the lattice spacing used in
lattice QCD calculations 
from equation~\ref{eq:manyFLOPS} is very large,
hence it is advantageous to use coarser lattice 
spacings~\cite{Alford:1995hw,Lepage:1998id}.

A standard technique from numerical analysis is to use derivatives
that are 
closer approximations to the continuum derivatives.  
For example the lattice derivative in 
equation~\ref{eq:better} 
should be more accurate with a larger lattice spacing than
derivative in equation~\ref{eq:OK}.
\begin{equation}
\frac{f(x+a)-f(x-a)}{2a} = 
\frac{df}{dx}+
  O(a^{2})
\label{eq:OK}
 \end{equation}
\begin{equation}
\frac{4}{3}
\{
  \frac{f(x+ a)-f(x- a)}{2a } 
- \frac{f(x+2a)-f(x-2a)}{16a}
\}
= \frac{df}{dx}+ O(a^{4})
\label{eq:better}
\end{equation}
However in a
quantum field theory there are additional complications, such as
the operators in equation~\ref{eq:better} mixing under 
renormalisation.

There is a formalism due to
Symanzik~\cite{Symanzik:1983dc,Symanzik:1983gh} called improvement
where new terms are added to the lattice action that cancel $O(a)$
terms (irrelevant operators) in a way that is 
consistent with quantum field theory.
The required terms in the improved Lagrangian can be
simplified by the use of field redefinitions in the
path integral~\cite{Sheikholeslami:1985ij}. 
A very elegant numerical procedure to improve the Wilson
fermion action has been developed by the 
ALPHA collaboration (see the review~\cite{Luscher:1998pe} by L\"{u}scher).

The Wilson fermion operator in
equation~\ref{eq:wilsonFERMION}
differs from the continuum Lagrangian
by $O(a)$ terms. 
The improvement scheme used in most lattice QCD calculations
with  Wilson fermions is called clover improvement.
The clover term~\cite{Sheikholeslami:1985ij}
is added to the 
Wilson fermion operator in equation~\ref{eq:wilsonFERMION}.
\begin{equation}
S_{f}^{clover}= S_{f}^{W} + c_{SW} \frac{i a \kappa }{2}\sum _{x}
(\overline{\psi _{x}}\sigma _{\nu \mu }F_{\nu \mu }\psi _{x})
\label{eq:cloverTERM}
\end{equation}
where $F_{\nu\mu}$  is the lattice field strength tensor.

If the $c_{SW}$ coefficient computed in perturbation theory is
used then the errors in the results from the lattice calculation
are $O(ag^{4})$. The 
ALPHA collaboration~\cite{Luscher:1997ug} 
have computed $c_{SW}$ to all orders in $g^{2}$
using a numerical technique. The result for $c_{SW}$ from ALPHA
is
\begin{equation}
c_{SW}= \frac{1-0.656g^{2}-0.152g^{4}-0.054g^{6}}{1-0.922g^{2}}
\label{eq:CSWnonpert}
\end{equation}
for $0 < g <1$, where $g$ is the coupling.
The estimate of $c_{SW}$, by ALPHA collaboration, agrees with the one
loop perturbation theory for $g<1/2$.

The clover improvement program for Wilson fermions has had many
practical successes.  Unfortunately, it is 
computationally very costly to reach light quark masses in quenched
or unquenched lattice calculations that use the clover fermion 
operator~\cite{Irving:2002fx}. Hence, attention has focused on
also improving the eigenvalue spectrum of the lattice 
representation of the Dirac operator.

The design of fermion operators on the lattice
has a deep connection with chiral symmetry and the
global chiral anomaly. The theoretical complications 
with transcribing the Dirac operator to the lattice are 
reviewed in many places~\cite{Gupta:1997nd,Rothe:1997kp}.
Our understanding of chiral symmetry on the lattice
has recently increased by the rediscovery of the 
Ginsparg-Wilson relation~\cite{Ginsparg:1982bj}
\begin{equation}
M \gamma_5 + \gamma_5 M = a M \gamma_5 M
\label{eq:GWeqn}
\end{equation}
where $M$ is the fermion operator in 
equation~\ref{eq:Faction} at zero mass.
Equation~\ref{eq:GWeqn} smoothly matches
onto the chiral symmetry equation in the continuum
as the lattice spacing is taken to zero.
Lattice fermion operators that obey the Ginsparg-Wilson relation
(equation~\ref{eq:GWeqn}) have a form of 
lattice chiral symmetry~\cite{Luscher:1998pq}.
Explicit solutions, such as overlap-Dirac~\cite{Neuberger:2001nb} 
or perfect 
actions~\cite{Hasenfratz:1998jp}, to
equation~\ref{eq:GWeqn} are known.  Actions that obey the
Ginsparg-Wilson relation are increasingly being used for quenched QCD
calculations~\cite{Hernandez:2001yd}.  
Domain Wall actions, that can loosely be thought of as being
approximate solutions to the Ginsparg-Wilson relation, have been used
in calculations~\cite{Noaki:2001un,Blum:2001xb} of the matrix elements
for the $\epsilon'/\epsilon$.

The main downside of fermion operators that obey the Ginsparg-Wilson
relation is that they are computationally expensive.  In a review of
the literature, Jansen argues~\cite{Jansen:2001fn} 
that overlap-Dirac type operators are
roughly 100 times more expensive computationally than calculations
with standard Wilson fermions. The development of new algorithms
should reduce this difference in computational cost~\cite{Gattringer:2003qx}. 
The various
versions of the overlap-Dirac operator are cheap enough to use for
quenched calculations. I speculate  that there will be an
increasing trend to use overlap-Dirac type operators for quenched
calculations. It will be some time before unquenched calculations,
that are useful for phenomenology,
are performed with overlap-Dirac operators.
 Unquenched calculations of
QCD with domain wall fermions have just started~\cite{Izubuchi:2002pt}.

A more pragmatic development in the design of light fermion actions is
the development of 
improved staggered fermion 
actions~\cite{Bernard:1999xx,Orginos:1998ue}.  
This class of action
is being used for unquenched lattice QCD calculations with very light
quarks (see table~\ref{tb:dynamicalPARAMS}) by the MILC collaboration.
The problem with standard
Kogut-Susskind quarks was that the formalism broke flavour symmetry.
So numerical calculations usually had fifteen pions split by 
a considerable amount from the Goldstone boson pion. 
The new variants of fermion operators
in the staggered formulations have much reduced flavour
symmetry breaking.
The improved staggered quark formalism is quite ugly compared to 
actions that are solutions of the Ginsparg-Wilson relation,
but lattice QCD is a pragmatic subject and utility wins out 
over beauty.
It is not understood why calculations using improved staggered quarks 
are much faster~\cite{Gottlieb:2001cf}
 than calculations using
Wilson fermions~\cite{Lippert:2002jm}.


\end{document}